\RequirePackage{rotating}
\documentclass[fleqn,usenatbib]{mnras}

\usepackage[T1]{fontenc}
\usepackage{units}

\DeclareRobustCommand{\VAN}[3]{#2}
\let\VANthebibliography\thebibliography
\def\thebibliography{\DeclareRobustCommand{\VAN}[3]{##3}\VANthebibliography}

\usepackage{graphicx}	% Including figure files
\usepackage{amsmath}	% Advanced maths commands
\usepackage{amssymb}	% Extra maths symbols
\usepackage{newtxtext,newtxmath}
\usepackage{multirow}
\usepackage{lscape}
\usepackage{makecell}
\usepackage{subfigure}
\usepackage{float}
\usepackage{newtxtext,newtxmath}
\usepackage{lineno}
\setcitestyle{notesep={ }}
\makeatletter
\newcommand{\labtext}[2]{%
  \@bsphack
  \csname phantomsection\endcsname % in case hyperref is used
  \def\@currentlabel{#1}{\label{#2}}%
  \@esphack
}

\usepackage{xcolor}
\usepackage{pifont}

\newcommand{\redcheck}{\textcolor{green!50!black}{\ding{51}}} % ✓
\newcommand{\redxmark}{\textcolor{red}{\ding{55}}} %

\setcitestyle{notesep={}}

\makeatother
\title[Gamma-Ray Periodicity Revision]{Gamma-Ray Periodicity in Jetted AGN: Revisiting Periodicity Candidates with $>$17 years of \textit{Fermi}-LAT Data}
\author[P. Pe\~nil et al.]{
P. Pe\~nil,$^{1}$\thanks{E-mail: ppenil@clemson.edu}
A. Rico,$^{1}$\thanks{E-mail: aricoro@clemson.edu}
A. Dom\'inguez,$^{2}$\thanks{E-mail: alberto.d@ucm.es}
M. Ajello,$^{1}$
S. Buson,$^{3,4}$
S. Adhikari,$^{1}$
\\
$^{1}$Department of Physics and Astronomy, Clemson University, Kinard Lab of Physics, Clemson, SC 29634-0978, USA\\
$^{2}$IPARCOS and Department of EMFTEL, Universidad Complutense de Madrid, E-28040 Madrid, Spain\\
$^{3}$Julius-Maximilians-Universität Würzburg, Fakultät für Physik und Astronomie, Emil-Fischer-Str. 31, D-97074 Würzburg, Germany\\
$^{4}$Deutsches Elektronen-Synchrotron DESY, Platanenallee 6, 15738 Zeuthen, Germany\\
}

\date{Accepted 2026 June 10. Received 2026 June 09; in original form 2026 May 18}

\pubyear{2026}

\begin{document}
\label{firstpage}
\pagerange{\pageref{firstpage}--\pageref{lastpage}}
\maketitle
% Abstract of the paper
\begin{abstract}
We reanalyze previously reported $\gamma$-ray periodicity candidates in jetted active galactic nuclei using more than 17 years of \textit{Fermi}-LAT observations. The updated data provide a robust test of whether earlier results correspond to persistent periodic behavior, to transient quasi-periodic oscillations (QPOs), or to fluctuations produced by stochastic variability. We apply complementary timing methods, including the Generalized Lomb--Scargle periodogram, Phase Dispersion Minimization, and Singular Spectrum Analysis, and estimate the test statistics using artificial light curves generated from different variability models, including simple and bending power-law power spectral densities, as well as ARIMA and ARFIMA autoregressive models. We find that most previously proposed candidates in literature are not confirmed when the longer baseline is considered, indicating that many reported periods were likely driven by limited temporal coverage and red-noise variability or transient QPO-like features rather than persistent periodic behavior. Only eight sources retain hints of periodic behavior at the $\geq$2$\sigma$ local test statistics. Among them, PG~1553+113 and S5~1044+71 remain the most significant cases, with local test statistics above 3$\sigma$, and with a global significance consistent with $\sim$0$\sigma$ (because of the large trial factor). In addition, we assess predictions from previous studies as an independent test of the proposed periods and find that some are consistent with the new observations.
\end{abstract}

\begin{keywords}
BL Lacertae objects: general - galaxies: active
\end{keywords}
\section{Introduction} \label{sec:intro}
Active galactic nuclei (AGN) hosting relativistic jets represent some of the most energetic and variable systems in the Universe. In these jetted AGN, non-thermal emission produced in collimated outflows dominates the observed radiation over a wide range of wavelengths, from radio to $\gamma$ rays \citep[e.g,][]{blandford_2019, otero_2023, penil_mwl_pg1553}. Additionally, these jets exhibit intrinsic variability, as $\gamma$-ray observations with the \textit{Fermi} Large Area Telescope \citep[LAT,][]{fermi_lat} have revealed through flux variations, spanning timescales from days to decades  \citep[e.g.,][]{sarkar_curved_jet, ren_s5_1044+71}. This variability is commonly described as stochastic and dominated by red-noise processes, reflecting the complex and turbulent nature of relativistic jets \citep[][]{vaughan_power_law, covino_negation, penil_2025_4fgl}. 

Despite this predominantly stochastic behavior, a growing number of studies have reported evidence for periodic patterns in the light curves (LCs) of jetted AGN, particularly in the $\gamma$-ray band \citep[e.g.,][]{ackermann_pg1553, prokhorov_set, ren_s5_1044+71}. In \citet{penil_2020}, we identified 24 periodicity candidates among a sample of $\sim$400 AGN, 19 of which were reported for the first time. Subsequently, we revisited these 24 candidates by extending the \textit{Fermi}-LAT time span by adding three additional years of observations, in order to assess the stability of the observed periodicity and their associated significances as more data became available \citep[][]{penil_24candidates_2025}. In \citet{penil_2025_4fgl}, we further expanded the search to a larger sample of $\sim$1500 AGN using the same 12-year \textit{Fermi}-LAT data set; however, no additional statistically significant periodicity candidates were identified. Finally, in \citet{alba_ssa}, a complementary analysis of the brightest $\sim$500 AGN was carried out using a novel technique, providing an independent assessment of periodic behavior in $\gamma$-ray LCs, expanding the number of periodicity candidates in literature by a factor of two.

Several physical scenarios have been proposed to explain the periodic variability observed in jetted AGN. These include models are based on  orbital motion of emitting plasma blobs within the jet, potentially driven or modulated by magnetic fields \citep[e.g.,][]{mohan_blobs}; twisted, helical, or otherwise inhomogeneous jet structures that lead to periodic changes in the Doppler factor \citep[e.g., ][]{raiteri2017}; and jet precession, commonly referred to as the lighthouse effect, which produces periodic modulation of the observed emission \citep[][]{camenzind1992}. Furthermore, modulation of the accretion rate has also been proposed as a possible origin of the periodic emissions \citep[e.g.,][]{farris_2014}. In addition, the detection of periodic emission may point to the presence of a supermassive black hole binary (SMBHB), either through orbital modulation of the accretion flow or induced jet precession \citep[e.g.,][]{sillanpaa_oj287, tavani_blazar, sagar_pg1553}.

However, detecting and characterizing periodicity in $\gamma$-ray LCs remains statistically challenging due to several well-known observational and methodological limitations. In particular, stochastic variability dominated by red noise can produce apparent periodic features that are not necessarily physically meaningful; therefore, any periodicity analysis must properly account for the underlying noise process \citep[][]{vaughan_criticism, covino_negation}. Strong flaring events, commonly observed in jetted AGN, can enhance the variability amplitude over limited time intervals, biasing also period searches or mimicking transient periodic behavior \citep[][]{penil_flares_2025}. Furthermore, uneven sampling and data gaps can affect the temporal analysis, thereby distorting potentially genuine periods and the corresponding statistical significance \citep[][]{penil_gaps_2025}.

In this paper, we analyze 52 jetted AGN exhibiting periodicity hints at a local significance level of $\geq$2$\sigma$, as reported in \citet{alba_ssa} and \citet{penil_24candidates_2025, penil_2025_4fgl}, using 17.3 years of \textit{Fermi}-LAT observations. This represents an increase of about 50\% in the temporal baseline relative to our previous systematic studies. The extended time coverage potentially increases the number of observed cycles, reduces uncertainties on the inferred periods, enhances sensitivity to multi-year signals, and improves the ability to distinguish genuine periodic behavior from stochastic variability.

Additionally, with these new observations, we test the predictions presented in \citet{alba_ssa} and \citet{penil_2025_4fgl} for the periodicity candidates reported in those studies. More generally, the periodic behavior observed in jetted AGN emission can be broadly classified as either quasi-periodic oscillations (QPOs) or genuine periodicity. QPOs are characterized by irregular or transient periodic behavior with an inherent stochastic component, which limits the ability to reliably predict future oscillations. In contrast, genuine periodicity implies a stable and coherent signal that allows for the prediction of future cycles. The extended temporal coverage provided by the new data therefore enables a direct test of our previous predictions: the persistence of periodic peaks supports earlier candidates, while their absence allows us to reject them or classify them as transient cases.

This paper is organized as follows. In $\S$\ref{sec:fermidata}, we describe the source sample analyzed in this work. In $\S$\ref{sec:methodology}, we present the periodicity-search methodology, including the methods employed and the procedure used to estimate the test statistics. In $\S$\ref{sec:calibration_signi}, we introduce the corrections applied to these test statistics. In $\S$\ref{sec:arima_arfima}, we investigate ARIMA and ARFIMA models, both to search for periodic signals and to characterize the stochastic variability, which is then used to estimate the test statistics. In $\S$\ref{sec:periodicity_results}, we present the periodicity results and discuss their main implications. In $\S$\ref{subsec:ssa_significance_tests}, we discuss the singular spectrum analysis decomposition properties and their impact on the resulting test statistics. In $\S$\ref{sec:predictions_results}, we evaluate predictions reported in previous works and test whether the forecasted events occur within the updated time span. In $\S$\ref{sec:periodicy_candidates}, we summarize the periodicity candidates identified in this study. Finally, in $\S$\ref{sec:summary}, we provide a summary of the main findings.

\section{Gamma-ray Sample} \label{sec:fermidata}
In this work, we reanalyze the sample of 52 jetted AGN using 17.3 years \textit{Fermi}-LAT observations. The sample is listed in Table \ref{tab:candidates_list}. It includes 19 BL Lac objects (BLLs; 36.6\%), 28 flat-spectrum radio quasars (FSRQs; 53.8\%), four blazar candidates of uncertain type (BCUs; 7.7\%), and one narrow-line Seyfert 1 galaxy (NLSy1; 1.9\%).  

The previous reported periods span the range 1.7–5 years, with most falling between 2 and 3 years ($\approx$50\%). Finally, the LCs in the sample exhibit a wide range of upper-limit (UL). According to the results of \citet{penil_gaps_2025}, periodicity analyses are significantly affected when the fraction of gaps (ULs) exceeds 50\%. We therefore adopt this value as a threshold and exclude sources with UL fractions above this limit. As a result, three objects are discarded from our study, thus, our sampled consists of 49 sources.

We use $\gamma$-ray data from the open-access \textit{Fermi}-LAT Light Curve Repository \citep[][]{fermi_repository}\footnote{\url{https://fermi.gsfc.nasa.gov/ssc/data/access/lat/LightCurveRepository/about.html}}, which provides uniformly processed LCs spanning more than 17 years of \textit{Fermi}-LAT observations. The repository LCs are produced with the standard \textit{Fermi}-LAT analysis tools (e.g.,  \texttt{P8R3\_SOURCE\_V3} instrument response functions) through a maximum-likelihood analysis \citep[][]{abdo_fermi_2009}. In each time bin, the source of interest is modeled together with the Galactic and isotropic diffuse backgrounds and the neighboring $\gamma$-ray sources in the region of interest. For this analysis, we select 30-day binned LCs, a binning choice that offers an optimal compromise between sensitivity to multi-year periodicities and computational efficiency. Such a bin size effectively suppresses short-term stochastic variability while preserving long-term information relevant for periodicity searches and reducing the number of upper limits. Furthermore, we adopt a fixed photon spectral index and require a minimum detection significance of 2$\sigma$ in each time bin.

\begin{table*}
\centering
\caption{List of periodicity results for the jetted AGN analyzed in this work. We include their \textit{Fermi}-LAT name, equatorial coordinates, AGN type, redshift, and association name. We obtain this information from the 4FGL-DR2 catalog \citep[][]{4fgl_dr2}. We include the percentage of upper limits (UL) in the LC. We also include the period (expressed in years) and the associated test statistics obtained in the previous works \citet{alba_ssa, penil_24candidates_2025, penil_2025_4fgl}. Additionally, we include the period (and test statistics) determined through the analysis in this work. The candidates are sorted according to the median of their test statistics \citep[][]{penil_2020, penil_24candidates_2025, penil_2025_4fgl}, for the bending-power law described by Equation \ref{eqn:bpl1}, as detailed in the results presented in Table \ref{tab:periodicity_results}. Note that this median significance does not have an actual statistical meaning; it is used as an arbitrary way that combines all test statistics to sort the candidates. \label{tab:candidates_list}}
{%
\begin{tabular}{ccccccccccc}
\hline
\hline
4FGL Name & RAJ2000 & DecJ2000 & Type & Redshift & Association Name & UL & Previous Results & Period This Work\\
 &  &  &  &  & & (\%) & [year] & [year] & \\
 \hline
\hline
J0043.8+3425 & 10.9678 & 34.42687 & fsrq & 0.966 & GB6 J0043+3426 & 11.5\% & 1.9 (4.1-2.7$\sigma$)  & 2.3$\pm$0.3 (0.8$\sigma$) \\
J0112.1+2245 & 18.0294 & 22.7515 & bll & 0.265 & S2 0109+22 & 0.5\% & 2.6-2.4 (4.1-2.1$\sigma$) & 2.7-1.7$\pm$0.2 (1.1$\sigma$) \\
J0137.0+4751 & 24.260 & 47.864 &  fsrq & 0.859 & OC 457 & 26.6\% & 1.8 (4.8$\sigma$) & 2.0-1.7$\pm$0.2 (0.9$\sigma$) \\
J0210.7$-$5101  & 32.695 & -51.022 & fsrq & 1.003 & PKS 0208$-$512 & 1.4\% & 2.5-3.8 (4.5-0.1$\sigma$) & 2.0-3.9$\pm$0.3 (0.8$\sigma$) \\
J0217.8+0144 & 34.4621 & 1.7346 & fsrq & 1.715 & PKS 0215+015 & 15.7\% & 3.4 (2.2$\sigma$) & 3.3$\pm$0.3 (2.0$\sigma$) \\
J0222.6+4302  & 35.670 & 43.036 & bll &  0.444 & 3C 66A & 0.5\% & 2.3 (4.6$\sigma$) & 2.2$\pm$0.3 (0.8$\sigma$) \\
J0252.8$-$2219  & 43.201 & -22.320 & fsrq & 1.419 & PKS 0250$-$225 & 16.2\% & 1.2 (2.7-1.7$\sigma$) & 1.2$\pm$0.4 (0.9$\sigma$) \\
J0303.4$-$2407 & 45.8625 & -24.12074 & bll & 0.266 & PKS 0301$-$243 & 3.3\% & 2.0 (3.3-2.0$\sigma$) & 2.1$\pm$0.3 (0.8$\sigma$) \\
J0405.6$-$1308 & 61.419 & -13.144 & fsrq & 0.571 & PKS 0403$-$13 & 56.6\% & 1.7 (2.3$\sigma$) & -- \\
J0407.0$-$3826 & 61.7627 & $-$38.4394 & fsrq & 1.285 & PKS 0405$-$385 & 16.2\% & 3.0-3.1 (4.8-2.5$\sigma$) & 2.6$\pm$0.4 (0.9$\sigma$) \\
J0427.3$-$3900  & 66.826 & -39.01 & bcu & -- & PMN J0427$-$3900 & 49.1\% & 2.8 (3.9$\sigma$) & 1.0-2.2$\pm$0.4 (0.7$\sigma$) \\
J0428.6$-$3756 & 67.1726 & -37.94081 & bll & 1.11 & PKS 0426$-$380 & 0.0\% & 3.5 (4.8-2.1$\sigma$) & 3.5$\pm$0.5 (1.6$\sigma$) \\ 
J0449.4$-$4350 & 72.3604 & -43.83719 & bll & 0.205 & PKS 0447$-$439 & 0.0\% & 1.9 (4.0-2.1$\sigma$) & 1.2-4.1$\pm$0.5 (1.1$\sigma$) \\ 
J0457.0$-$2324 & 74.2609 & -23.41384 & fsrq & 1.003 & PKS 0454$-$234 & 0.0\% & 3.6 (4.8-2.8$\sigma$) & 3.4$\pm$0.5 (1.8$\sigma$) \\
J0501.2$-$0158  & 75.302 & -1.975 & fsrq & 2.291 & S3 0458$-$02 & 12.8\% & 2.3-3.8 (4.8-1.4$\sigma$) & 4.0$\pm$0.4 (2.0$\sigma$) \\        
J0521.7+2113 & 80.4437 & 21.21369 & bll & 0.108 & TXS 0518+211 & 0.5\% & 2.8-3.1 (4.8-2.6$\sigma$) & 3.1$\pm$0.3 (2.0$\sigma$) \\
J0526.2$-$4830 & 81.5714 & -48.5151 & fsrq & 1.3 & PKS 0524$-$485 & 19.4\% & 2.0 (4.8-2.4$\sigma$) & 2.0$\pm$0.2 (0.9$\sigma$) \\
J0533.3+4823 & 83.3313 & -55.8247 & bcu & -- & PMN J0533$-$5549 & 24.6\% & 2.8 (2.0$\sigma$) & 2.1-4.0$\pm$0.4 (0.5$\sigma$) \\
J0719.3+3307  & 109.840 & 33.123 & fsrq & 0.779 & B2 0716+33 & 18.4\% & 2.3 (2.2$\sigma$) & 2.1$\pm$0.3 (0.5$\sigma$) \\
J0721.9+7120 & 110.4888 & 71.34127 & bll & 0.127 & S5 0716+714 & 0.0\% & 2.9 (4.5-2.8$\sigma$) & 2.5$\pm$0.3 (1.4$\sigma$) \\
J0811.3+0146 & 122.8641 & 1.77344 & bll & 1.148 & OJ 014 & 9.0\% & 4.6-4.1 (4.8-2.8$\sigma$) & 4.2$\pm$0.4 (2.5$\sigma$) \\
J0948.9+0022  & 147.244 & 0.372 & nlsy1 & 0.585 & PMN J0948+0022 & 29.9\% & 1.4 (4.0$\sigma$) & 1.2$\pm$0.5 (0.6$\sigma$) \\
J1007.6$-$3332  & 151.912 & -33.543 & fsrq & 1.837 & PKS 1005$-$333 & 60.1\% & 2.6 (4.8$\sigma$) & -- \\
J1033.1+4115  & 158.275 & 41.262 & fsrq & 1.117& S4 1030+41 & 42.1\% & 2.3 (2.0$\sigma$) & 2.4$\pm$0.3 (1.8$\sigma$) \\
J1033.9+6050 & 158.4849 & 60.8493 & fsrq & 1.401 & S4 1030+61 & 16.2\% & 2.9 (2.0$\sigma$) & 4.1$\pm$0.5 (1.5$\sigma$) \\
J1044.6+8053 & 161.1638 & 80.8941 & fsrq & 1.254 & S5 1039+81 & 40.5\% & 3.5 (2.0$\sigma$) & 1.2-3.4$\pm$0.5 (0.6$\sigma$) \\
J1048.4+7143 & 162.1067 & 71.7297 & fsrq & 1.15 & S5 1044+71 & 5.8\% & 3.1 (4.8-2.5$\sigma$) & 3.1$\pm$0.3 (3.7$\sigma$) \\
J1146.9+3958  & 176.740 & 39.978 & fsrq & 1.089 & S4 1144+40 & 6.6\% & 3.2 (4.8-2.5$\sigma$) & 3.1$\pm$0.2 (2.2$\sigma$) \\
J1222.5+0414  & 185.627 & 4.239 & fsrq & 0.964 & 4C +04.42 & 28.4\% & 2.2 (2.7$\sigma$) & 2.5$\pm$0.5 (1.0$\sigma$) \\
J1223.8+8039  & 185.971 & 80.660 & bll & -- & S5 1221+80 & 25.3\% & 2.7 (3.5$\sigma$) & 2.6$\pm$0.5 (1.2$\sigma$) \\
J1253.2+5301  & 193.307 & 53.017 &  bll & -- & S4 1250+53 & 0.9\% & 2.3 (2.6$\sigma$) & 2.2$\pm$0.3 (1.5$\sigma$) \\ 
J1310.5+3221 & 197.6324 & 32.3547 & fsrq & 0.997 & OP 313 & 14.7\% & 5.7 (2.0$\sigma$) & 5.1-1.3$\pm$0.6 (0.4$\sigma$) \\
J1312.8$-$0425  & 198.217 & -4.420 & fsrq & 0.825 & PKS B1310$-$041 & 38.8\% & 2.4 (3.6$\sigma$) & 4.6-2.3$\pm$0.6 (1.4$\sigma$) \\
J1321.1+2216  & 200.296 & 22.281 & fsrq & 0.943 & TXS 1318+225 & 24.2\% & 1.2 (3.8$\sigma$) & 5.0$\pm$0.4 (1.5$\sigma$) \\
J1427.6$-$3305  & 216.913 & -33.094 & bll & -- & PKS 1424$-$328 & 0.0\% & 1.3 (2.9$\sigma$) & 1.5$\pm$0.1 (1.6$\sigma$) \\
J1454.4+5124  & 223.625 & 51.409 & bll & -- & TXS 1452+516 & 1.9\% & 2.1 (4.2-0.7$\sigma$) & 2.1-5.0$\pm$0.4 (0.5$\sigma$) \\
J1512.8$-$0906  & 228.215 & -9.106 & fsrq & 0.360 & PKS 1510$-$089 & 0.0\% & 1.7 (2.5$\sigma$) & 1.2-3.8$\pm$0.4 (0.9$\sigma$) \\
J1522.1+3144  & 230.545 & 31.740 & fsrq & 1.489 & B2 1520+31 & 12.3\% & 3.4 (4.8$\sigma$) & 3.7-1.2$\pm$0.4 (1.0$\sigma$) \\
J1532.7$-$1319  & 233.197 & -13.326 & bcu & 1.37 & TXS 1530$-$131 & 42.3\% & 1.4 (4.8$\sigma$) & 2.8-1.4$\pm$0.3 (0.6$\sigma$) \\ 
J1555.7+1111 & 238.9316 & 11.18768 & bll & 0.433 & PG 1553+113 & 0.0\% & 2.2 (4.8-4.5$\sigma$) & 2.1$\pm$0.2 (3.7$\sigma$) \\
J1637.7+4717  & 249.434 & 47.291 & fsrq & 0.735 & 4C +47.44 & 58.1\% & 3.2 (4.8$\sigma$) & -- \\
J1649.4+5238 & 252.352 & 52.58336 & bll & -- & 87GB 164812.2+524023 & 46.4\% & 2.8 (2.2$\sigma$) & 3.2$\pm$0.6 (2.0$\sigma$) \\
J1657.7+4808  & 254.438 & 48.137 & fsrq & 1.669 & 4C +48.41 & 33.1\% & 1.5 (4.8$\sigma$) & 5.4$\pm$0.4 (1.1$\sigma$) \\
J1723.6$-$7714  & 260.922 & -77.238 & bcu & -- & PKS 1716$-$771 & 23.7\% & 2.3 (2.9$\sigma$) & 2.2$\pm$0.3 (1.4$\sigma$) \\
J1903.2+5541 & 285.8085 & 55.67557 & bll & -- & TXS 1902+556 & 0.9\% & 3.3 (1.5$\sigma$) & 3.2$\pm$0.3 (1.4$\sigma$) \\
J1913.0$-$8009  & 288.27 & -80.157 & fsrq & 1.756 & PKS 1903$-$80 & 31.7\% & 2.4 (4.8$\sigma$) & 4.0$\pm$0.6 (1.1$\sigma$) \\
J2012.0+4629  & 303.020 & 46.488 & bll & -- & 7C 2010+4619 & 19.9\% & 3.2 (4.2$\sigma$) & 2.6-3.7$\pm$0.7 (0.7$\sigma$) \\
J2139.4$-$4235 & 324.8546 & -42.5895 & bll & -- & MH 2136$-$428 & 1.9\% & 1.8 (4.8-2.5$\sigma$) & 1.9-3.5$\pm$0.3 (1.1$\sigma$) \\
J2158.8$-$3013 & 329.714 & -30.22556 & bll & 0.116 & PKS 2155$-$304 & 0.0\% & 1.7 (4.5-3.3$\sigma$) & 1.7$\pm$0.1 (1.7$\sigma$)\\
J2201.5$-$8339  & 330.379 & -83.663 & fsrq & 1.865 & PKS 2155$-$83 & 39.3\% & 4.3 (4.8$\sigma$) & 2.7-4.7$\pm$0.5 (0.9$\sigma$) \\	
J2202.7+4216  & 330.695 & 42.282 & bll & 0.069 & BL Lacertae & 0.0\% & 2.0 (4.1$\sigma$) & 5.6$\pm$0.5 (0.9$\sigma$) \\ 
J2236.3+2828  & 339.096 & 28.483 & fsrq & 0.790 & B2 2234+28A & 9.5\% & 2.2 (4.1$\sigma$) & 1.4-2.9$\pm$0.4 (0.5$\sigma$) \\   
\hline
\hline
\end{tabular}%
}
\end{table*}

\section{Methodology} \label{sec:methodology}
This section outlines the analytical framework adopted to identify and assess periodic signals. The methodology is specifically designed to account for irregular sampling, resulting of removing the ULs presented in each LCs, and stochastic variability. The analysis is structured around two main components: (i) the identification of candidate periodicities using a set of complementary analysis techniques, and (ii) the quantitative evaluation of their statistical significance through well-defined test statistics.

\subsection{Periodicity Analysis Methods}\label{sec:methods_methodology}
To perform the periodicity analysis search, we employ three different methods. The first of them is the Generalized Lomb-Scargle Periodogram \citep[GLSP;][]{lomb_gen}. The GLSP is an extension of the classical Lomb-Scargle Periodogram \citep[][]{lomb_1976, scargle_1982} that explicitly incorporates measurement uncertainties into the computation of the periodogram. The second is the Phase Dispersion Minimization \citep[PDM;][]{pdm_stellingwerf}. PDM searches for dominant frequencies in a signal evaluating frequencies through the scatter of phase-folded data. Finally, we apply the pre-processing technique Singular Spectrum Analysis \citep[SSA;][]{ssa_greco, SSA_algorithm}. SSA decomposes the time series into a set of components, which include one associated with the oscillatory structure of the LC, whereas others capture the noise. By reconstructing the dominant oscillatory component and filtering out noise or distorting factor as flares or trends, SSA provides a “cleaned’’ LC on which we apply the LSP in order to obtain the source period, following the approach of \citet{alba_ssa}. The SSA decomposition relies on one parameter, the window length ($WL$)\footnote{In \citet{ssa_greco} and 
\citet{alba_ssa}, the window length is represented by $L$; here, we adopt $WL$ to improve clarity for the reader}, which sets the maximum variability timescale that can be resolved. For periodicity studies, commonly adopted values are a $WL$ of $0.4$ (expressed as a fraction of the total time series length) used in \citet[][]{alba_ssa}. This configuration has been shown to be well suited for isolating long-term oscillating variability while mitigating the impact of stochastic events such as strong flares and the presence of data gaps \citep[][]{penil_flares_2025, penil_gaps_2025}. These values should be considered as guidelines rather than fixed assumptions, as the optimal choice may vary depending on the intrinsic properties of a given LC, including its noise characteristics, variability amplitude, and sampling pattern \citep[][]{alba_ssa, penil_gaps_2025}.

\subsection{Test Statistics Estimation}\label{sec:test_statistics}
The test statistics associated with the inferred periods identified by the methods described above are re-estimated using artificial and stochastic LCs that reproduce the statistical properties of the original data. Specifically, following the procedure of \citet{emma_lc}, we generate 150,000 simulated LCs that preserve the original sampling pattern, power spectral density (PSD), and probability density function for each source. 

\begin{figure}
	\includegraphics[width=\columnwidth]{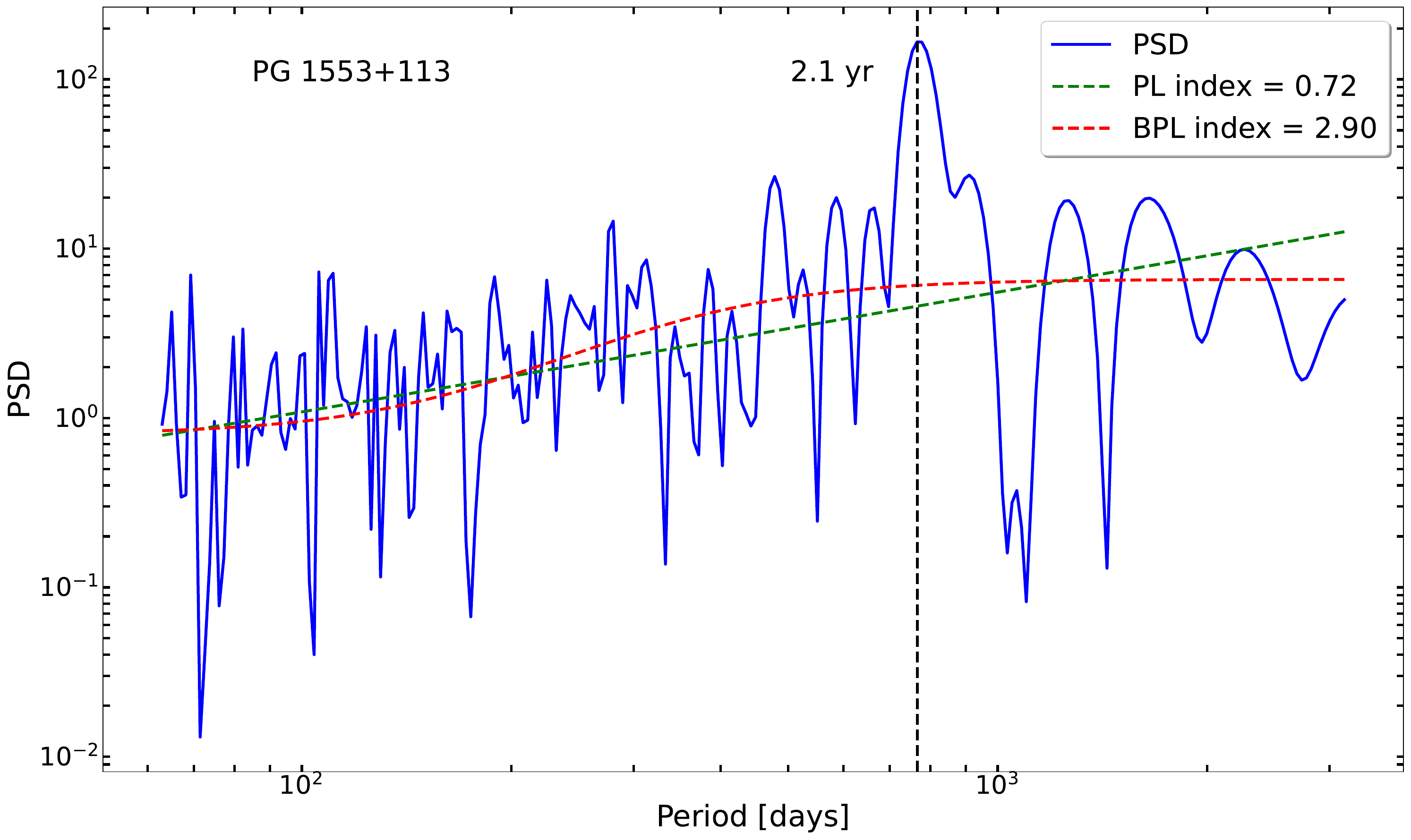}
	\caption{PSD of the PG 1553+113 LC as a function of period. The blue curve shows the PSD, while the green and red dashed curves represent the best-fit PL and BPL models, respectively. The vertical dashed black line marks the dominant peak at 2.1 yr. The fitted spectral indices are reported in the legend.}
	\label{fig:psd_example}
\end{figure}

As the underlying PSD model, we adopt a bending power-law (BPL), which provides a more realistic description of blazar variability over timescales ranging from weeks to years \citep{chakraborty_bending_power_law} and has been shown to improve the modeling of AGN PSDs \citep[][]{penil_2025_4fgl}. The BPL is fitted using the formulation of \citet{chakraborty_bending_power_law},
\begin{equation} \label{eqn:bpl1} 
  P(\nu) = A \left( 1 + \left\{ \frac{\nu}{\nu_{b}} \right\}^{\beta} \right)^{-1} + C, 
\end{equation} 

where $A$ is the normalization, $\nu_{b}$ represents the break frequency, $\beta$ denotes the spectral index, and $C$ is the Poisson (white) noise (see Figure \ref{fig:psd_example}). These parameters are estimated using a combination of maximum-likelihood optimization and Markov Chain Monte Carlo (ML–MCMC) sampling.\footnote{Utilizing the Python package emcee.}

In addition, we use a simple power-law (PL) model as a complementary approach to estimate the test statistics. This model provides a convenient baseline description of the stochastic variability, as it captures the red-noise behavior commonly observed in blazar LCs \citep[][]{vaughan_power_law}.  

\begin{equation}\label{eqn:pl_adhoc}
P(\nu) = A\,\nu^{-\alpha} + C,
\end{equation}

where $A$ is the normalization, $\alpha$ is the spectral index, and $C$ represents the Poisson noise level (see Figure \ref{fig:psd_example}). The PL parameters are estimated using the same fitting procedure adopted for the BPL model. 

\section{Correction and the Calibration of the Test Statistics}\label{sec:calibration_signi}
We quantify how uncertainties in the fitted PSD parameters (for both the PL and BPL models) propagate into the derived test statistics, following the approach of \citet[][]{benkhali_power_spectrum}. For each source, we recompute the test statistics under three parameter realizations: the best-fit value, and the best-fit value shifted by $\pm$uncertainty. These $\pm$uncertainty perturbations are applied to each PSD parameter to assess the sensitivity of the test statistics to plausible variations in the underlying noise model. We restrict this analysis to the candidate periodic sources with test statistics $\gtrsim 2\sigma$ which are reported in $\S$\ref{sec:periodicy_candidates}. 

Table~\ref{tab:correction_noise_models} shows that propagating the uncertainties of the adopted variability models to generate the artificial LCs, it results in only modest changes in the inferred test statistics, although the impact depends on both the source and the PSD model. In general, the BPL-based corrections are more stable than the PL-based ones, with variations typically limited, indicating that once a characteristic bend is included, the test statistics estimates become less sensitive to small changes in the PSD parameters. By contrast, the PL modeling can lead to larger shifts, particularly for sources such as OJ~014 and S4~1144+40. This behavior is consistent with the fact that under a pure PL model, small changes in the slope and normalization directly modify the amount of low-frequency power which can result in mimicking a previous periodic pattern, and therefore alter the chance probability of detecting spurious long-term periods \citep[][]{benkhali_power_spectrum}. The most robust cases are PG~1553+113 and S5~1044+71, which remain significant under both PL and BPL models.

\subsection{Global Significance}\label{sec:global_significance}
We correct for the look-elsewhere effect by converting the test statistics ("local" significance) obtained as explained in \S\ref{sec:test_statistics} for each method into a "global" significance. In this context, the local p-value, quantifies the probability of obtaining a fluctuation at a pre-specified period, whereas the global p-value, accounts for the probability of finding a comparable excess anywhere within the scanned period range \citep{Gross_Vitells_Trial}. Assuming
$N$ effectively independent trials, we compute:
\begin{equation}\label{eq:global}
p_{\mathrm{global}}=1-(1-p_{\mathrm{local}})^{N},
\end{equation}

The effective trial factor $N$ reflects two distinct sources of multiplicity. First, we search for periodic behavior across the full sample that we analyzed in \citet{penil_2025_4fgl}, $B=$1492, without prior knowledge of which objects-if any-should exhibit periodicity. Second, for each source we scan a range of periods, and the absence of prior information on the true period introduces an additional multiplicity associated with the number of effectively independent frequencies, $P$. We therefore write
\begin{equation}
\label{eq:trial}
 N=P \times B,    
\end{equation}

We search for periods in the range 1 to 6 yr \citep[][]{alba_ssa, penil_2025_4fgl}, considering over a grid of 100 frequencies.

\begin{figure}
	\includegraphics[width=\columnwidth]{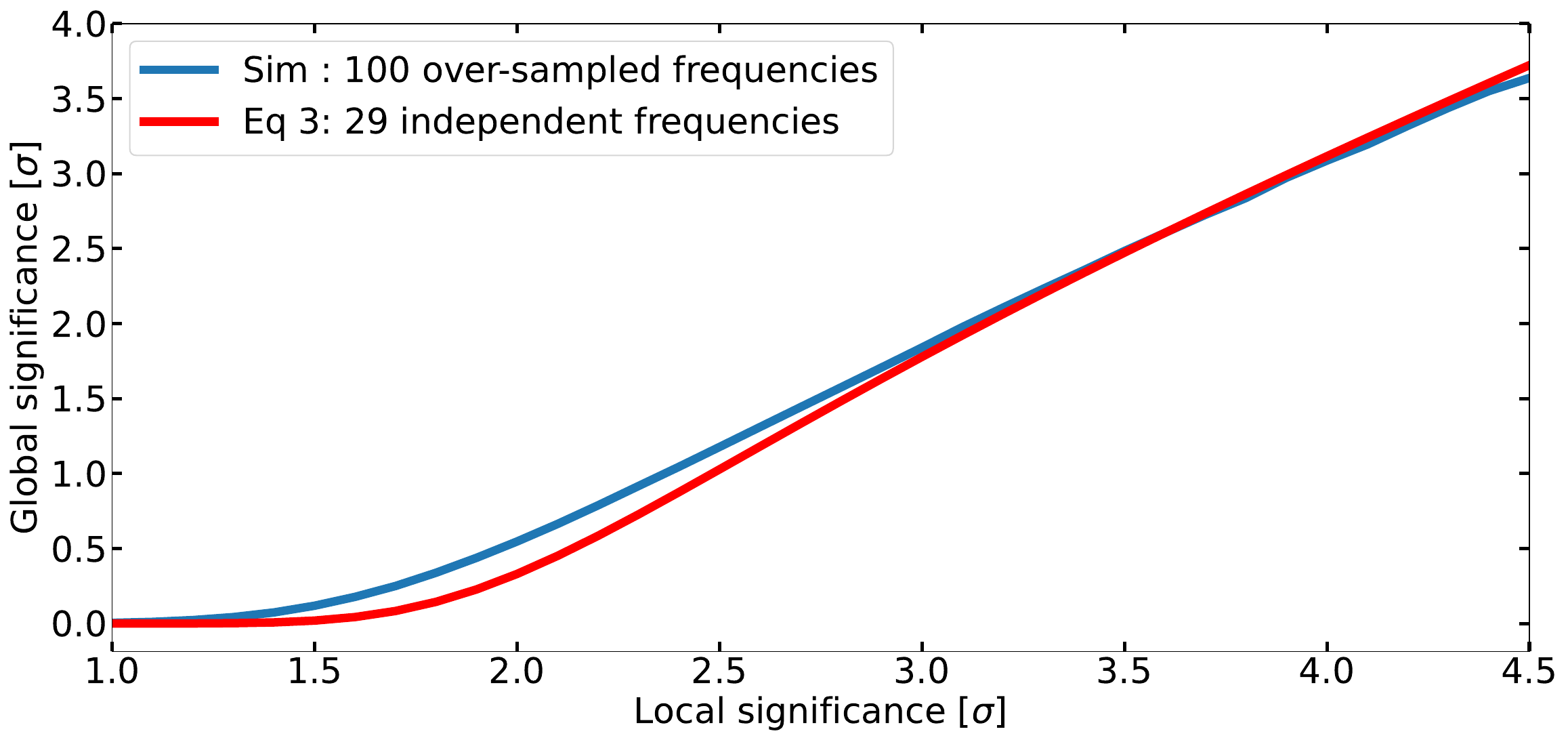}
		\caption{Estimation of the independent frequencies {\it P} according to the the experimental relationship between local-global significance. {\it Eq. 3} represents the results obtained by applying Eq. \ref{eq:trial} for a particular number of independent frequencies, reporting 29. The $P$ is chosen to adjust the test statistics of $\approx$3.5$\sigma$, thereby improving the correction of the most significant blazars listed in Table \ref{tab:final_candidates}.} \label{fig:trials}
\end{figure}

We estimate $P$ using Monte Carlo simulations following the procedure of \citet[][]{penil_24candidates_2025, penil_2025_4fgl}. Using $10^{8}$ simulated LCs generated with the Timmer-Koenig method \citep[][]{timmer_koenig_1995}, we empirically calibrate the relation between local and global significance (blue curve in Figure~\ref{fig:trials}). We then vary $P$ in Equation~\ref{eq:trial} to obtain the best agreement with this empirical relation. The adopted value of $P$ is chosen such that the local significances are adjusted to yield a global threshold of $\approx$3.5$\sigma$, which improves the identification of the most significant jetted AGN in the sample (see $\S$\ref{sec:periodicy_candidates}). 

Figure~\ref{fig:trials} indicates that an effective value of $P$=29 provides the closest match to the empirically calibrated local-to-global significance relation. Adopting this value, and accounting for the $B=$1492 sources in the survey, the corresponding effective trials factor becomes $N$$=$$PB$$=$29$\times$1492$=$43268. When the look-elsewhere correction is applied using this $N$ (Equation~\ref{eq:global}), for the most significant periods, $\sim$3.5$\sigma$, the "global" significance is $\approx$0$\sigma$ for the candidates of $\S$\ref{sec:periodicy_candidates}.

The global significance close to 0$\sigma$ indicates that such peaks are not unexpected once the full search space is considered. For the adopted trial factor, the expected number of red-noise fluctuations with comparable local significance in the full survey is $\simeq$10. Therefore, these signals can be interpreted as fluctuations produced by the red-noise null hypothesis within the survey. Consequently, additional tests are required to assess whether the candidate periods are genuine, including checks of their consistency with temporal predictions from the obtained periods.

\section{Autoregressive Models}\label{sec:arima_arfima}
In addition to the traditional models based on fitting the PSD to describe the variability properties (e.g., PL and BPL), there are alternative approaches for characterizing the $\gamma$-ray variability of jetted AGN. In particular, autoregressive models provide a framework for describing variability in astronomical time series \citep[][]{scargle_1981, caceres_arima}. In these models, the value of a process at a given time is expressed as a linear combination of its past values and a stochastic term, allowing temporal correlations to be quantified directly in the time domain. 

The simplest case is given by ARMA($p$, $q$) models, which describe short-memory processes with correlations that decay exponentially over time \citep[][]{feigelson_arima}. In this framework, the autoregressive component of order (complexity) $p$ accounts for correlations between successive observations by expressing the current flux as a function of a finite number of past measurements. This term captures local persistence in the time series, such as gradual flux changes or correlated sequences of points associated with evolving variability features. The moving-average component of order, $q$, accounts for correlations introduced by recent stochastic perturbations, describing how the impact of short-lived variability episodes propagates over a limited number of time bins \citep[][]{feigelson_arima}. ARMA models are well-suited for modeling stationary variability dominated by a limited range of timescales. However, they are unable to reproduce the long-range dependence and red-noise-dominated variability of the jetted AGN LCs \citep[][]{scargle_1981,feigelson_arima}. 

To face these ARMA limitations, alternative autoregressive models have been used in the analysis of astronomical LCs. The autoregressive integrated moving-average (ARIMA) model incorporates a differencing component \citep[][]{chatfield_arima, feigelson_arima}. In $ARIMA(p,d,q)$ models, the parameter $d$ (degree of differencing) allows for integrated, random-walk-like behavior that produces long-term drifts in the LCs. The case $d$=0 indicates stationarity, meaning that the LC is described by a constant mean and variance over time. 

While ARIMA models assume that temporal correlations decay exponentially, many astrophysical time series exhibit long-memory behavior, characterized by correlations that persist over a wide range of timescales \citep[][]{vaughan_power_law}. To account for this behavior, autoregressive fractionally integrated moving-average (ARFIMA) model extends the ARIMA framework by allowing the differencing parameter $d$ to take non-integer values \citep[][]{feigelson_arima}. The fractional differencing in ARFIMA models produces a slow, power-law decay of correlations, making them particularly suitable for describing long-range dependence and self-similar variability \citep[][]{penil_2025_4fgl}. This property is especially relevant for jetted AGN LCs, where stochastic variability is often well described by PL or BPL noise processes \citep[][]{vaughan_power_law, chakraborty_bending_power_law}. As a result, ARFIMA models provide a framework for modeling intrinsic variability and for generating realistic artificial LCs used in significance testing of periodic oscillations \citep[][]{tarnopolski20}. 

\subsection{Application to Irregular Distributed LCs}
ARIMA and ARFIMA models are formally defined for regularly sampled time series. Nevertheless, as discussed by \citet[][]{feigelson_arima}, moderately irregular astronomical time series may be treated as evenly spaced sequences with missing observations, provided that the fraction of gaps is limited and does not dominate the temporal structure of the data. Under these conditions, the underlying stochastic process can still be reliably characterized without resorting to interpolation, which may otherwise introduce artificial correlations.

To quantify the regime in which the adopted autoregressive model remains valid for our cadence and period range, we performed a series of injection tests. The periodicity search in this work targets timescales of approximately 1–6 years, corresponding to approximately 12–60 bins for the adopted 30-day binned LCs. At these scales, an accurate characterization of red and long-memory stochastic variability is essential for a reliable periodicity analysis.

Starting from gap-free LCs in our sample, we introduced random fractions of missing data (0-30\%) and generated multiple realizations. For each case, we evaluated whether the temporal correlation structure of the LC is preserved after data removal. This was assessed by comparing correlations between flux measurements separated by lags ranging from months to several years with those of the original gap-free data. In addition, we also examined how the long-term variability of the LC changed as the gap fraction increased, without adopting any specific model for the underlying noise process.

These tests explicitly verify that the temporal correlation properties assumed by both ARIMA and ARFIMA models, ranging from short-memory to long-memory behavior, are not artificially suppressed or distorted by moderate levels of missing data. We find that the correlation structure remains largely unchanged for gap fractions up to approximately 10–15\% of the LC. Based on these results, we restrict the ARIMA/ARFIMA analysis to LCs with $\leq$10\% missing data and without extended contiguous gaps ($<$6 consecutive bins), ensuring that the ARIMA/ARFIMA models preserve the statistical properties. After applying these constraints to our sample, the remaining subsample analyzed with ARIMA/ARFIMA models consists of 18 sources. 

\subsection{AGN Behavior Modeling}
We analyze the variability of the 18 sources by modeling their LCs using the autoregressive models introduced above. Model selection is performed using the Bayesian Information Criterion \citep[BIC;][]{bic_schwarz}, which balances goodness of fit against model complexity. Lower BIC values indicate a more adequate description of the data. We explore whether the variability is more consistently described by ARIMA ($p$, $d$, $q$), or by ARFIMA($p$, $d$, $q$). 

The selected model for each LC is subsequently validated using the Ljung–Box test \citep{ljung_test}, which probes the presence of residual autocorrelation \citep[e.g.,][]{feigelson_arima}. A fit is considered valid when the test returns a p-value$\geq$0.05, implying that the residuals are consistent with a white-noise process.

\paragraph*{ARFIMA Modeling}
For all the sources, positive $\Delta\mathrm{BIC}$ values, defined as the BIC difference between the ARFIMA and ARIMA, fits indicate that ARFIMA models provide a statistically better description of the data than ARIMA models (Table \ref{tab:arima_arfima}). This preference reflects the presence of long-range correlations dominants the temporal variability of the LCs, which are captured through fractional differencing parameters typically in the range $d$$\sim$ 0.2–0.8 \citep[e.g.,][]{baillie_long_memory_arfima_1996, long_memory_arfima_1980}. In practical terms, such long-memory stochastic behavior manifests as variability patterns in which flux states separated by several months or even years remain statistically correlated \citep[e.g.,][]{baillie_long_memory_arfima_1996}. Observationally, this corresponds to LCs that exhibit persistent high- or low-flux states, slow modulations, or envelope-like trends extending over long timescales, rather than rapid fluctuations that decorrelate quickly \citep[][]{uttley_long_memory_2005, soldi_varibility_2014}. 

Additionally, mostly of the fits results in $d<$0.5 denoting that they are stationary and exhibits long-memory behavior \citep[][]{hosking_fractional_differencin_1981}. In this regime, the autocorrelation function decays as a power law, implying that flux measurements separated by many time bins remain statistically correlated \citep[][]{hosking_fractional_differencin_1981}. In the cases of 0.5$\leq$$d<$1 (e.g., PKS 0208$-$512 and PKS 0301$-$243, see Figure \ref{fig:other_sources}), the process becomes non-stationary but mean-reverting \citep[][]{baillie_long_memory_arfima_1996}. A mean-reverting process allows deviations from the mean level but exhibits a statistical tendency to return toward its mean over sufficiently long timescales. This scenario can arise from large flares, or from epochs during which the flux remains systematically higher or lower than average, e.g., several flares occurring within a short time interval \citep[][]{uttley_long_memory_2005, emma_lc}. 

In some cases, the LC is best described by pure fractional models, ARFIMA (0, $d$, 0), as for instance, S2 0109+22 and TXS 1902+556 (see Figure \ref{fig:other_sources}). In these cases, this model indicates that the variability does not require explicit short-memory structure \citep[][]{feigelson_arima}. Observationally, this suggests LCs in which variability is dominated by slow, persistent modulations and long-term flux states \citep[][]{uttley_long_memory_2005}, without strong localized correlations between immediately adjacent data points. In such sources, short-timescale fluctuations appear consistent with stochastic noise superimposed on a long-memory stochastic process \citep[][]{vaughan_power_law}, rather than being organized into coherent patterns. 

Sources best described by ARFIMA models with $p$=0 and nonzero $q$ (e.g., PKS~0447$-$439 and PG~1553+113) emphasize the role of short-lived variability structure superimposed on a long-memory stochastic process \citep[][]{baillie_long_memory_arfima_1996, feigelson_arima}. In real LCs, this behavior is associated with discrete variability episodes, such as isolated flares or groups of flares occurring within a short time interval \citep[][]{uttley_long_memory_2005, emma_lc}, which induce correlated responses over a limited number of subsequent time bins. When combined with long-memory correlations, such short-lived variability structure can give rise to transient, quasi-oscillatory patterns or brief sequences of peaks driven by individual stochastic events, producing a QPO-like signal. While these features can, over limited intervals, resemble periodic behavior, they do not by themselves establish the presence of a coherent periodic signal and therefore require confirmation through dedicated periodicity analyses \citep[][]{vaughan_power_law, caceres_arima}.

Conversely, sources best described by models with  $p=1$ and $q=0$ (e.g., PKS 0426$-$380 and PKS 2155$-$304, see Figure \ref{fig:other_sources}), the flux at a given epoch is closely related to the immediately preceding measurement \citep[][]{macleod_short_live_2010, andrae_short_live_2013}. Observationally, this is reflected in LCs with smooth variations, gradual rises or declines, and short intervals where the flux changes steadily rather than through isolated, impulsive events. When such structure is combined with long-memory correlations in ARFIMA models \citep[][]{baillie_long_memory_arfima_1996, feigelson_arima}, the LCs show smooth bin-to-bin changes occurring within longer intervals of slowly varying flux, rather than abrupt transitions.

\paragraph*{ARIMA Modeling}
The ARIMA results nonetheless provide complementary insight into the short-memory structure of the variability. In several cases, relatively high $p$ and/or $q$ orders (typically $p$ or $q$ $\geq$3-4) are favored (e.g., PKS 0301$-$243 and S4 1250+53, see Figure \ref{fig:other_sources}), indicating that the flux at a given time is significantly correlated with a limited number of preceding observations \citep[][]{feigelson_arima, tarnopolski20}. In observational terms, this reflects localized persistence in the LCs, such as correlated sequences of data points, flare profiles extending over several time bins, or clustered variability episodes whose influence decays on short timescales \citep[][]{emma_lc}. 

In the cases $q$$=$0 (e.g., PKS 0426$-$380 and PG 1553+113), the flux at a given time depends only on a finite number of preceding measurements \citep[][]{tarnopolski20}. Observationally, this corresponds to LCs that evolve smoothly from one time bin to the next \citep[][]{kelly_ar_2009, feigelson_arima}, with gradual rises or declines and correlated sequences of points, rather than being dominated by isolated, abrupt flux changes. ARIMA models with $p$$=$0 (e.g., PKS 0447$-$439 and PG 1553+113) indicates that the flux at a given epoch is primarily influenced by recent disturbances rather than by past flux levels. Observationally, this behavior is associated with variability dominated by discrete, short-lived events \citep[][]{emma_lc, tarnopolski20}, such as isolated flares or groups of closely spaced flux enhancements, whose effects extend over a limited number of subsequent time bins before decorrelating.

\subsection{Comparing Test Statistics}
Table~\ref{tab:periodicity_results} highlights method-dependent differences that primarily trace back to the adopted null variability model used to calibrate the test statistic. When comparing PSD-based  (PL/BPL) with time-domain (ARIMA/ARFIMA) results, two consistent patterns emerge.

First, PL- and BPL-based test statistics are generally similar, with typically modest differences across GLSP, PDM, and SSA. This suggests that the dominant requirement is a null hypothesis that reproduces the red-noise correlations. The additional PSD curvature introduced by a bending model (pure PL versus BPL) usually produces some slight differences in test statistic; however, BPL fits more accurate than the PL \citep[][]{chakraborty_bending_power_law, penil_2025_4fgl}.

Second, ARFIMA-based test statistics are typically closer to the PL/BPL results than the ARIMA-based ones. In contrast, ARIMA often yields systematically larger test statistics, presumably because its short-memory null hypothesis underestimates the persistent low-frequency fluctuations that characterize the variability on multi-year timescales. As a result, ARIMA-based artificial LCs tend to produce less persistent multi-year structure than the PL, BPL, or ARFIMA null models. This can make large periodogram peaks in the observed LC appear less common under the ARIMA null, leading to higher inferred significances. By contrast, ARFIMA allows long-range dependence, which increases the probability that purely stochastic LCs exhibit multi-year coherent-looking structure. This typically lowers the inferred test statistics, being into closer agreement with PL/BPL-based red-noise artificial LCs.

Overall, PL/BPL and ARFIMA behave as comparatively conservative null hypotheses, whereas ARIMA tends to inflating nominal test statistics. In practice, this means that artificial LCs generated from either a frequency-domain model (red-noise-based PSD) or a time-domain model (through fractional integration and long-memory persistence) lead to comparable null distributions for the test statistic. Consequently, either PL/BPL or ARFIMA are suitable to asses the test statistics estimation of long term $\gamma$-ray periodicity studies of jetted AGN.  

\section{Periodicity Analysis Results} \label{sec:periodicity_results}
We obtained two sources with a significant period ($\geq$3$\sigma$ local test statistics), S5~1044+71 and PG~1553+113. For a $\simeq$78\% of the sample, the updated periods remain compatible (within the uncertainty of the period) to those previously reported \citep[][]{penil_24candidates_2025,alba_ssa,penil_2025_4fgl}. This indicates that the periods obtained in previous works are often persistent. However, even when the same period is confirmed, the associated test statistics frequently decreases to $\lesssim1$–$2\sigma$, implying that the additional epochs do not reinforce the period persistence. In practice, the longer baseline supports the existence of periodic oscillations, but in many cases these behave more like quasi-periodic features embedded in red-noise variability than like robust periodic signals.

A second outcome is that, for several objects, the previous observed period shifts toward longer period once the extended LCs are included (e.g., S4~1030+61, TXS~1318+225, BL~Lacertae). We interpret this behavior primarily as a red-noise effect: as the temporal baseline increases, the analysis becomes sensitive to progressively lower frequencies where red-noise processes carry more power, so stochastic components can increasingly dominate the period-search statistics. Consequently, peaks found at shorter periods in earlier datasets may weaken if the added data do not repeat coherently. Non-stationary episodes such as strong flares (e.g., BL~Lacertae, Figure \ref{fig:other_sources}) can further accentuate this tendency of resulting in higher periods \citep[][]{penil_flares_2025}.

Across the sample, GLSP, PDM, and SSA generally identify candidate periods that are mutually consistent for a given source, but the test statistics for those period is, in most cases, no significant enough. To summarize the heterogeneous test statistics results in Table~\ref{tab:periodicity_results}, we compute for each source the median test statistics over all available entries. This median (see Table \ref{tab:candidates_list} and Table \ref{tab:final_candidates}), without any statistics meaning, helps to obtain conclusions about the results our analysis \citep[][]{penil_2020, penil_24candidates_2025, penil_2025_4fgl}. Specifically, these results show that the typical source is below the conventional hint threshold: most of the sources has test statistics $<$2$\sigma$ while only 8 sources has test statistics $\geq$2$\sigma$.

In general, the periods obtained in this work are broadly consistent with those reported in previous studies. For instance, for PKS 0405$-$385, \citet{gong_pks_0405_385} reported a period of $\sim$2.8 yr, in agreement to the value of $\sim$2.6 yr found in our analysis. In contrast, for MH 2136$-$428, \citet{gong_mh2136} reported a period of $\sim$1.8 yr, whereas our analysis yields a longer period of $\sim$3.5 yr. The reported test statistics also differ: while the previous studies found statistically significant periods ($>$3$\sigma$), in our analysis, the periods are $\approx$1$\sigma$. Such differences may arise from several factors, including the temporal baseline of the LCs, the presence of gaps, or the estimation of the underlying PSD.

\paragraph*{Conclusions}
The newly available \textit{Fermi}-LAT data tend to lower the inferred test statistics. In several earlier works, high nominal test statistics (often $\sim4$–$5\sigma$) were obtained from shorter LCs with only a few observed cycles; in that regime, red-noise dominated variability and clustered flaring can more easily produce extreme peaks that look periodic by chance. When additional epochs are included, a genuine periodicity should remain phase-consistent and the peak should persist (or strengthen). Instead, for most of the candidates the preferred peak becomes less stable across estimators and/or less concentrated in frequency, which reduces the calibrated $S/N$ values in Table~\ref{tab:periodicity_results} and drives down the median test statistics used  in Table~\ref{tab:candidates_list}. This ``regression'' is visible in multiple sources that previously reached $\gtrsim3\sigma$ but now fall below $\sim1\sigma$ (e.g., GB6~J0043+3426, PMN~J0948+0022, and 4C~+48.41).

The extended LCs can amplify the impact of non-stationarity (e.g., strong flaring episodes) and gaps, as longer observational baselines are intrinsically more likely to include additional flares and missing intervals, thereby increasing the complexity of the variability pattern. Both effects can dilute or distort periodic signatures by broadening peaks, reducing cross-method consistency \citep[][]{penil_flares_2025, penil_gaps_2025}. For example, BL Lacertae was previously reported to show a $\sim2.0$ yr period at $\sim4\sigma$, whereas our updated analysis favors a much longer period of $\sim5.5$ yr with low significance ($\sim0.5\sigma$). Such a discrepancy could be consistent with a transient flaring-dominated interval around MJD 59000 (Figure \ref{fig:other_sources}) that could imprint an increase of the power in the signal at low-frequencies \citep[][]{penil_flares_2025}.

Gaps can have a similarly important impact by degrading the test statistics by reducing the number of well-sampled cycles \citep[][]{penil_gaps_2025}. An example could be OC 457 (Figure \ref{fig:other_sources}): a $\sim1.8$ yr period was reported at $4.8\sigma$ using a LC with no gaps, while we recover a comparable period but with only $\sim1\sigma$ in a LC with $\sim25\%$ gaps. This illustrates that, even when the candidate period remains similar, with high peaks coincidence with the period inferred, incomplete sampling could substantially weaken the statistical support by increasing the probability that red-noise realizations plus the observational window produce comparable peaks \citep[][]{penil_gaps_2025}. 

\subsection{ARIMA/ARFIMA Results}
Table~\ref{tab:arima_arfima} shows the results associated of the period inferred from applying residual-based search to each ARIMA/ARFIMA model \citep[][]{zhang_pks0301, caceres_arima}. Most peaks remain below $\sim$3$\sigma$ in both ARIMA and ARFIMA residuals, indicating that the periodic components, if present, are weak relative to the stochastic variability. The highest values in Table~\ref{tab:arima_arfima} occur for PG~1553+113 (ARIMA: 2.8~yr at 2.9$\sigma$), PKS~2155$-$304 (ARIMA: 1.6~yr at 2.9$\sigma$), and BL~Lacertae (ARIMA: 1.5~yr at 2.7$\sigma$). For several sources, the characteristic period is broadly stable across ARIMA and ARFIMA fits (e.g., PKS~0301$-$243 at 2.0~yr, PKS~1510$-$089 at $\sim$4~yr), suggesting that these periodicities are comparatively robust to the details of the stochastic model. 

\begin{figure}
	\includegraphics[width=\columnwidth]{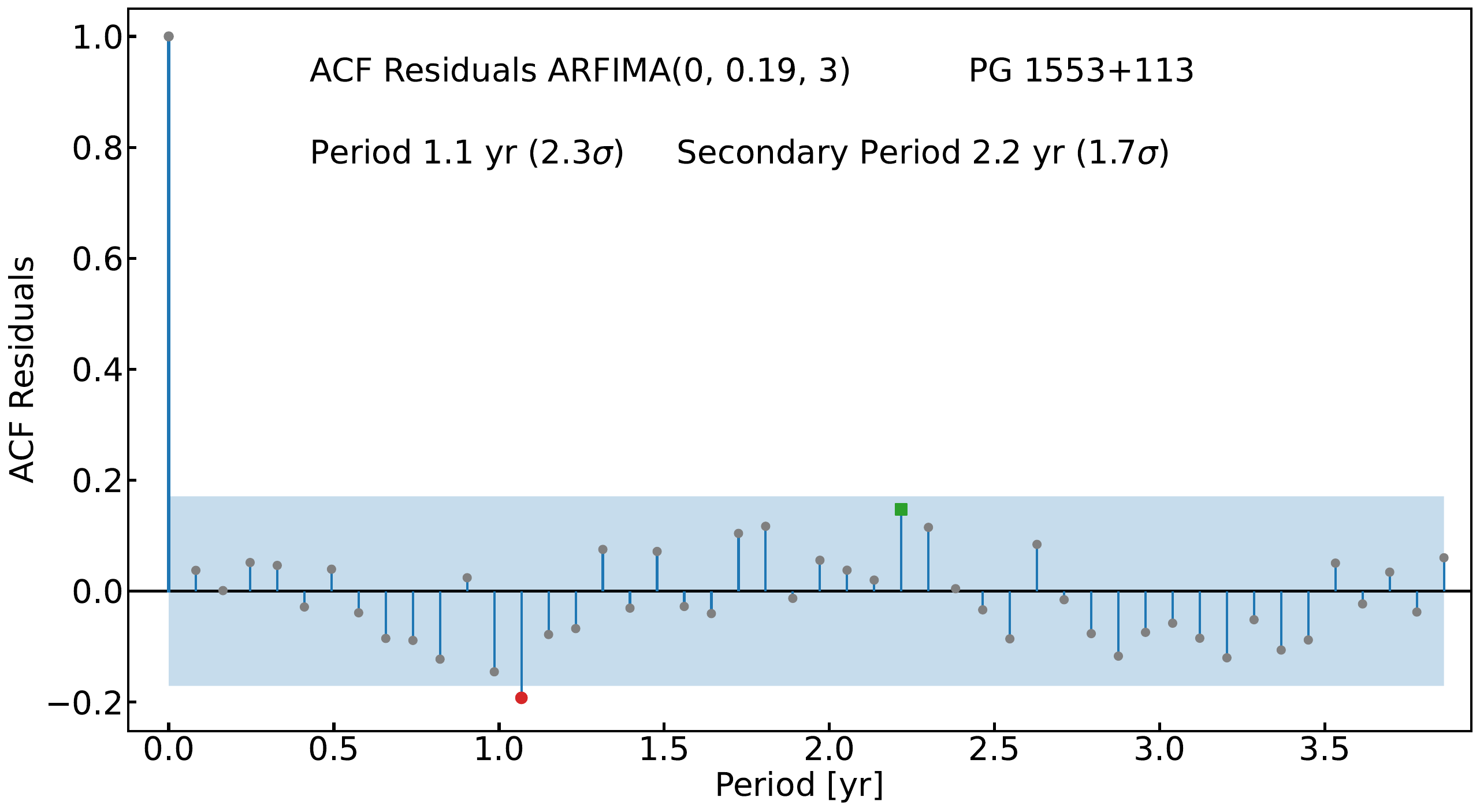}
	\caption{Autocorrelation function of the residuals obtained after subtracting the ARFIMA ($p=0$, $d=0.19$, $q=3$) model from the PG 1553+113 LC. The shaded region indicates the 2$\sigma$ confidence interval. The peak at zero lag corresponds to the autocorrelation of the residuals with themselves. The dominant candidate period is marked with a red circle, while the secondary candidate period is marked with a green square.}
	\label{fig:arfima}
\end{figure}

PG~1553+113 illustrates the potential sensitivity of ARIMA/ARFIMA of absorbing the long-memory component, redistributing variance across timescales and reducing the prominence of multi-year peaks. In Table~\ref{tab:arima_arfima}, the ARIMA residuals yield a prominent $\sim$2.8~yr peak (2.9$\sigma$), while the ARFIMA residuals favor a shorter $\sim$1.1~yr periodicity (2.3$\sigma$). Importantly, we also infer an additional (secondary) peak at $\sim$2.2~yr at 1.7$\sigma$ (not listed in Table~\ref{tab:arima_arfima} but shown in Figure \ref{fig:arfima}). The coexistence of two comparably significant multi-year features, combined with the reduction of the $\sim$2--3~yr significance under ARFIMA, supports a scenario in which the LC contains intermittent or multi-component quasi-periodic variability superposed on a strongly correlated stochastic background. 

\paragraph*{Comparison of ARIMA/ARFIMA with GLSP/PDM/SSA Periods}
Restricting to the sources common to Table~\ref{tab:arima_arfima} and Table~\ref{tab:periodicity_results}, we compare the period estimates obtained from the residual analysis of the best-fit ARIMA/ARFIMA models and the direct period searches using GLSP, PDM, and SSA. Overall, the two approaches recover broadly consistent periods for a subset of sources, while others show method-dependent shifts that suggest either multiple competing periods or sensitivity to harmonics/aliases.

A first group shows good agreement in the dominant period across ARIMA/ARFIMA and GLSP/PDM/SSA. For example, PKS~0301$-$243 exhibits $\sim$2.0~yr in both ARIMA and ARFIMA residuals, consistent with GLSP and SSA ($\sim$2.2~yr); but longer PDM value (4.2~yr) is approximately twice this value, suggesting that it may represent a harmonic, rather than a distinct physical period, consistent with the behavior previously reported by \citet{penil_2020}. TXS~0518+211 shows residual periods of $\sim$3.4~yr (ARIMA) and $\sim$3.0~yr (ARFIMA), matching the consistent GLSP/PDM/SSA periods ($\sim$3.1–3.2~yr). 3C~66A shows ARIMA/ARFIMA periods near $\sim$2.0–2.2~yr, consistent with the other methods ($\sim$2.2-1.9~yr). 

A second scenario displays partial consistency but with method dependence among GLSP, PDM, and SSA relative to ARIMA/ARFIMA. S2~0109+22 has ARIMA/ARFIMA periods at $\sim$2.6–2.9~yr, while GLSP returns $2.7\pm0.2$~yr (consistent), but PDM and SSA favor shorter $\sim$1.7–1.9~yr. S4~1250+53 differs more strongly: ARIMA/ARFIMA analysis favors $\sim$1.3~yr, while GLSP/PDM/SSA cluster near $\sim$2.2–2.3~yr. PKS~2155$-$304 shows a period of $\sim$1.6~yr under ARIMA, while GLSP/PDM/SSA consistently return $\sim$1.6–1.7~yr, indicating good correspondence. However, ARFIMA period is $\sim$3.9~yr.

Several sources show clear discrepancies between the residual-based and the GLSP/PDM/SSA periods, often accompanied by substantial spread among GLSP, PDM, and SSA themselves. PKS~0447$-$439 has a residual period of $\sim$2.1~yr, whereas GLSP yields 1.2~yr, PDM gives 2.4~yr, and SSA+LSP favors a longer 4.1~yr. TXS~1452+516 shows a ARIMA period at $\sim$4.8~yr (ARFIMA reports a period of $\sim$1.5)  GLSP/PDM return $\sim$2.1–2.2~yr while SSA+LSP returns $5.0\pm0.4$~yr, suggesting multiple periods with different sensitivity across estimators. 

The comparison also highlights cases where ARIMA and ARFIMA residuals analysis disagree with each other while the GLPS-PDM-SSA methods converge on a single period. The most relevant example is PG~1553+113: the ARIMA residuals favor $\sim$2.8~yr, whereas ARFIMA residuals favor $\sim$1.1~yr, and a secondary residual peak appears at $\sim$2.1~yr. In contrast, GLSP, PDM, and SSA all return a consistent 2.1~yr period. This configuration indicates that the $\sim$2.1~yr period is the most stable across the GLPS-PDM-SSA methods, while the residual-based period can shift depending on the adopted stochastic model, with the residual secondary peak providing the match.

Finally, BL~Lacertae has ARIMA/ARFIMA periods at $\sim$1.5~yr, whereas GLSP/PDM/SSA converge near $\sim$5.4–5.7~yr, implying that the ARIMA/ARFIMA residual analysis is capturing a different characteristic period than the dominant peak in GLSP/PDM/SSA. This differences could indicate that either multiple competing oscillating components or method-dependent sensitivity to the presence of flaring states at approximately MJD 59000 (Figure \ref{fig:other_sources}).

\section{Origin of Method-Dependent Test Statistics in PG~1553+113}\label{subsec:ssa_significance_tests}
We adopt PG~1553+113 as a reference case because it is one of the most robust periodicity candidates in the sample and has been extensively discussed in previous works \citep[][]{ackermann_pg1553, penil_mwl_pg1553, sagar_pg1553, magic_pg1553_2024}, making it particularly suitable for a detailed comparison of how different methods and noise prescriptions affect the test statistics estimates \citep[][]{penil_2020, penil_flares_2025, penil_gaps_2025}. 

For PG~1553+113, GLSP and PDM yield test statistics of order $\sim$3.5-4$\sigma$ when the artificial LCs are generated using PL, BPL, ARIMA and ARFIMA, whereas the SSA+LSP approach yields systematically lower values, typically $\sim$2$\sigma$ (PL and BPL) to $\sim$3$\sigma$ (ARIMA and ARFIMA). To investigate the origin of this discrepancy, we performed a set of diagnostic tests based on (i) the distribution of recovered periods from the artificial LCs and (ii) the distribution of the maximum Lomb-Scargle power that defines our test statistics estimator.

\subsection{Period Distributions from Artificial LCs}
We first computed the distribution of periods inferred from artificial LCs generated under four noise models: PL, BPL, ARIMA, and ARFIMA, and analyzed them with GLSP, PDM, and SSA+LSP. Overall, the artificial LCs produce broad period distributions (with standard deviations comparable to the mean values). Thus, the red-noise realizations do not preferentially reproduce a unique period bias, but instead results in apparent periodicities across the frequency-search interval (Figure~\ref{fig:period_distribution_tests}).

For GLSP, all the noise models concentrate the mean recovered periods around $\sim$2.5--2.9~yr and yield a non-negligible fraction of realizations with periods in the $2.1\pm0.2$~yr window ($\sim$8--12\%; Figure~\ref{fig:period_distribution_tests}). In contrast, PL artificial LCs shift the mean toward longer periods ($\sim$4.1~yr), consistent with enhanced low-frequency power producing longer apparent periods (Figure~\ref{fig:period_distribution_tests}). For PDM, the recovery fractions near 2.1$\pm$0.2~yr are generally smaller than for GLSP (particularly for PL and ARIMA), while BPL remains comparatively higher and therefore provides the most conservative benchmark among the tested noise models (Figure~\ref{fig:period_distribution_tests}). These results further indicate that the observed differences reflect the combined effect of the stochastic variability model adopted for the artificial LCs and the period-search method used in the analysis. 

For SSA+LSP, only a small fraction of realizations reproduce the period within the PG 1553+113 period window. The fraction of artificial LCs with recovered periods within 2.1$\pm$0.2~yr is $\sim$2.8\% (ARIMA), $\sim$5.5\% (ARFIMA), $\sim$8.5\% (BPL), and $\sim$6.2\% (PL; Figure~\ref{fig:period_distribution_tests_ssa}). Nevertheless, the SSA+LSP period distributions remain broad (typical standard deviations of $\sim$1.7--2.2~yr) and include a substantial fraction of periods shorter than the distribution mean (roughly $\sim$50\%--66\% depending on the noise model). This indicates that SSA frequently extracts oscillatory reconstructions on comparatively short periods, which correspond to many cycles over an 17-yr baseline.

\subsection{Analysis of the SSA Test Statistics}
Recovering the period 2.1 yr is not the only relevant information for the diagnosis of the results. Our test statistics is obtained by measuring how many of the artificial LCs produces a higher power than the original LC. We therefore also examined the subset of artificial LCs for which the maximum SSA+LSP power exceeded the maximum power measured in the original LC. Specifically, after extracting the oscillatory component for each artificial LC and computing its LSP, we retained only those realizations whose maximum periodogram power over the searched range satisfies are higher than the corresponding oscillatory component of the original PG 1553+113 LC. We then analyzed the distribution of the corresponding recovered periods in Figure~\ref{fig:power_distribution_tests_ssa}.

These distributions show that the SSA analysis producing a higher power LSP than the original LC are not primarily produced by artificial LCs reproducing the PG~1553+113 period. Instead, exceedance are dominated by shorter periods (typically $\sim$1--2~yr), particularly for the BPL model, with additional contributions near the long-period boundary for some noise models (PL and ARFIMA: Figure~\ref{fig:power_distribution_tests_ssa}). Even among the artificial LCs whose maximum SSA+LSP power exceeds that of the original LC, only a modest fraction of realizations have their strongest period within 2.1$\pm$0.2~yr ($\sim$6.5\%--11.5\% depending on the noise model; Figure~\ref{fig:power_distribution_tests_ssa}). This shows that the reduced SSA test statistics is driven mainly by artificial LCs that generate large maximum LSP peaks at any frequency within the searched band, rather than by a preferential reproduction of the candidate period itself. Shorter periods are particularly effective in this respect because they correspond to many cycles over the baseline, allowing the periodogram to accumulate power more efficiently and increasing the probability of large peak values in noise-only realizations. Nevertheless, this mechanism alone cannot fully account for the differences in the test statistics obtained.

\subsection{Exploring the Impact of the SSA Window Length}
As noted in $\S$\ref{sec:methods_methodology}, our baseline SSA configuration adopted a $WL$ of $0.4$, corresponding to 40\% of the data points in each LC, following the implementation in \citet{alba_ssa}. This choice provides a convenient default when performing a homogeneous, systematic search for periodic behavior across a heterogeneous sample, where no a priori information is available on whether individual sources host a QPO signal. However, recent analyses \citep[e.g, ][]{penil_gaps_2025} emphasize that $WL$ cannot be treated as a universal standard rule: the optimal embedding depends on LC properties such as the gap fraction, which can alter the separability between oscillatory and red-noise components \citep[][]{penil_gaps_2025}. The tests show that the choice of $WL$ can affect the behavior of noise-dominated LCs, with shorter windows reducing the tendency to produce spurious periodic features under highly gapped conditions.

In SSA, $WL$ sets the embedding dimension of the trajectory matrix and therefore controls the trade-off between component separability and statistical robustness \citep[][]{SSA_algorithm}. In practice, $WL$ regulates how efficiently SSA isolates the oscillatory component of the signal from low-frequency variability (trends and red noise) and high-frequency variability (short-timescale fluctuations, including white-noise-dominated variations). Because the inferred test statistics can be sensitive to $WL$, we implemented an optimization procedure for the SSA $WL$ tailored to the candidate period.

To select the $WL$ without a posteriori tuning, we adopt an a priori choice tied to the candidate period $T$. A commonly used in SSA applications is that oscillations reliably resolved by an embedding of length $M$\footnote{In \citet{ssa_greco} and \citet{alba_ssa}, the embedding length is represented by $N$; here, we adopt $M$ to avoid confusion with $N$, which in our notation denotes the number of trials used to estimate the global significance.} fall approximately within the period range $(M/5,\,M)$ \citep[e.g., ][]{krishnamurthy_shukla_2007_ssa}, following \citet{plaut_vautard_1994_ssa}. Enforcing $T\in(M/T,M)$ implies $T\lesssim M\lesssim 5T$. We therefore set $T$ proportional to the candidate period in sampling units, $M = k\,T/\Delta t$, with a fixed factor $k$ applied uniformly across sources; in practice $k\simeq$1-2 provides a conservative compromise between representing at least one full cycle within the trajectory matrix and maintaining statistical stability in finite, red-noise dominated LCs.

We set the SSA $WL$ proportional to the candidate period, following standard SSA practice \citep[e.g.,][]{plaut_vautard_1994_ssa, ghil_2002_ssa}. Using $O$ as the number of points in the LC, $O$$=$212 points in our case, and the candidate period $T$, we define an a priori choice
\begin{equation}
M_{\rm opt} \simeq \frac{k\times T}{\Delta t},
\end{equation}
where $\Delta t$ is the binning of the LC, 30 days in our case. When we consider one cycle ($k$$=$1), using $T$$=$2.1*365.25 days, $M_{\rm opt}$$=$26, and for $k$$=$2, $M_{\rm opt}$$=$52. In fractional terms, this choice corresponds to
\begin{equation}
{\rm WL}_{\rm opt}\equiv \frac{M_{\rm opt}}{O}
\end{equation}
resulting in $0.12$ and $0.24$, respectively. 

In addition to adopting a specific ${\rm WL}_{\rm opt}$, we performed a controlled grid exploration to quantify the sensitivity of SSA+LSP results to the $WL$ choice. Specifically, we explored values that span $\sim$1--3 cycles of the candidate oscillation. As a practical test bench within, we adopted the discrete grid as percentage of the total data length
\begin{equation}
{\rm WL}\in [0.12,\,0.16,\,0.20,\,0.24,\,0.28,\,0.32,\,0.36,\,0.40]~,
\end{equation}
and repeated the full SSA reconstruction and surrogate-based test statistics calculation at each value. 

Table~\ref{tab:ssa_wl_pg1553} summarizes the test statistics obtained from the SSA+LSP pipeline for PG~1553+113 as a function of the SSA $WL$ and for four alternative noise models used to build the artificial LCs (PL, BPL, ARIMA, and ARFIMA). Two systematic dependencies are immediately apparent. First, for all null models the inferred test statistics decreases monotonically with increasing $WL$: the smallest explored value ($WL=$0.12) yields the largest $S/N$ (3.1--4.0$\sigma$ depending on the noise model), coincident with one of the ${\rm WL}_{\rm opt}$, whereas $WL=$0.40 yields the lowest values (1.9--3.1$\sigma$). This monotonic behavior indicates that the reported test statistics is not solely determined by the data, but is also conditioned by how $WL$ redistributes variance among reconstructed components. For relatively small $WL$, limited frequency resolution favors mixing between the target oscillation and nearby components, while for excessively large $WL$ the finite-sample stability of the decomposition degrades \citep[][]{groth_ssa_2015}. In practice, both extremes can contaminate the reconstructed oscillatory component with low-frequency (trend/red-noise) and high-frequency (short-timescale stochastic) variability, thereby biasing the inferred LSP peak power and test statistics (Figure \ref{fig:ssa_descompositions_pg1553}). 

\begin{table}
\centering
\caption{Test statistics of applying the approach SSA+LSP, considering different window length (WL) for the SSA oscillatory decomposition, and for different models for the noise: simple power-law (PL), bending power-law (BPL), ARIMA, and ARFIMA. The result shows the impact of the WL selected for inferring the test statistics associated to the period of PG 1553+113. The $WL$ in bold is the ${\rm WL}_{\rm opt}$, which is the optimal $WL$ according to the properties of the LC and the period of the signal. \label{tab:ssa_wl_pg1553}}
{%
\begin{tabular}{c|cccccccccc}
\hline
\hline
WL & PL & BPL & ARIMA & ARFIMA  \\
   & [S/N] & [S/N] & [S/N] & [S/N] \\ 
\hline
\hline
\textbf{0.12} & 3.2$\sigma$ & 3.1$\sigma$ & 4.0$\sigma$ & 3.9$\sigma$ \\
0.16 & 2.9$\sigma$ & 2.7$\sigma$ & 3.7$\sigma$ & 3.4$\sigma$ \\
0.20 & 2.8$\sigma$ & 2.6$\sigma$ & 3.7$\sigma$ & 3.4$\sigma$ \\
\textbf{0.24} & 2.7$\sigma$ & 2.5$\sigma$ & 3.7$\sigma$ & 3.3$\sigma$ \\
0.28 & 2.5$\sigma$ & 2.3$\sigma$ & 3.5$\sigma$ & 3.2$\sigma$ \\
0.32 & 2.3$\sigma$ & 2.0$\sigma$ & 3.3$\sigma$ & 3.0$\sigma$ \\
0.36 & 2.2$\sigma$ & 1.9$\sigma$ & 3.3$\sigma$ & 2.9$\sigma$ \\
0.40 & 2.0$\sigma$ & 1.9$\sigma$ & 3.1$\sigma$ & 2.7$\sigma$ \\
\hline
\hline
\end{tabular}%
}
\end{table}

We repeated the same SSA $WL$ exploration for the other significant source, S5~1044+71, extending the scan over $WL\simeq0.09$–$0.49$ to provide a more complete view within the range typically recommended for SSA decompositions \citep[$WL\lesssim$0.5;][]{ssa_greco}. In this case, the ${\rm WL}_{\rm opt}$ according to its period are, 0.17 and 0.34. The results are summarized in Table~\ref{tab:ssa_wl_s51044}. 

In contrast to PG~1553+113 (Table~\ref{tab:ssa_wl_pg1553}), S5~1044+71 does not show a monotonic decrease of the inferred test statistics with increasing $WL$. Instead, the test statistics exhibits a non-monotonic dependence. The results of both ${\rm WL}_{\rm opt}$ produced comparable test statistics ($\sim$2.1--2.3$\sigma$). The apparent difference between PG~1553+113 and S5~1044+71 does not imply distinct underlying mechanisms, but rather different responses of the same SSA variance-partition effect to the finite length of the LC. In both cases, the inferred SSA+LSP test statistics depends on $WL$ because changing the embedding window modifies the allocation of variance between the oscillatory component and high-frequency variability (Figure \ref{fig:ssa_descompositions_s51044}), which in turn affects the amplitude and coherence of the reconstructed signal used for the LSP. The distinction is that PG~1553+113 (T$\sim$2.1 yr; $\sim$9 cycles over $\sim$17 yr) lies in a more stable separability regime, yielding a smooth monotonic decrease of the test statistics with increasing $WL$, whereas S5~1044+71 (T$\sim$3.1 yr; $\sim$6 cycles) lies in a weaker-separability regime, resulting in a non-monotonic $WL$ dependence.

We complete these tests with two additional sources, PKS 2155$-$304, and  OJ 014. For PKS~2155$-$304 (Table~\ref{tab:ssa_pks2155}, Figure \ref{fig:ssa_descompositions_pks2155}), the SSA+LSP test statistics remain low under any noise models for all explored $WL$, typically $\lesssim$1.3$\sigma$, and decrease further at larger $WL$. The two highlighted values, $WL$=0.09 (approximately one cycle per window) and $WL$=0.18 (approximately two cycles), yield comparable and modest test statistics. Overall, the absence of a pronounced peak or systematic enhancement at ${\rm WL}_{\rm opt}$ indicates that the candidate periodicity in PKS~2155$-$304 is not strongly supported by the SSA+LSP analysis.

For OJ~014 (Table~\ref{tab:ssa_wl_oj014}, Figure \ref{fig:ssa_descompositions_oj014}), the behavior is qualitatively different. The test statistics increases from small $WL$ up to $WL$$\sim$0.21--0.24, and then gradually decreases toward larger $WL$. The primary $WL=$0.24 (approximately one cycle per window) lies near the plateau of maximum test statistics, while the larger highlighted value $WL=$0.48 (approximately two cycles) yields systematically lower test statistics. This pattern suggests that the oscillatory component in OJ~014 is moderately separable at window lengths close to one cycle.

\subsection{Conclusions}
From these tests, we can obtain different conclusions about the use of SSA. The $WL$ should be used as a controlled methodological choice rather than a value optimized on the same data. To not have the confirmation bias, we recommend to set $WL$ a priori, the ${\rm WL}_{\rm opt}$, instead of selecting the value that maximizes the observed SSA+LSP test statistics. When several window lengths are explored, the scan is used only as a robustness diagnostic, while the quoted test statistics is reported for the pre-defined $WL$.

\section{Predictions}\label{sec:predictions_results}
Periodicity studies are fundamentally limited by the difficulty of determining whether a observed period is genuine. Red-noise variability, flares, and data gaps can all affect period searches and produce spurious detections \citep[][]{mcquillan_trend_fake_detection, vaughan_criticism,covino_negation,penil_flares_2025, penil_gaps_2025}. Specifically, these effects can obscure genuine periodic signals and distort their inferred properties by shifting or broadening periodogram peaks, altering the recovered period, and reducing the apparent significance of the period. In addition to that, they can result in fake detection of a period pattern. As a result, test statistics alone, even when derived from dedicated stochastic null tests, may not be sufficient to establish periodicity, and further analyses can provide additional support for a periodic interpretation.

In this context, we performed a prediction-based analysis. A genuinely periodic process should be predictable in the sense that the inferred cycle can be extrapolated to anticipate future emission peaks. By contrast, QPOs typically include a stochastic component, which reduces coherence and can suppress or even eliminate a predictive behavior. Following this principle, we performed forward predictions for the periodicity candidates reported in \citet{alba_ssa, penil_2025_4fgl}.

\subsection{Sample Validation of Predicted Maxima}
We first reassess the peak-time predictions of \citet{alba_ssa} and \citet{penil_2025_4fgl} using the newly available $\sim$6~yr of monitoring beyond the original $\sim$12~yr dataset. This is an out-of-sample timing test: the predicted epochs of maxima from the 12-yr analysis (Table~\ref{tab:predictions}) are treated as fixed, and we check whether high states occur near those times in the extended LCs.

Using the period $T$, we compute the predicted maxima epochs $t_{\rm pred}$ that fall within the test interval. Each prediction is evaluated within a symmetric window $[t_{\rm pred}-W,t_{\rm pred}+W]$ with $W=0.20T$. We choose this relatively broad window because 30-day binning and observational gaps can shift the sampled maximum away from the true turning point; a narrower window could miss the relevant peak due to incomplete phase coverage. A prediction is considered evaluable only if the window contains at least $N_{\min}=2$ measurements.

For each evaluable prediction, we define the observed cycle-associated peak as the maximum flux within the window, occurring at $t_{\rm obs}$. We classify the prediction as compatible with the new data if the peak timing satisfies $|t_{\rm obs}-t_{\rm pred}|\le \Delta t_{\rm tol}$, with $\Delta t_{\rm tol}=\max(60~{\rm d},\,0.20T)$, which accounts for the 30-day cadence while allowing modest phase jitter and sampling-induced offsets, yet still requires agreement within $\sim$20\% of a cycle.

\subsection{Results}
Table~\ref{tab:predictions} summarizes the sample check of previously published peak forecasts against the extended LCs. In total, we evaluated 59 predicted high states across 30 objects; 19/59 ($\simeq32\%$) are supported by the newly available data according to our criteria. At the source level, 7/30 objects show full agreement, meaning that all predicted peaks within the analyzed time span are confirmed, whereas 15/30 show no confirmed predictions (e.g., GB6~J0043+3426, S4~1030+41, PKS~1510$-$089, TXS~1530$-$131, and S4~1250+53). The remaining 8/30 display mixed behavior, with at least one forecast supported and at least one forecast not supported (e.g., PKS~0447$-$439,  S4~1144+40, PKS 2155$-$304). For PKS~0454$-$234, OJ~014, S5~1044+71, and OP~313, however, only one predicted peak falls within the time span tested here, which limits the strength of any conclusion. With the exception of OP~313, these objects are associated with periods having test statistics of $\gtrsim 2\sigma$ (Table~\ref{tab:candidates_list}). 

For S4~1144+40 and PG~1553+113, two and three predicted high-flux states, respectively, fall within the time span analyzed here. For B2~2234+28A, both predicted peaks are confirmed. At the same time, the periodicity analysis identifies two possible periods, 1.4 and 2.9 yr, each with low test statistics ($0.5\sigma$). Nevertheless, the LC in Figure~\ref{fig:other_sources} reveals an evident sequence of oscillation-like variations. This makes the source a promising candidate for continued monitoring, which will be necessary to determine whether this apparent pattern persists over longer timescales. 

\begin{table*}
\centering
\caption{List of periodicity candidates with test statistics $\geq$3$\sigma$. The table also reports the global significance estimated following the procedure described in $\S$\ref{sec:global_significance}, which is $\sim$0$\sigma$ for all objects. We further provide predictions for the next two future peaks using the two methods presented in $\S$\ref{sec:periodicy_candidates}: phase folding and SSA-R. The symbol -- in the SSA predictions indicates that the corresponding LC contains too many gaps for this method to be applied.\label{tab:final_candidates}}
{%
\begin{tabular}{c|cccccccccc}
\hline
\hline
Name & Period & Global Significance & Next Cycles & Next Cycles \\
     & [Year] & [S/N] & [Phase Folding] & [SSA] \\ 
\hline
\hline
PG 1553+113 & 2.1$\pm$0.2 (3.7$\sigma$) & $\approx$0$\sigma$ & (2027-06, 2029-07) & (2027-03, 2029-05) \\
S5 1044+71 & 3.1$\pm$0.3 (3.7$\sigma$) & $\approx$0$\sigma$ & (2026-04, 2029-05) & (2026-03, 2029-05) \\
\hline
OJ 014 & 4.2$\pm$0.4 (2.5$\sigma$) & $\approx$0$\sigma$ & (2027-09, 2031-09) & (2028-05, 2032-10) \\
S4 1144+40 & 3.1$\pm$0.2 (2.2$\sigma$) & $\approx$0$\sigma$ & (2027-05, 2030-08) & (2027-08, 2030-12) \\
PKS 0215+015 & 3.3$\pm$0.3 (2.0$\sigma$) & $\approx$0$\sigma$ & (2029-02, 2032-06) & -- \\
S3 0458$-$02 & 4.0$\pm$0.4 (2.0$\sigma$) & $\approx$0$\sigma$ & (2026-10, 2030-09) & -- \\
TXS 0518+211 & 3.1$\pm$0.3 (2.0$\sigma$) & $\approx$0$\sigma$ & (2026-03, 2028-11) & (2028-08, 2031-03) \\
87GB 164812.2+524023 & 3.2$\pm$0.6 (2.0$\sigma$) & $\approx$0$\sigma$ & (2026-10, 2030-03) & -- \\
\hline
\hline
\end{tabular}%
}
\end{table*}

\section{Periodicity Candidates}\label{sec:periodicy_candidates}
\begin{figure*}
	\centering
 	\includegraphics[scale=0.2195]{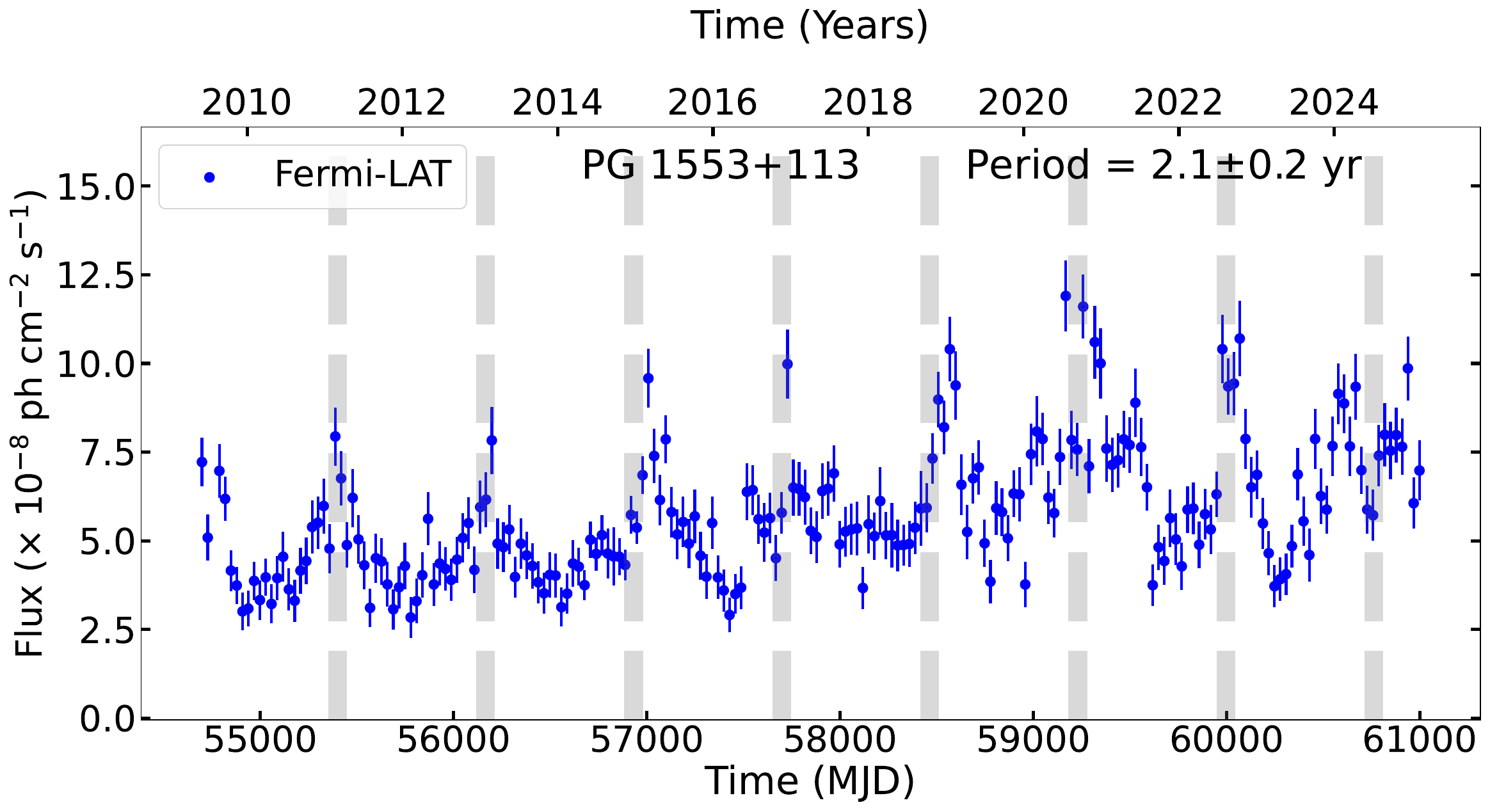}
 	\includegraphics[scale=0.2195]{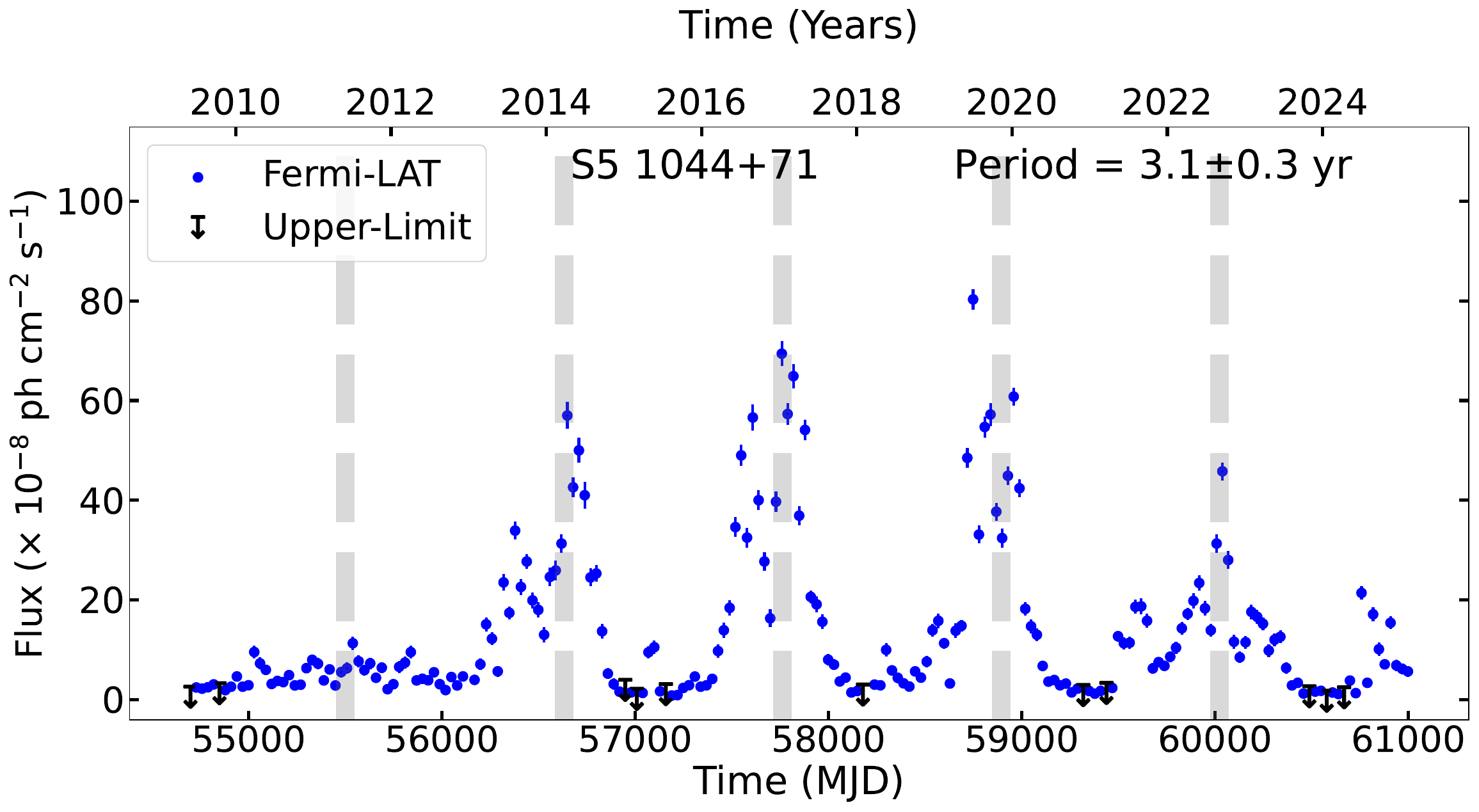}
	\caption{\textit{Left}: Light curve of the blazar PG 1553+113. \textit{Right}: Light curve of the blazars S5 1044+71. The gray vertical bars approximate high-flux periods inferred by the methodology for the given blazar. The width of the gray bars indicates the uncertainty in the period value.}
	\label{fig:lc_candidates}
\end{figure*}

After considering the results of previous sections ($\S$\ref{sec:periodicity_results} and $\S$\ref{sec:predictions_results}), we define a sample of jetted AGN that show evidence of periodic behavior. The selected sources are listed in Table~\ref{tab:final_candidates}. Two objects, PG~1553+113 and S5~1044+71 (Figure \ref{fig:lc_candidates}), reach significant test statistics ($\geq$3$\sigma$). In both cases, the candidate period is supported across the methods and remains significant under the different noise models. They also show overall consistency with published peak-time forecasts. For PG~1553+113, the three predicted peak epochs are supported by the extended LCs, while for S5~1044+71 the single prediction evaluated is consistent with the new data.  

We also identify a second group of candidates with a hint for periodicity ($\geq$2$\sigma$): OJ~014, PKS~0215+015, S3~0458$-$02, TXS~0518+211, S4~1144+40, and 87GB~164812.2+524023 (Figure~\ref{fig:low_sgnificance}). In these cases, the candidate period is typically recovered consistently. Consequently, these sources are candidates for continued monitoring.

Among the sources included in Table~\ref{tab:final_candidates}, OJ~014 is consistent with its single predicted peak time, while S4~1144+40 is consistent with both forecast peaks. Nevertheless, the oscillation amplitudes of S4~1144+40 are irregular (Figure~\ref{fig:low_sgnificance}), with a prominent flaring state around MJD~57000. Such  flaring activity may influence the inferred test statistics, as discussed by \citet{penil_flares_2025}. In contrast, the predicted peak for 87GB~164812.2+524023 is not supported by the updated LC. The remaining objects in this subset, PKS~0215+015, S3~0458$-$02, and TXS~0518+211, are not listed in Table~\ref{tab:predictions}, and therefore their forecast evaluation is not reported there.

For this sample of Table \ref{tab:final_candidates}, we compute new peak-time predictions using two complementary approaches. The first is SSA-based forecasting SSA-R\footnote{We use the implementation available at \url{https://github.com/AndrewSukhobok95/ssa} and \url{www.ucm.es/blazars/ssa}.}
 following \citet{ssa_forecast}. In practice, this method is sensitive to data gaps and is therefore applied only when the LC has sufficiently low missing-data fraction. Here we adopt a conservative requirement of a maximum gap fraction of 10\%.

\begin{figure}
    \centering
    % Top panel
    \includegraphics[width=0.49\textwidth]{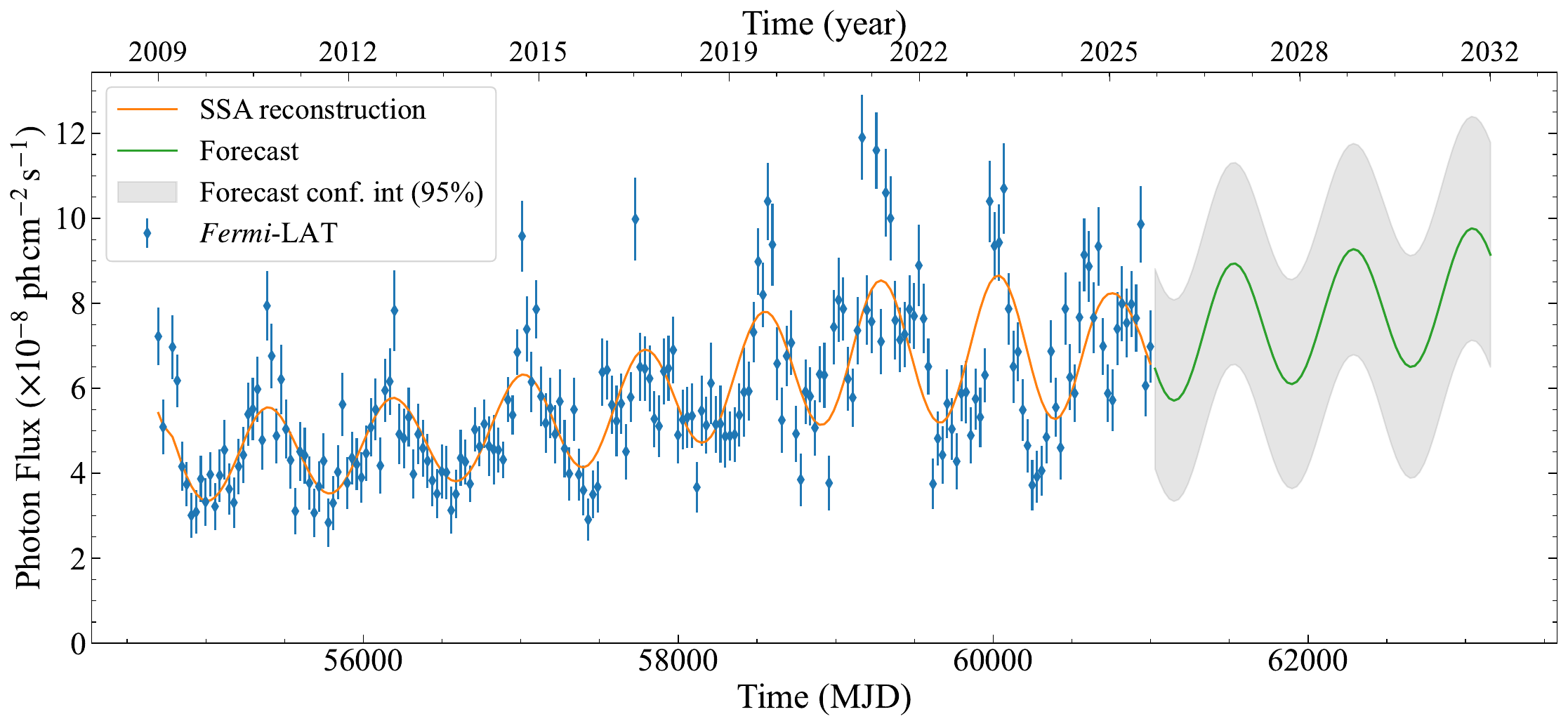}    
    \vspace{0.2cm}
    % Bottom panel
    \includegraphics[width=0.49\textwidth]{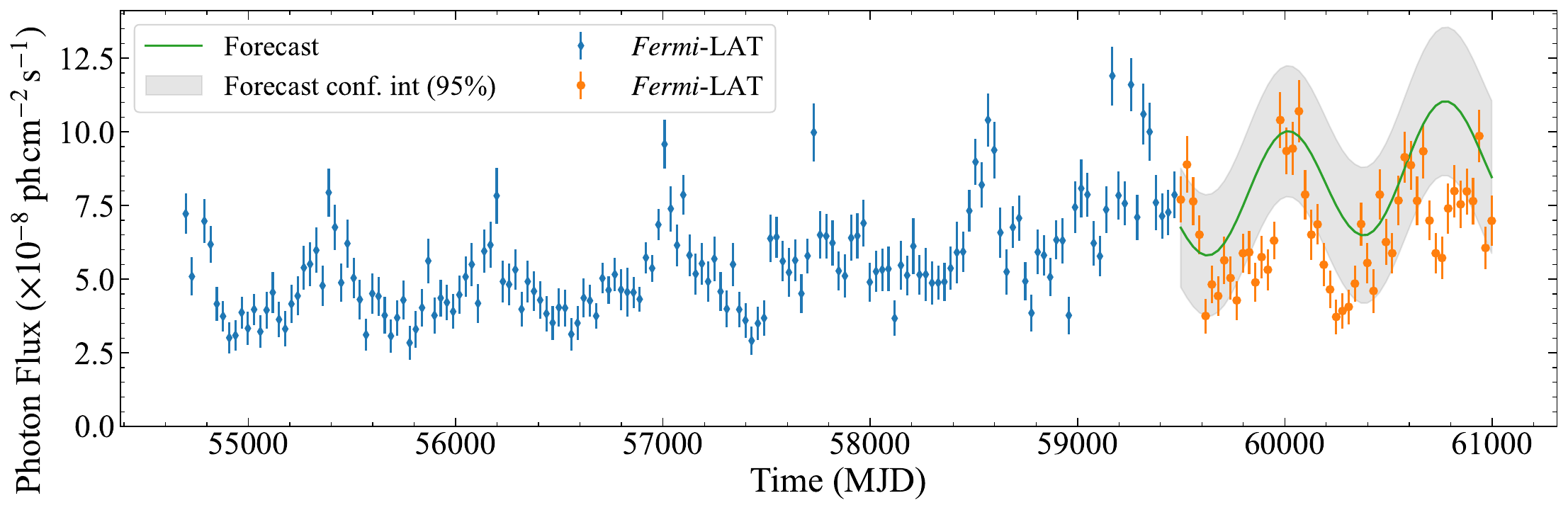}
    \caption{
    {\it Top}: SSA-R prediction for the expected emission of PG 1553 over the next 6 yr.
    {\it Bottom }: Validation test in which one cycle is removed from the end of the LC, the forecasting is repeated, and the predicted cycle is compared with the observed data. 
    }
    \label{fig:forecast_pg}
\end{figure}

For each source, we predicted the next two cycles ($\sim$7 years) using the SSA reconstructed components associated with the inferred period. The $WL$ was chosen within the range of optimal values (e.g. see Table \ref{tab:ssa_wl_pg1553}, Table \ref{tab:ssa_wl_s51044}, and Table  \ref{tab:ssa_wl_oj014}), ensuring an stable reconstruction and forecasting robustness. The 95\% confidence interval was estimated from the residual components, which represents the variability not accounted for by the reconstructed signal (e.g. noise structure). Figure \ref{fig:forecast_pg} top panel shows the prediction obtained for the next three cycles for PG 1553+113 using the full available LC. The bottom panel shows a test in which the last two cycles are removed from the end of the PG 1553 LC \citep[to reproduce the LC analyzed in ][]{alba_ssa}. Considering this truncated LC, we then apply the SSA-R forecasting model to predict the next two cycles. The resulting prediction is then compared with the full observed data to assess its agreement with the observed behavior. For all sources listed in Table~\ref{tab:final_candidates}, the forecasts obtained after removing one or two cycles remain in good agreement with the observed data, with an overlap greater than 80\% considering flux uncertainties (Figure \ref{fig:ssa_forecasting}).

The second approach relies on phase folding of the LC \citep[][]{sagar_pg1553, penil_2025_4fgl}. In this method, we search for the period that minimizes the variance of the phase-binned profile. We then fix the best-fit period and model the folded modulation with a sinusoid, sampling the offset, amplitude, and phase with an MCMC procedure. The resulting forecast peak epochs from both methods are reported in Table~\ref{tab:predictions}.

In addition, we note a small subset of sources that show only marginal hints in our periodicity reanalysis (test statistics $\sim$1.8$\sigma$), while also displaying at least one peak time that is consistent with a published forecast (Table~\ref{tab:predictions}). In particular, this applies to PKS~0454$-$234 and PKS~2155$-$304. Although the current test statistics do not indicate a hint of periodic signal, the partial agreement with earlier timing predictions makes these objects worth keeping under continued monitoring. Their LCs exhibit substantial intrinsic variability, which could plausibly dilute a periodic component and thus contribute to the reduced the test statistics we obtain in the extended dataset \citep[][]{penil_flares_2025}.

\section{Summary}\label{sec:summary}
In this work, we reanalyze 52 previously reported $\gamma$-ray periodicity candidates using more than five additional years of \textit{Fermi}-LAT data, for a total baseline of $>$17~yr. We search for periodic behavior with three complementary techniques, GLSP, PDM and SSA+LSP, and we assess statistical significance under four alternative descriptions of the intrinsic stochastic variability: a simple PL PSD, a BPL PSD, and time-domain ARIMA and ARFIMA processes.

Overall, ARFIMA$(p,d,q)$ models provide the most adequate autoregressive representation of the analyzed LCs, consistent with the presence of long-memory variability in many sources. Importantly, the resulting test-statistic calibrations are broadly consistent between PSD-based (PL/BPL) and ARFIMA-based artificial LCs, indicating that these approaches are equivalent as null hypotheses for modeling red-noise dominated variability in multi-year \textit{Fermi}-LAT LCs.

Under these calibrations, we identified eight sources with a local test statistics $\geq$2$\sigma$, including two sources, S5~1044+71 and PG~1553+113, that reach $\geq$3$\sigma$. The global significance associated is $\sim$0$\sigma.$ Finally, we provide updated forecasts for the epochs of future high-state peaks for these eight candidates, offering a direct test of periodic predictability with forthcoming \textit{Fermi}-LAT monitoring.

\section{Acknowledgements}
P.P and M.A acknowledge funding under NASA contract 80NSSC20K1562. 
This work was supported by the European Research Council, ERC Starting grant \emph{MessMapp}, S.B. Principal Investigator, under contract no. 949555, and by the German Science Foundation DFG, research grant “Relativistic Jets in Active Galaxies” (FOR 5195, grant No. 443220636).

\section*{Data Availability}

All the data used in this work are publicly available or available on request to the responsible for the corresponding observatory/facility.

\section{Software}
\begin{enumerate}
    \item arfima R-software \citep{arima_hyndman_2008, arima_hyndman_2024},
	\item astroML \citep{astroml},
	\item astropy \citep{astropy_2013, astropy_2018, astropy_2022}, 
	\item emcee \citep {emcee}, 
	\item PyAstronomy \citep{PyAstronomy},	  
	\item rpy2 \url{https://rpy2.github.io/doc/latest/html/index.html},
	\item stats R-software \citep{r_manual},  
	\item statsmodels \citep[][]{stats_scipy_2010},, 
	\item SciPy \citep {SciPy},
	\item Simulating light curves \citep{connolly_code},
    \item Singular Spectrum Analysis \url{https://www.kaggle.com/code/jdarcy/introducing-ssa-for-time-series-decomposition}
\end{enumerate}

\bibliographystyle{mnras}
\bibliography{literature} % if your bibtex file is called example.bib
\appendix
\setcounter{table}{0}
\setcounter{figure}{0}
\renewcommand{\thetable}{A\arabic{table}}
\renewcommand{\thefigure}{A\arabic{figure}}
\section*{Appendix}\label{sec:appendix}
This appendix presents additional diagnostic material supporting the analysis described in the main text. The supplementary figures are organized into four groups: tests based on artificial light curves used for significance estimation, SSA reconstructions obtained for different window lengths, light curves of secondary or lower-significance candidates, and SSA-R forecasting results. The supplementary tables provide the corresponding numerical results, including PSD-model parameters, ARIMA/ARFIMA fits, period estimates, SSA window-length tests, and prediction checks.

\subsection{Supplementary Figures}
This section presents several figures that support some analyses discussed in this paper. 

\subsubsection{Light curves of sources discussed in the text}
Figures~\ref{fig:other_sources} present the light curves of additional blazars discussed in the main text.  The group includes sources mentioned in the discussion of the broader sample and of the comparison with previous studies or related with the analysis of the ARIMA/ARFIMA modeling.

\begin{figure*}
	\centering
        \includegraphics[scale=0.2195]{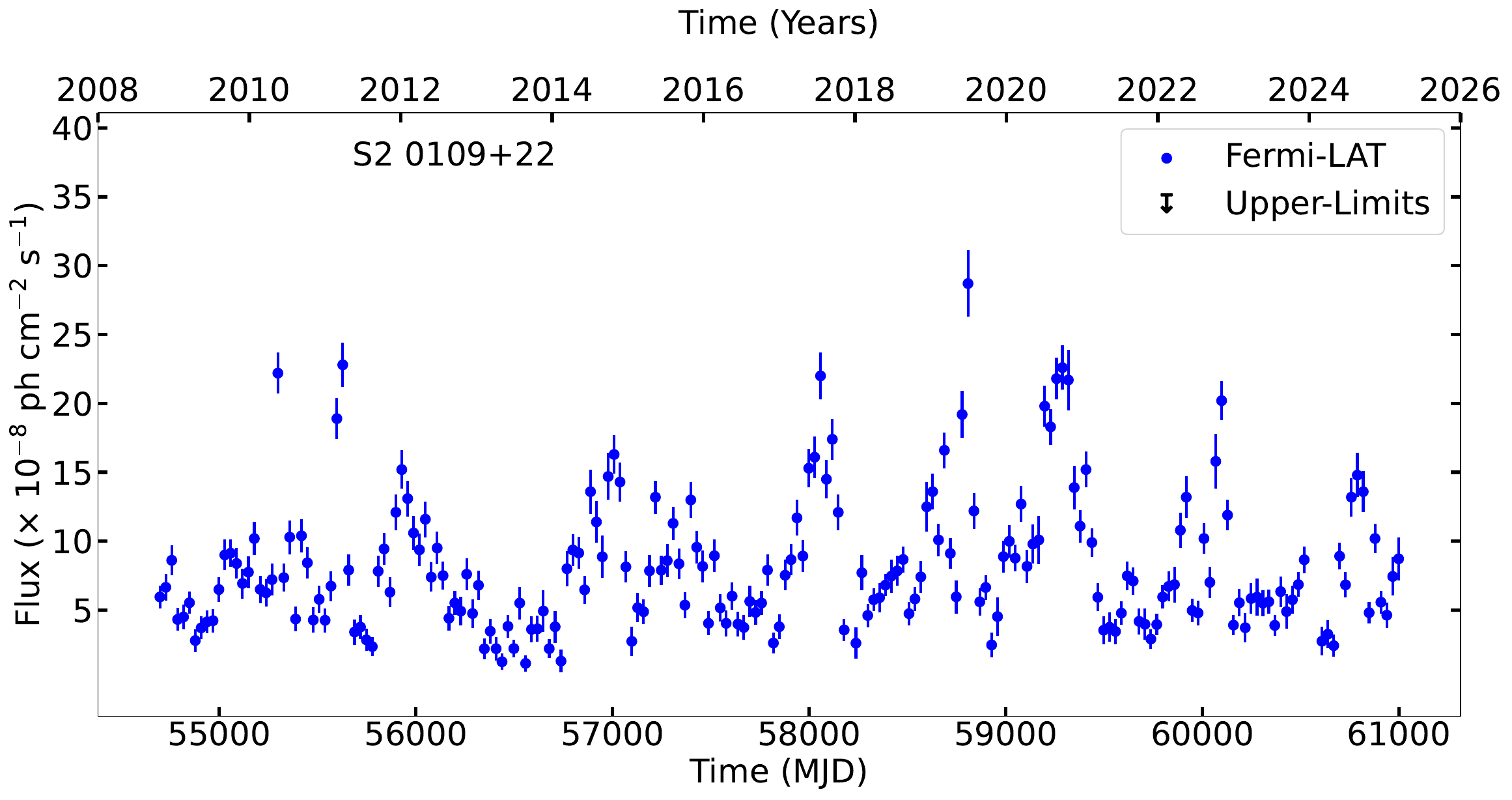}
        \includegraphics[scale=0.2195]{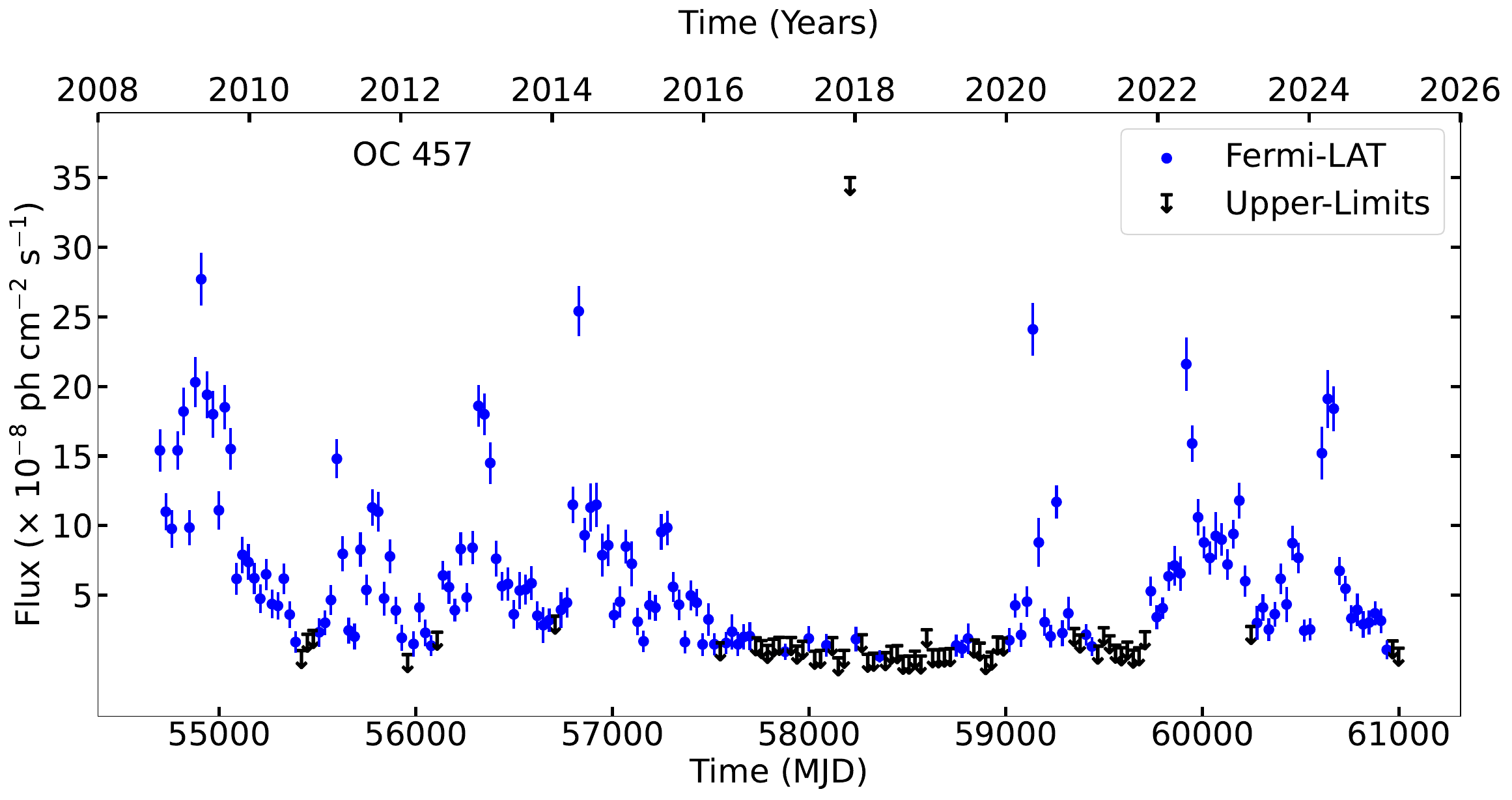}
        \includegraphics[scale=0.2195]{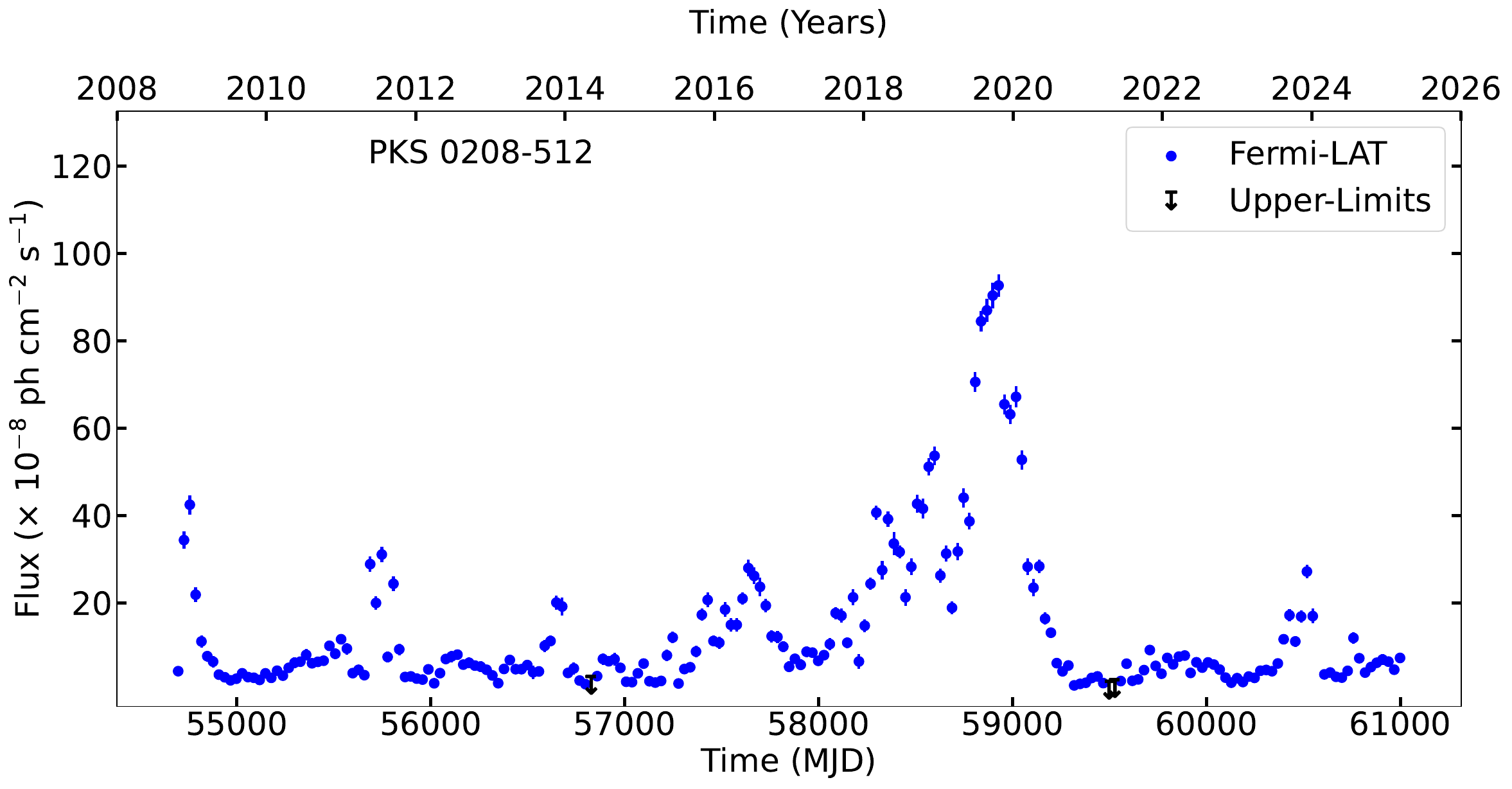}
        \includegraphics[scale=0.2195]{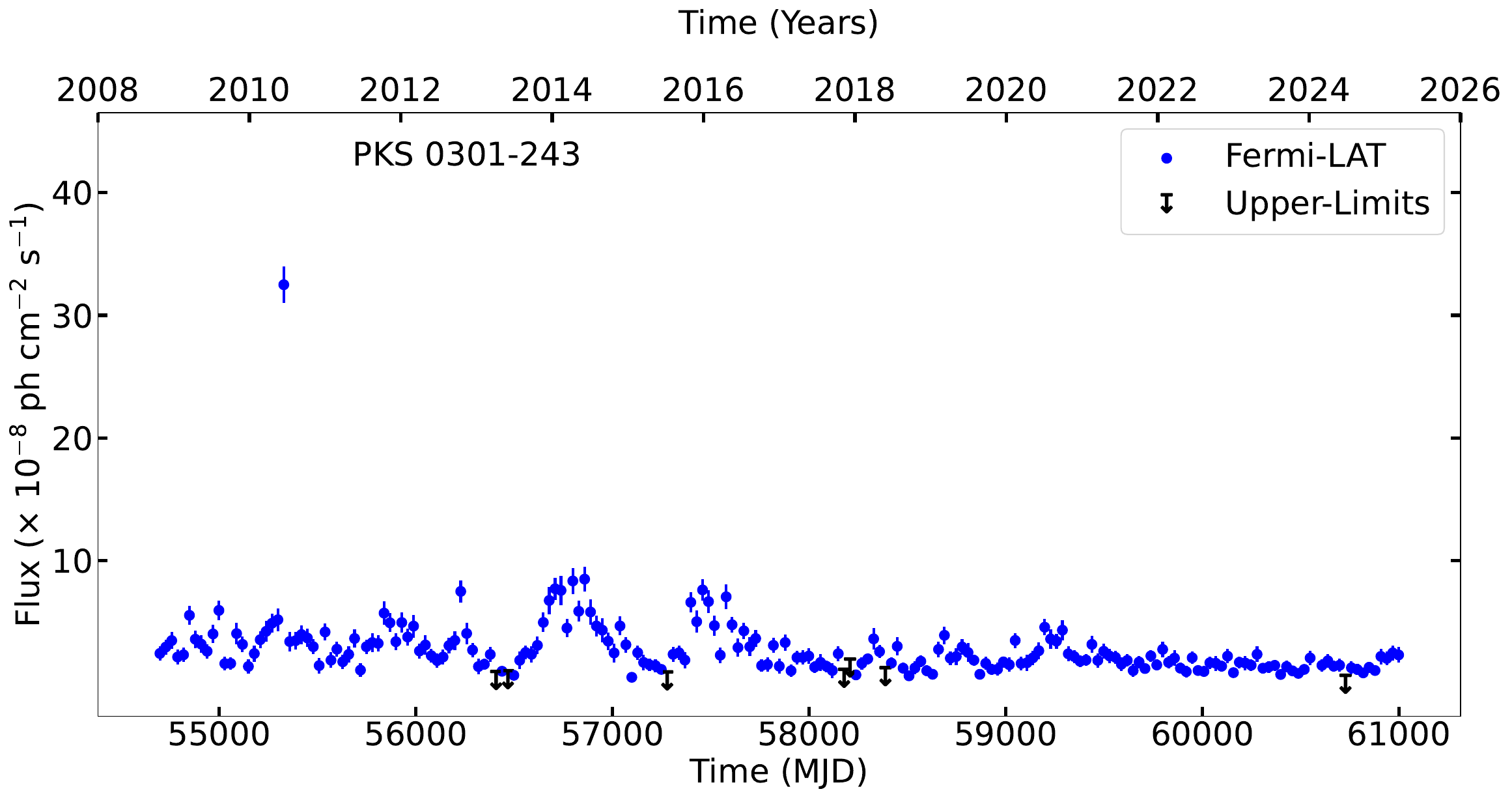}
        \includegraphics[scale=0.2195]{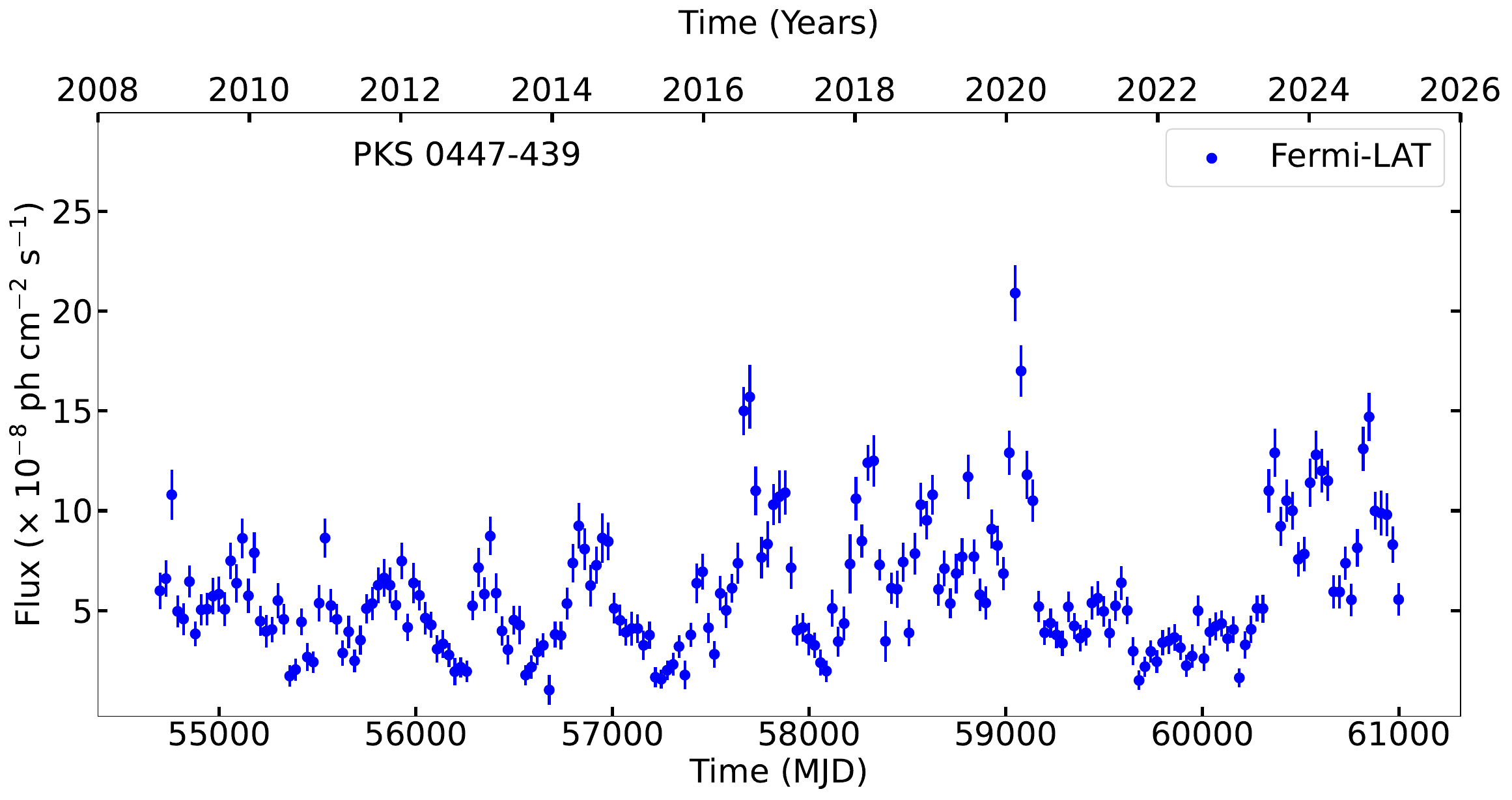}
        \includegraphics[scale=0.2195]{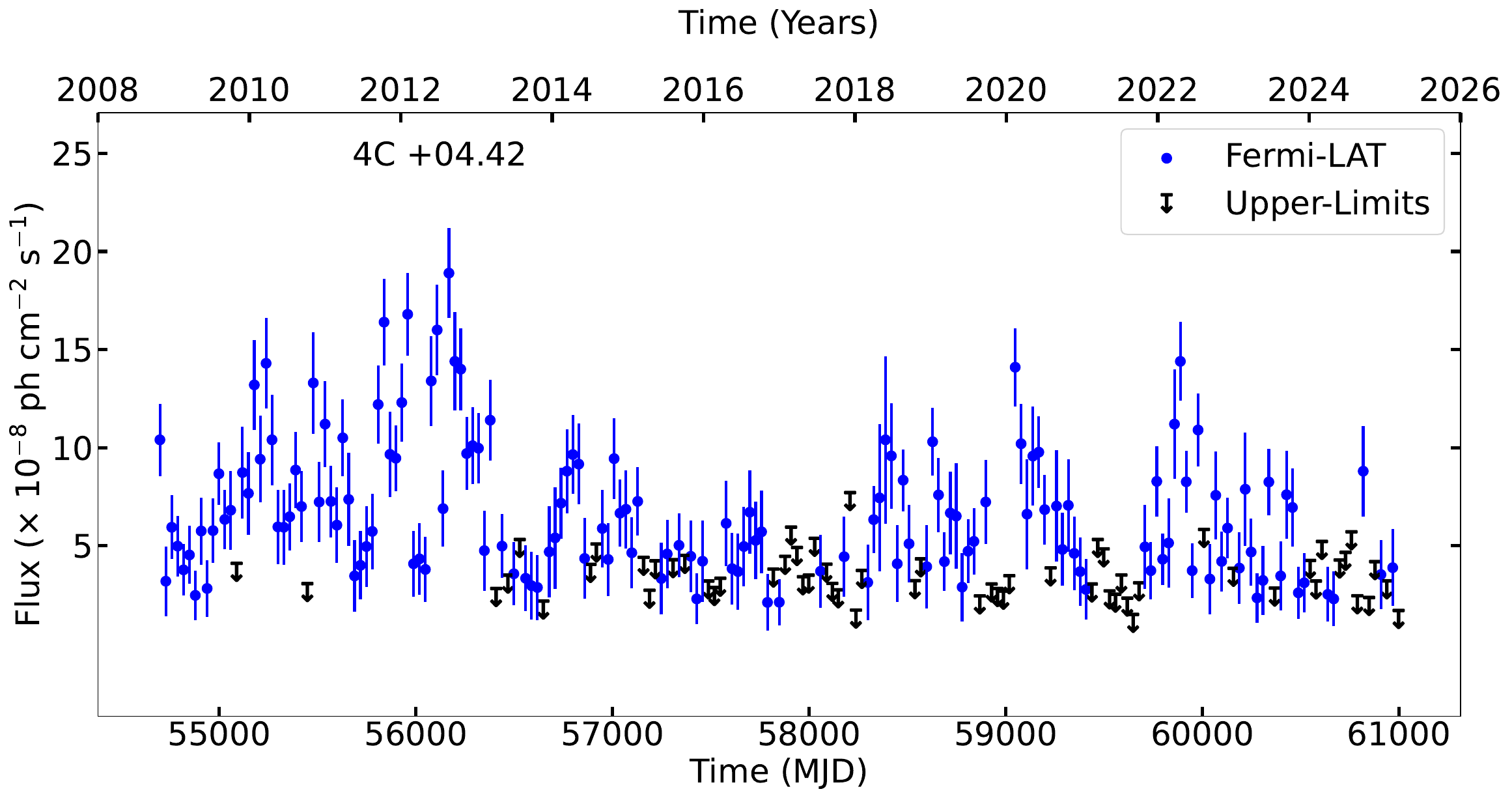}
        \includegraphics[scale=0.2195]{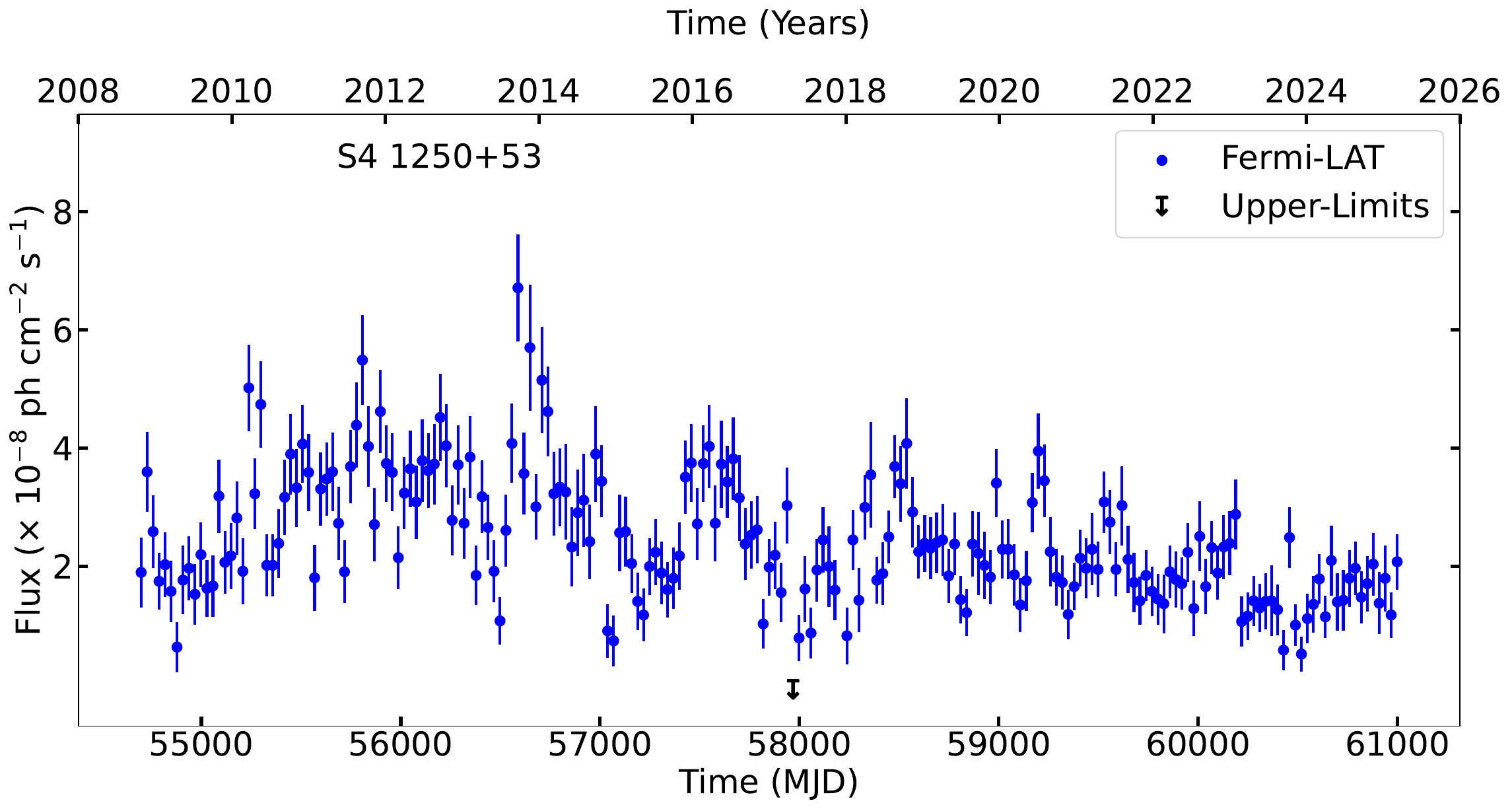}
        \includegraphics[scale=0.2195]{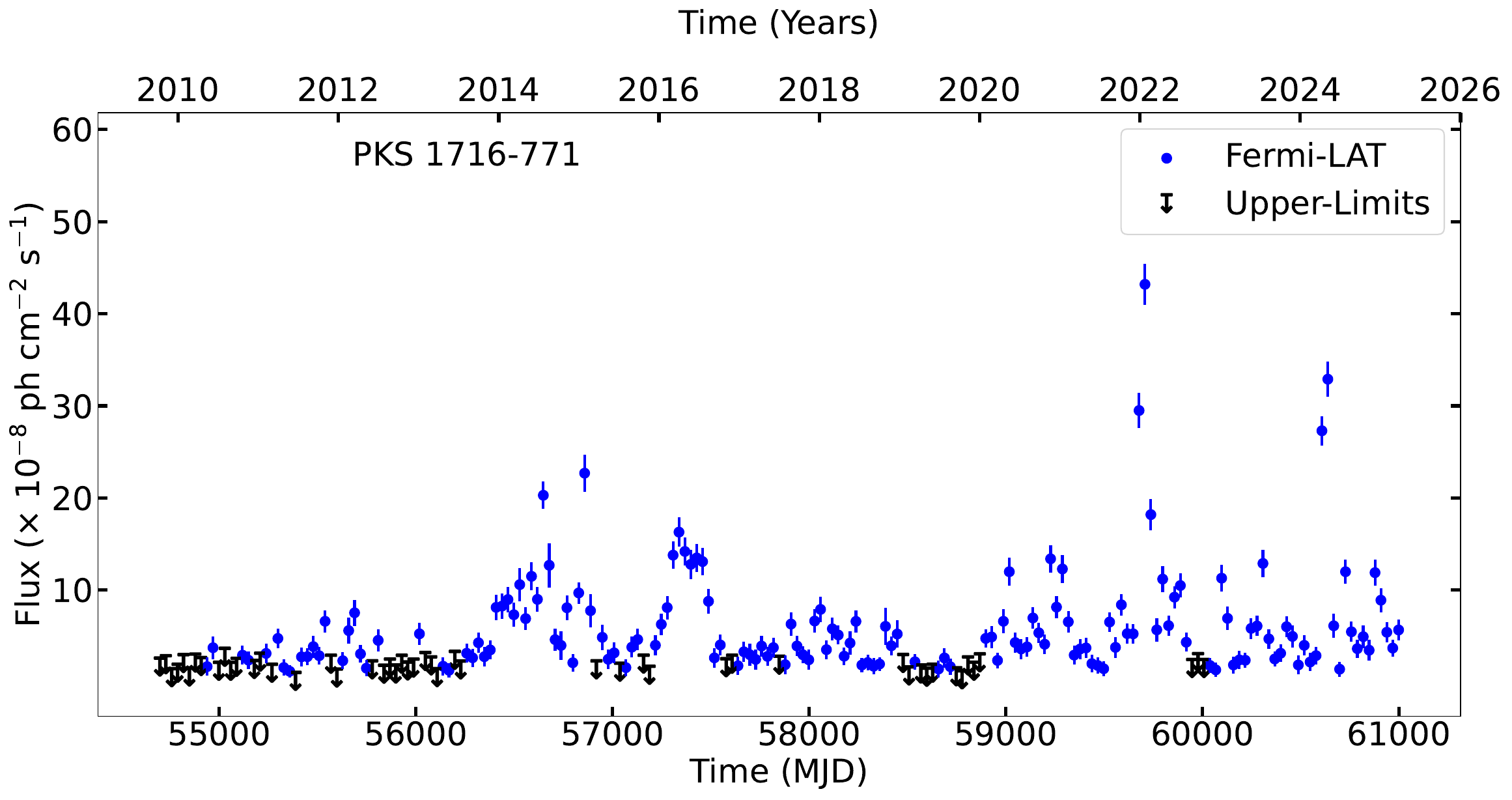}
         \caption{Light curves of several sources mentioned in the main text in the context of the different results presented in this paper. \label{fig:other_sources}}
\end{figure*}

\setcounter{figure}{0}
\begin{figure*}
	\centering
        \includegraphics[scale=0.2195]{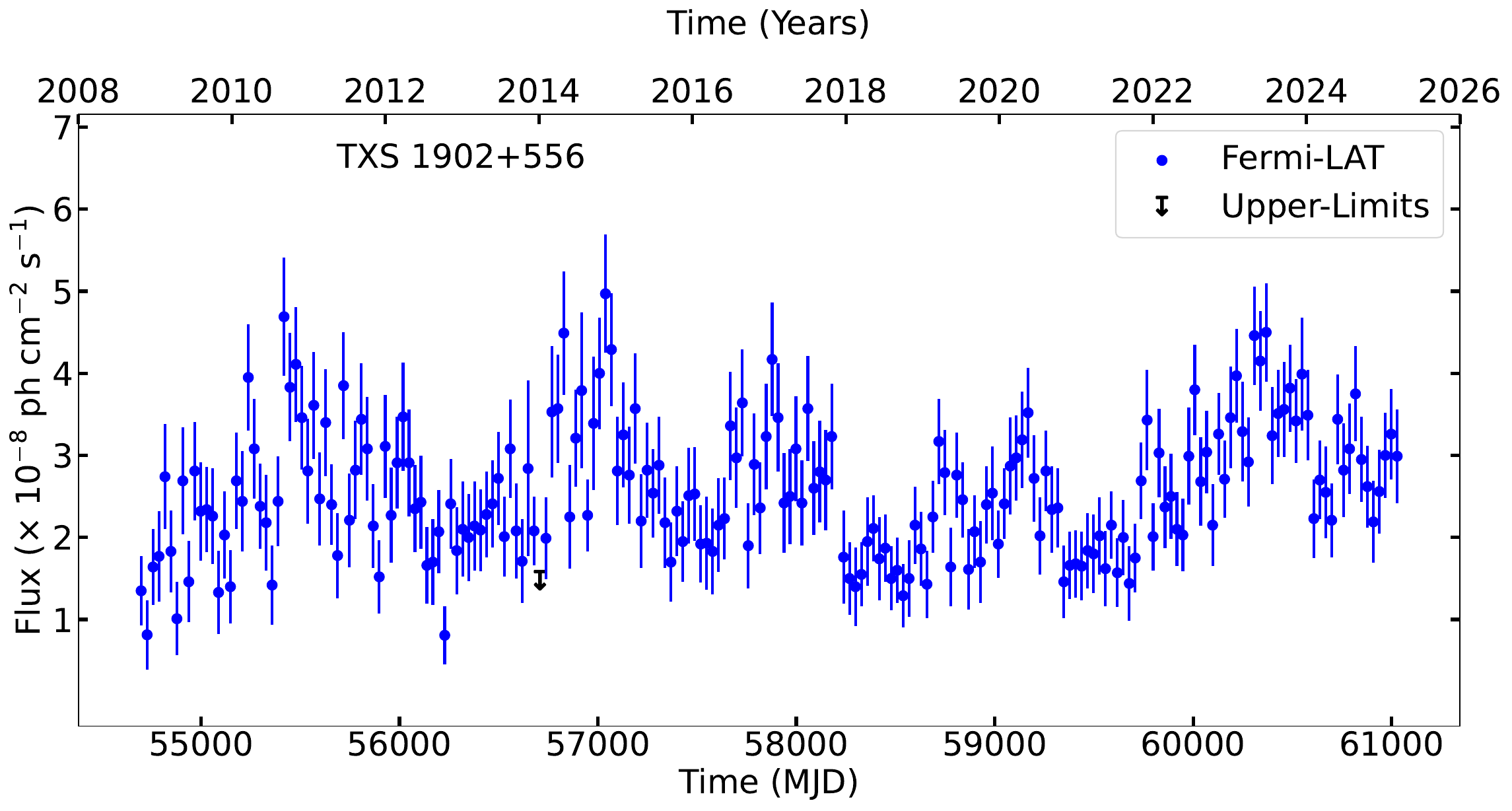}
        \includegraphics[scale=0.2195]{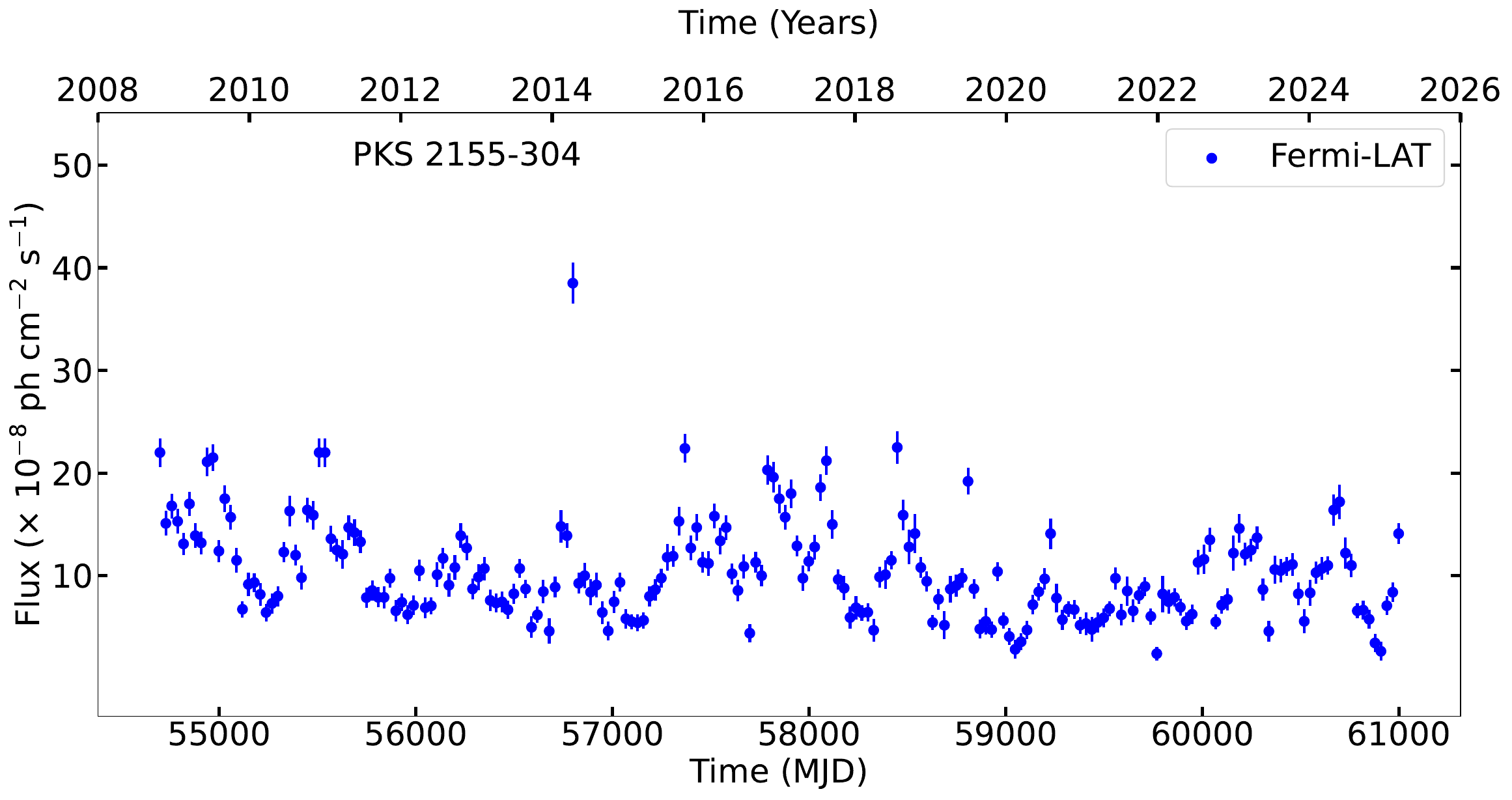}
        \includegraphics[scale=0.2195]{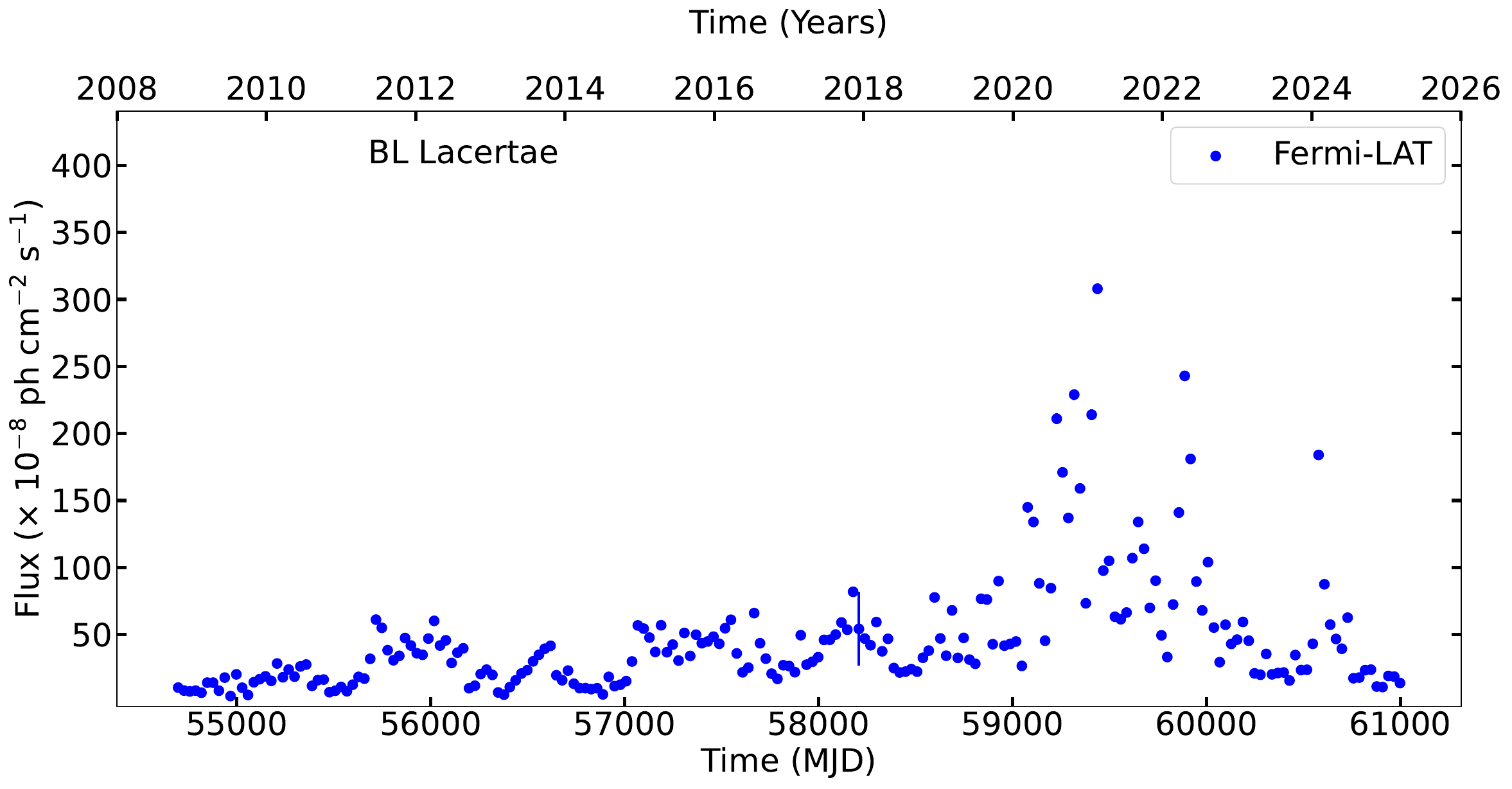}
        \includegraphics[scale=0.2195]{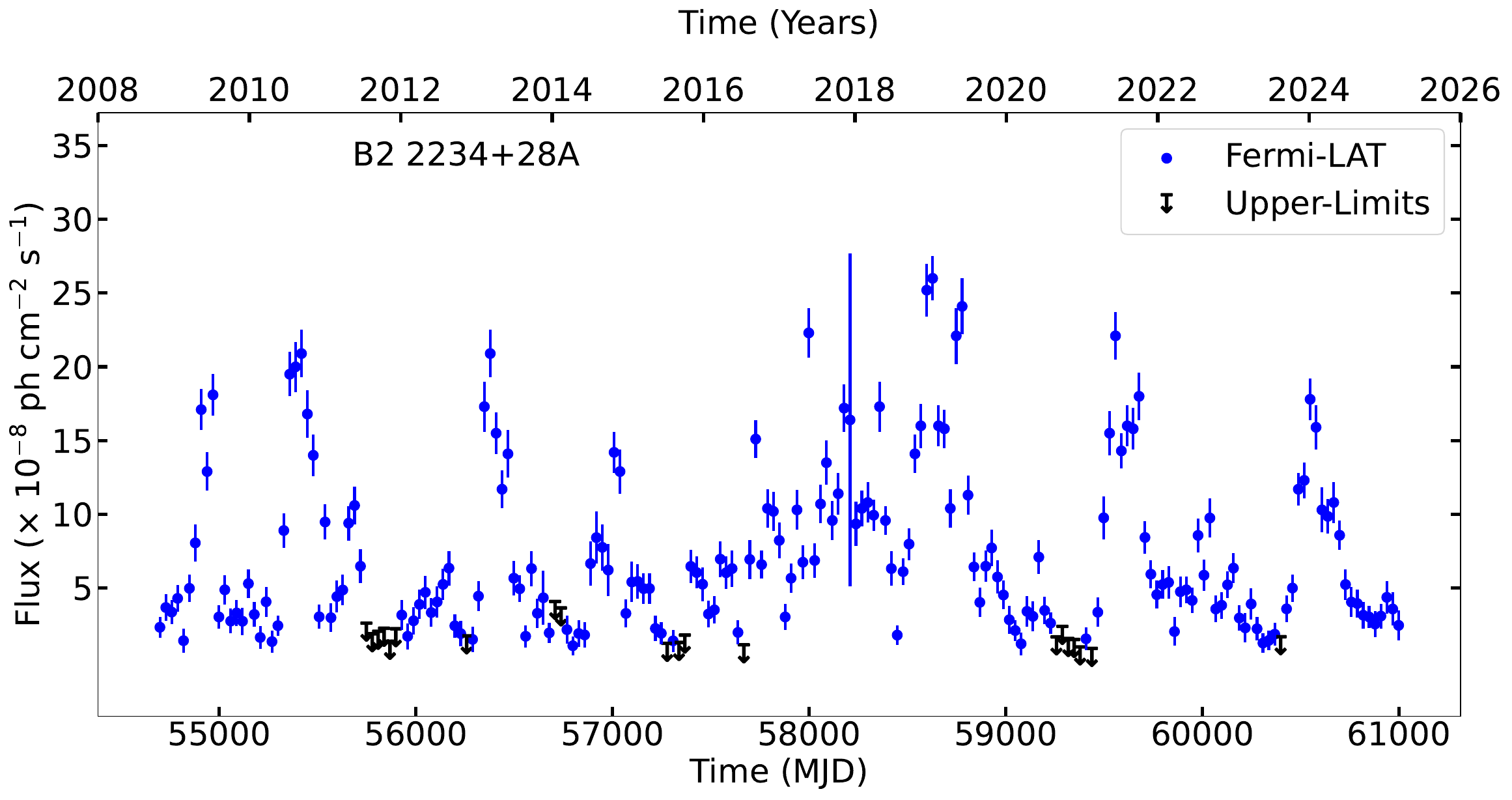}
         \caption{(Continued)).}
\end{figure*}

\subsubsection{Artificial light-curve tests for PG~1553+113}
\begin{figure*}
	\centering
         \includegraphics[scale=0.255]{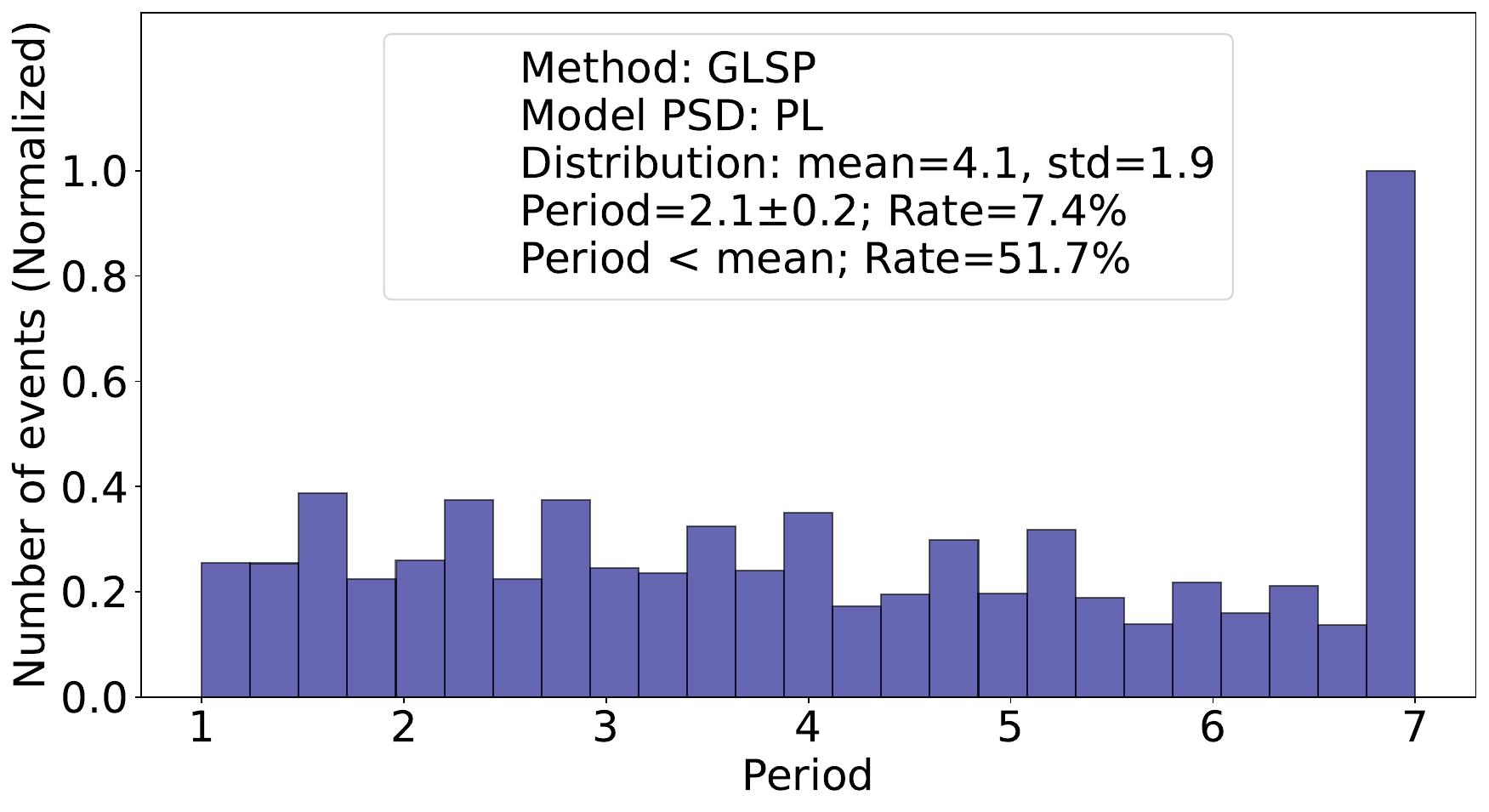}
         \includegraphics[scale=0.255]{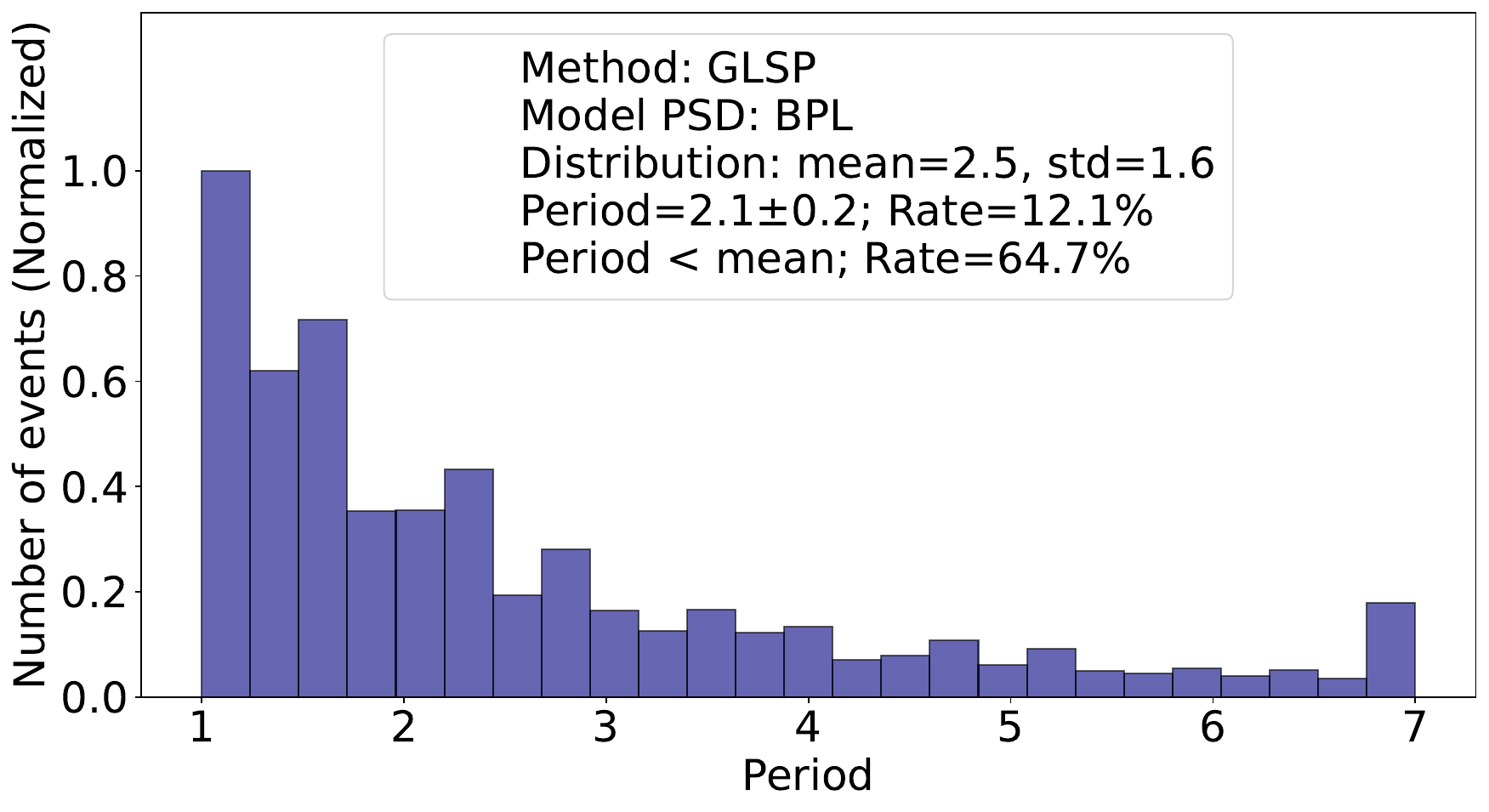}
         \includegraphics[scale=0.255]{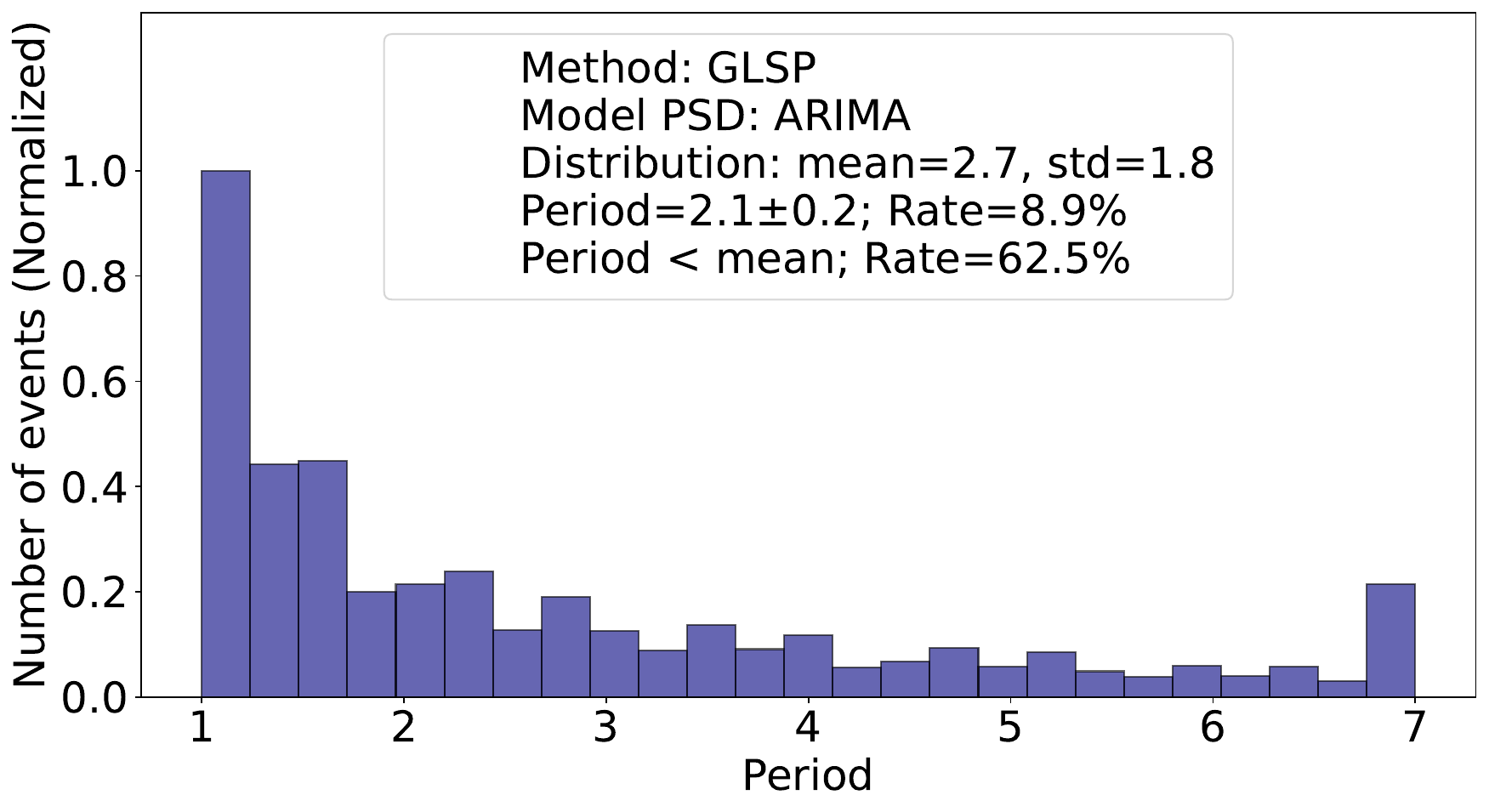}
         \includegraphics[scale=0.255]{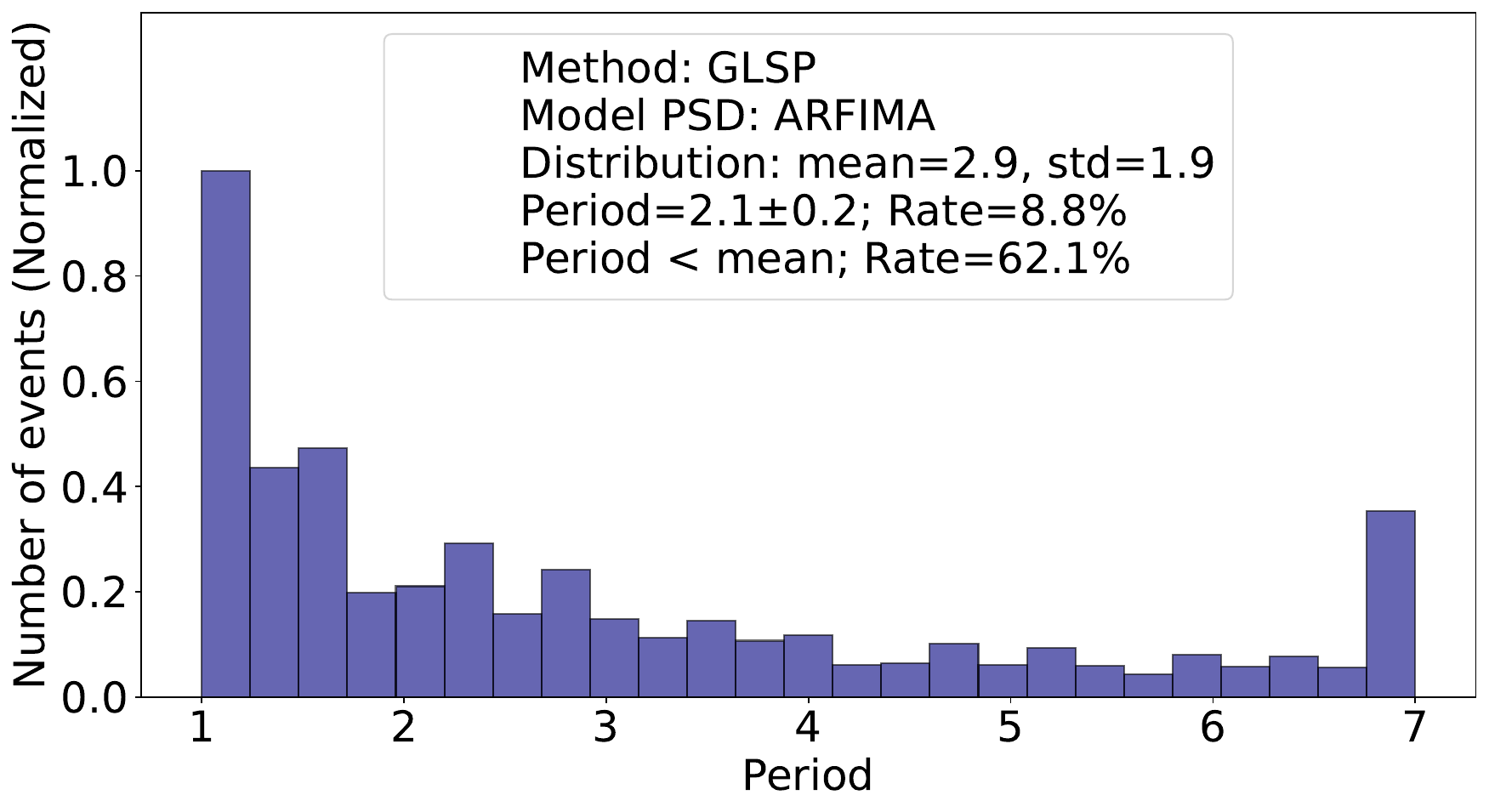}
         \includegraphics[scale=0.255]{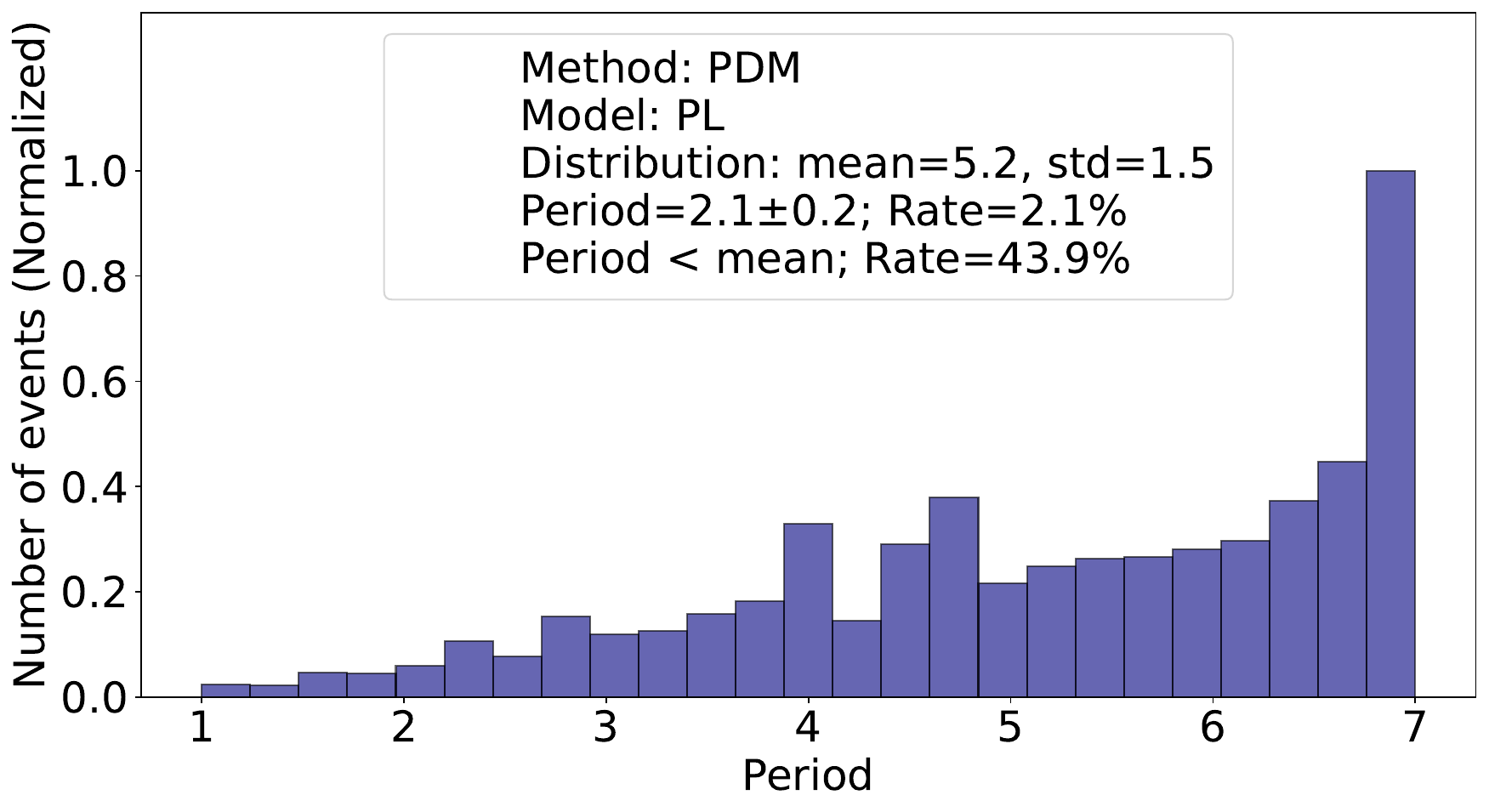}
         \includegraphics[scale=0.255]{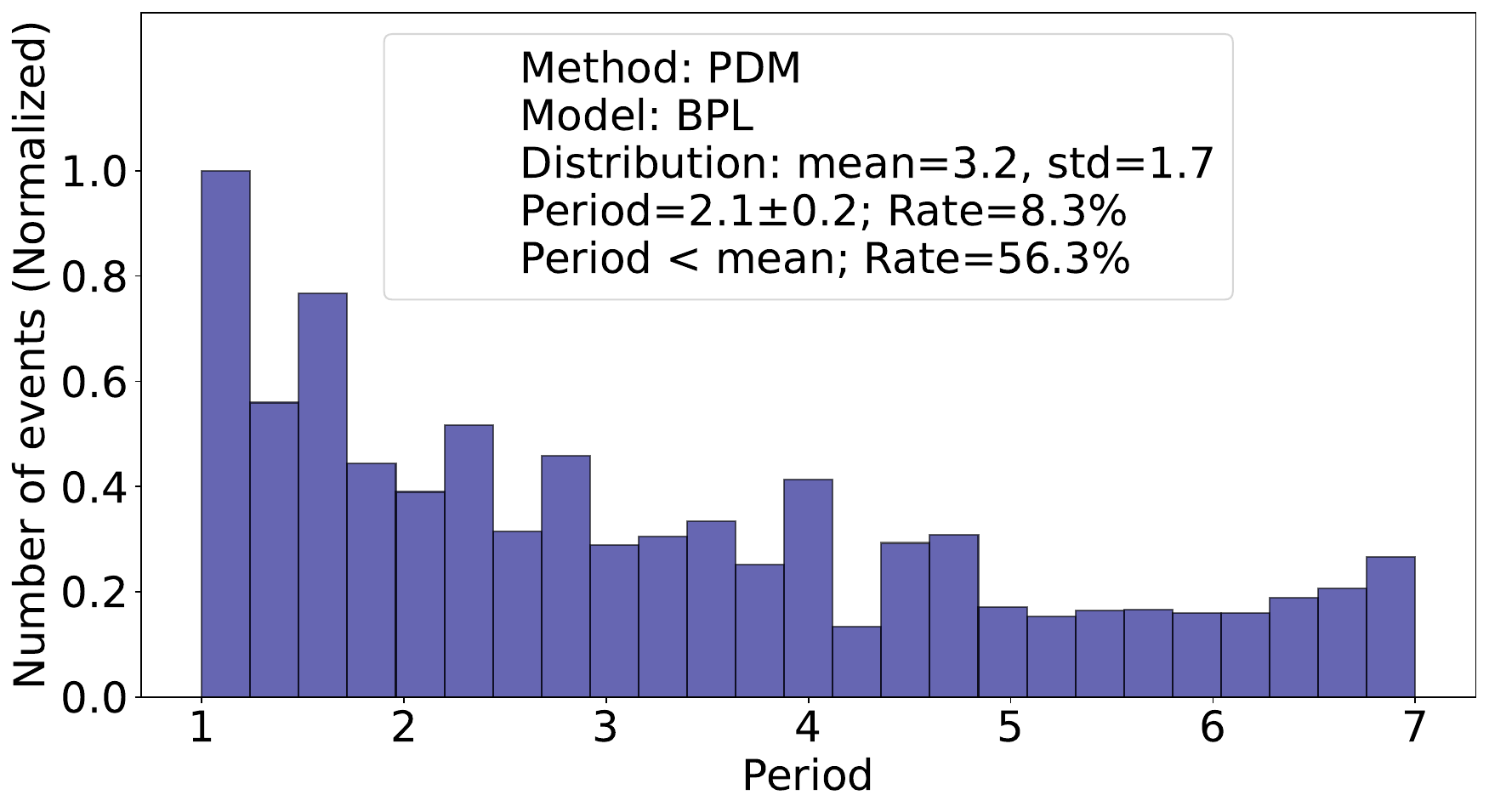}
         \includegraphics[scale=0.255]{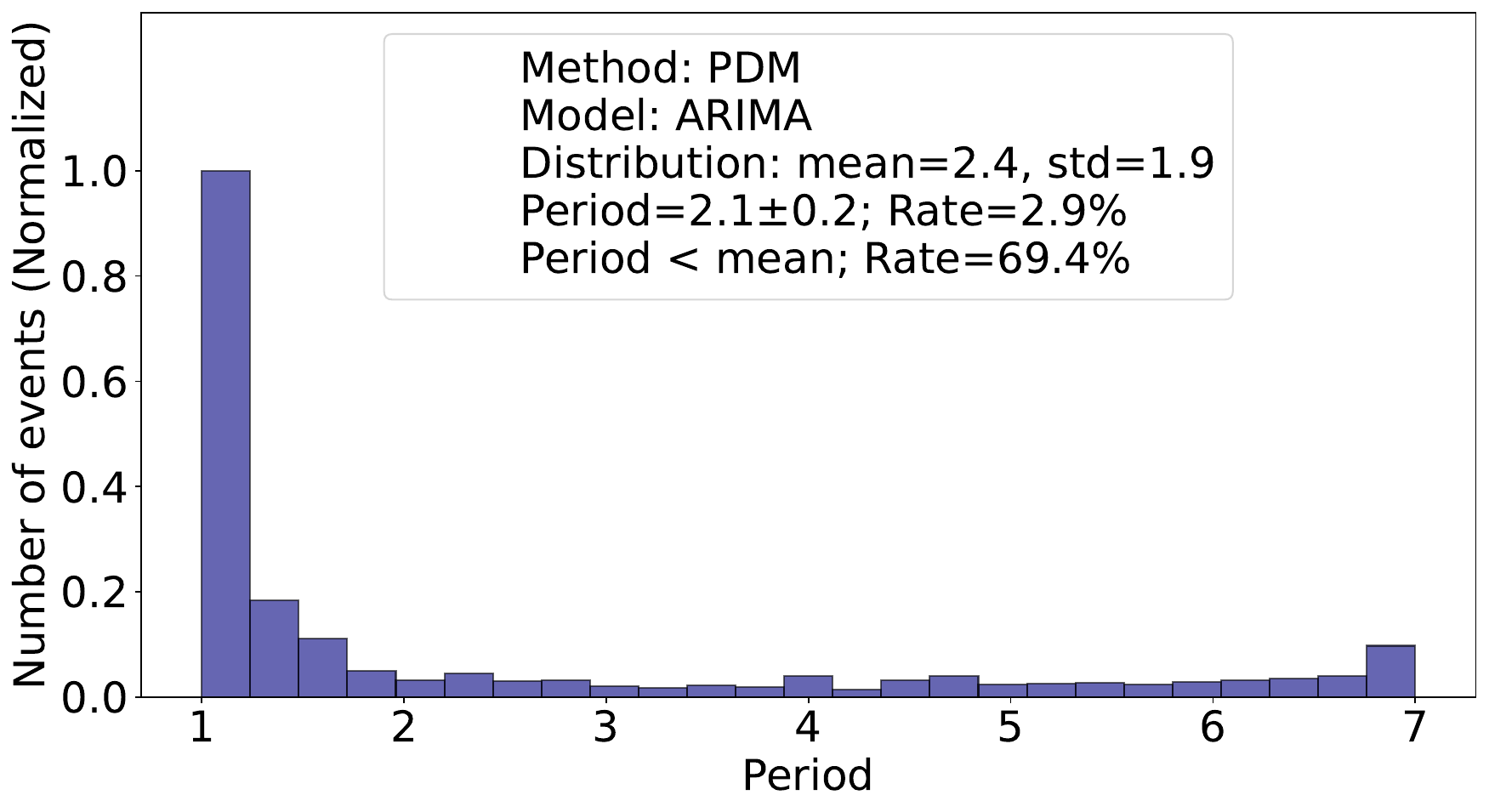}
         \includegraphics[scale=0.255]{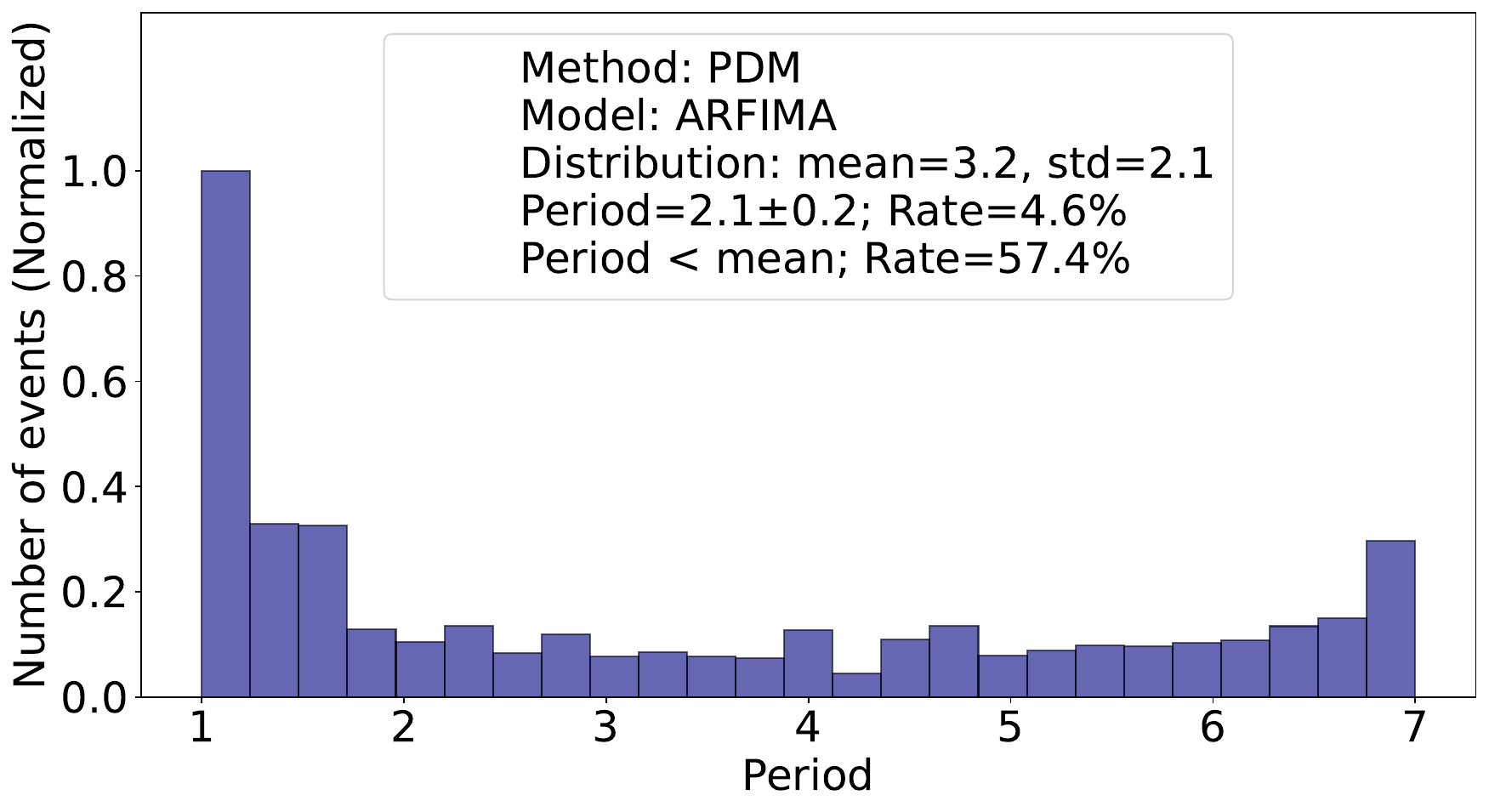}
         \caption{Distribution of periods recovered from artificial LCs generated with four stochastic-noise models: a simple power law (PL), a bending power law (BPL), ARIMA, and ARFIMA. For each surrogate ensemble, we apply the GLSP and PDM and report the mean period and standard deviation of the recovered-period distribution. We also indicate (i) the fraction of surrogates whose recovered period falls within the candidate-period interval of PG 1553+113, 2.1$\pm$0.2 yr, and (ii) the fraction with recovered periods shorter than the ensemble mean. Overall, PL surrogates yield an approximately uniform distribution across the searched range, whereas BPL, ARIMA, and ARFIMA preferentially populate periods $\lesssim$ mean period. The period is in years. \label{fig:period_distribution_tests}}
\end{figure*}

\begin{figure*}
	\centering
         \includegraphics[scale=0.255]{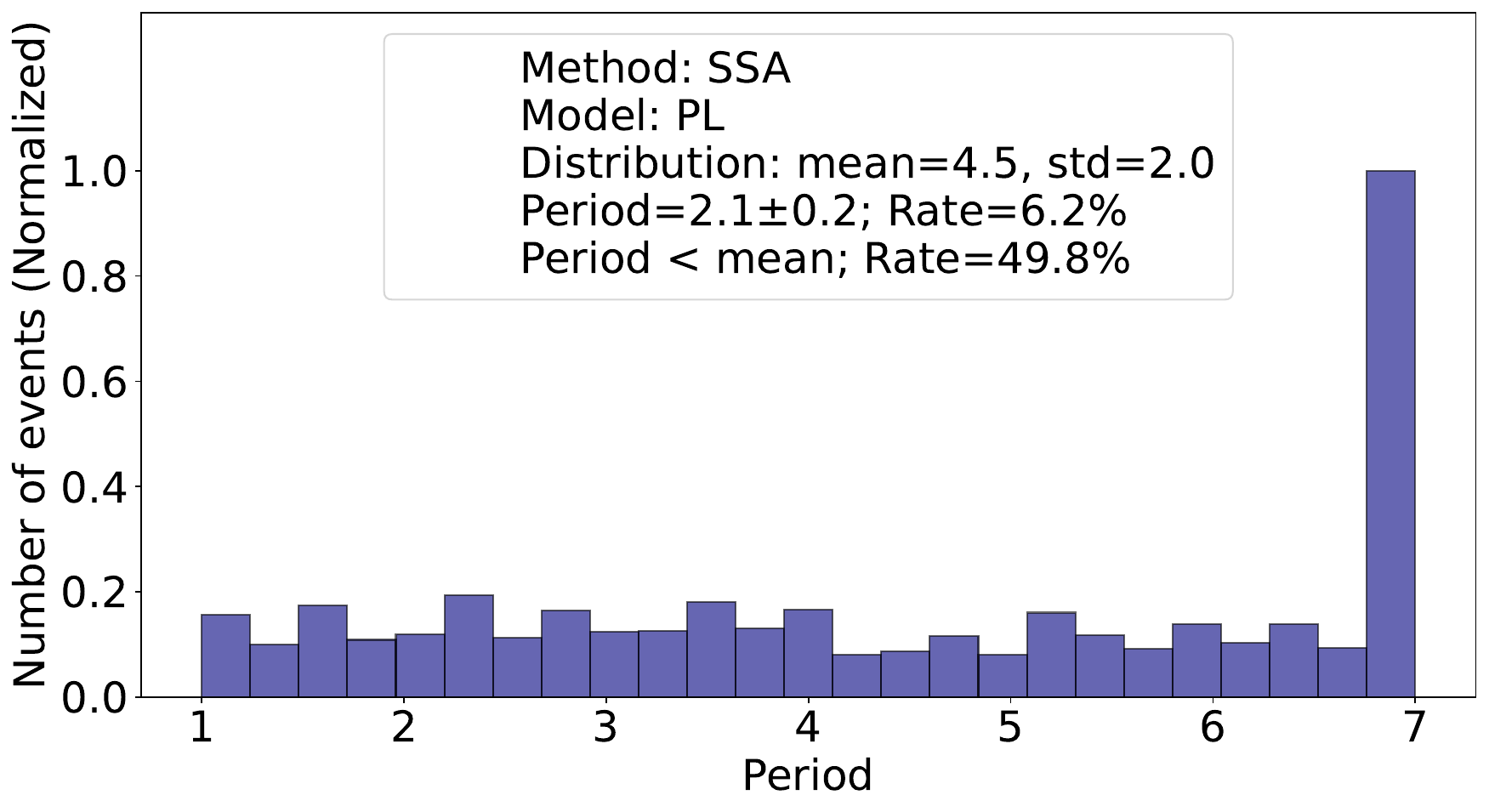}
         \includegraphics[scale=0.255]{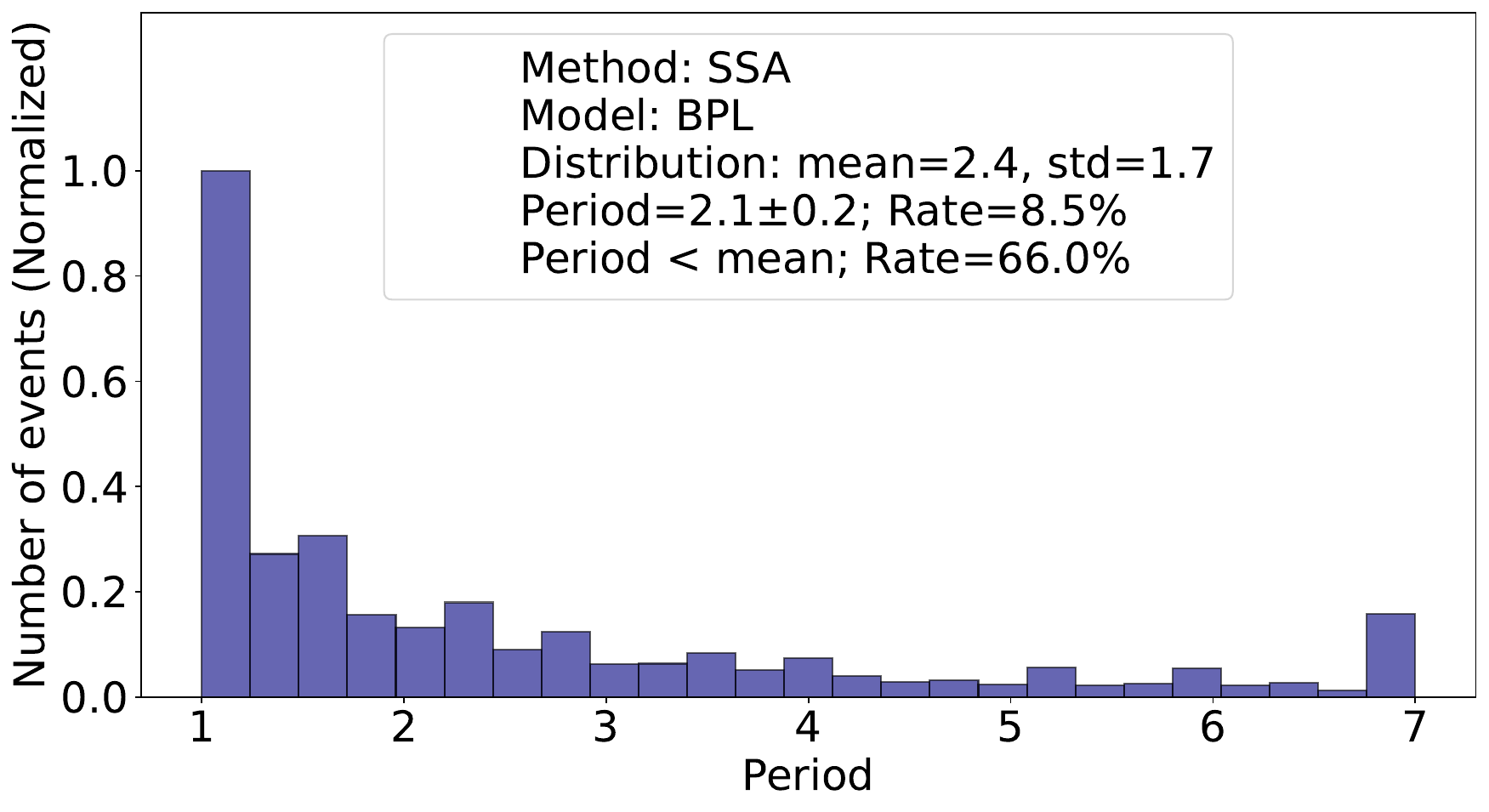}
         \includegraphics[scale=0.255]{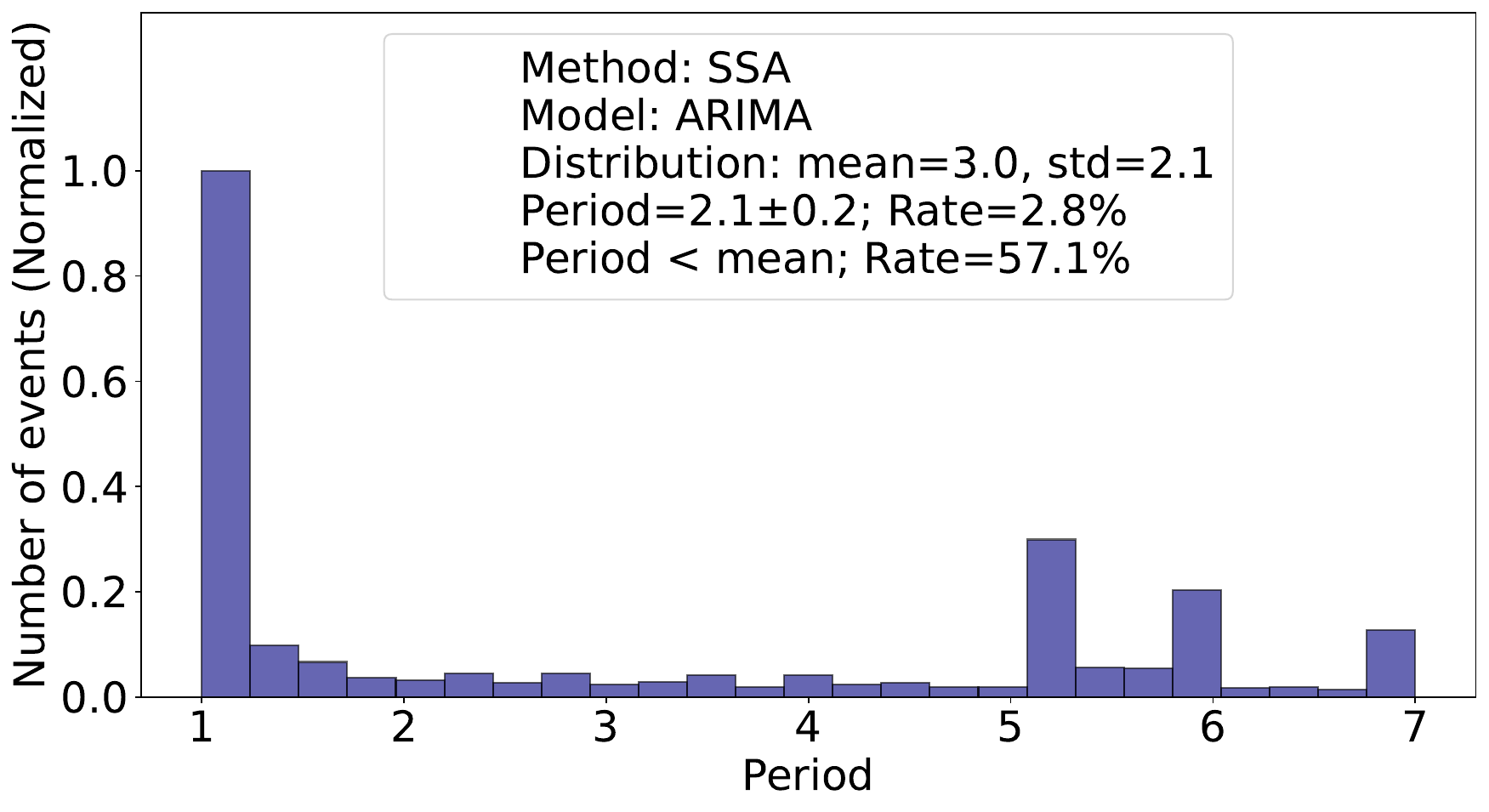}
         \includegraphics[scale=0.255]{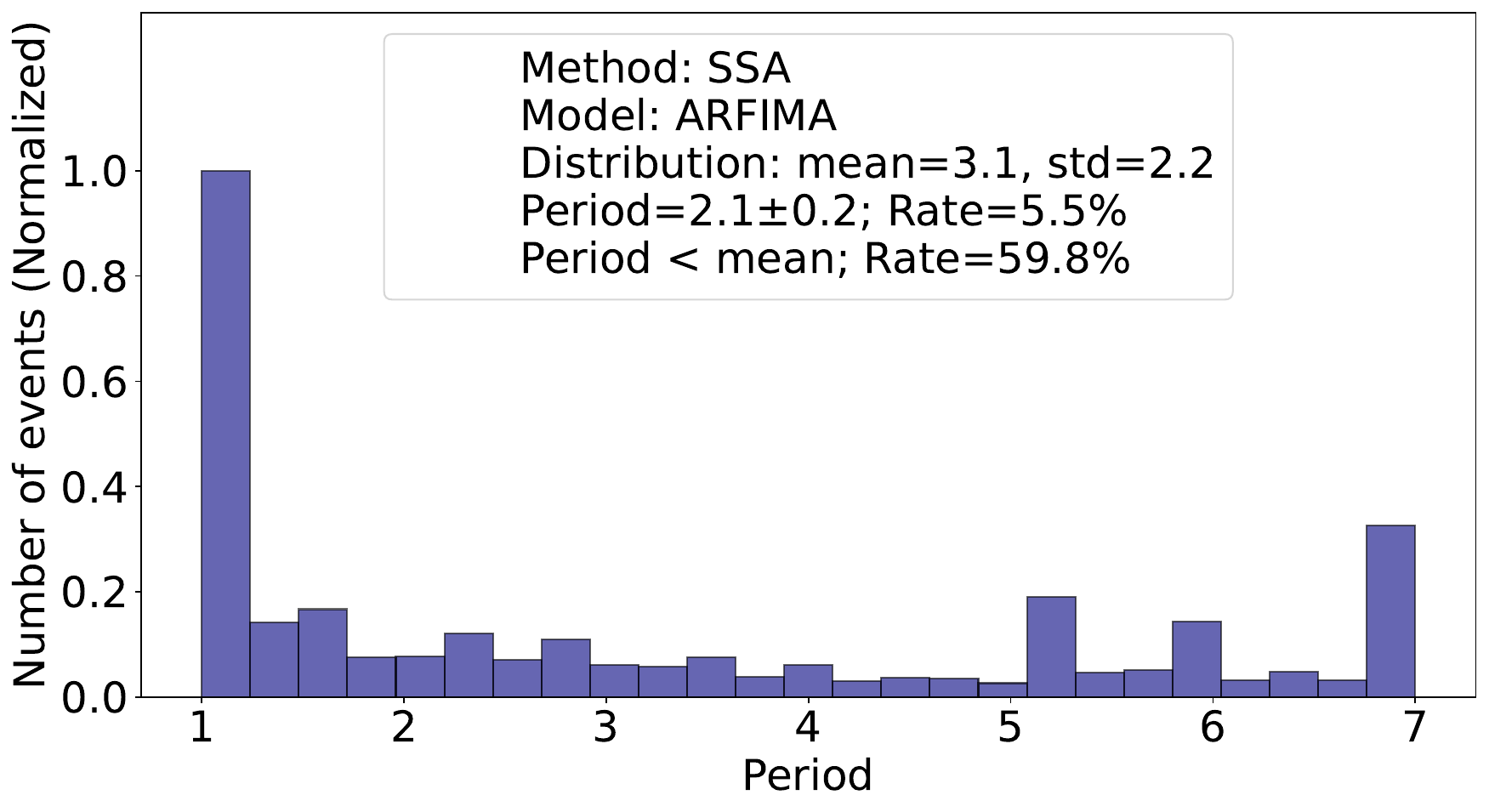}
         \caption{Distribution of periods recovered from artificial LCs generated with four stochastic-noise models: a simple power law (PL), a bending power law (BPL), ARIMA, and ARFIMA. For each surrogate ensemble, we apply the SSA combined with the LSP and report the mean period and standard deviation of the recovered-period distribution. We also indicate (i) the fraction of surrogates whose recovered period falls within the candidate-period interval of PG 1553+113, 2.1$\pm$0.2 yr, and (ii) the fraction with recovered periods shorter than the ensemble mean. Overall, PL surrogates yield an approximately uniform distribution across the searched range, whereas BPL, ARIMA, and ARFIMA preferentially populate periods $\lesssim$ mean period. The period is in years.\label{fig:period_distribution_tests_ssa}}
\end{figure*}

\begin{figure*}
	\centering
         \includegraphics[scale=0.285]{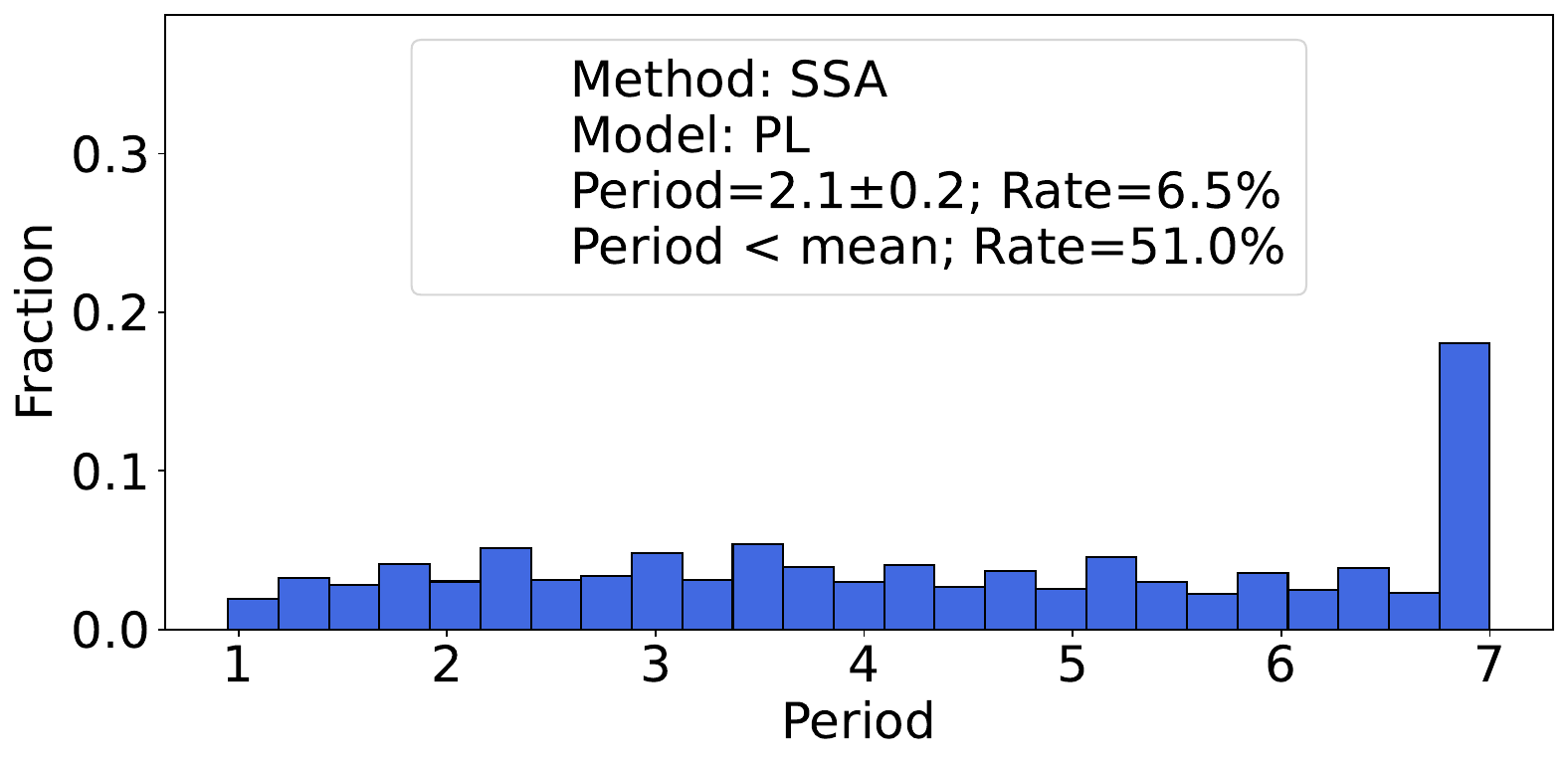}
         \includegraphics[scale=0.285]{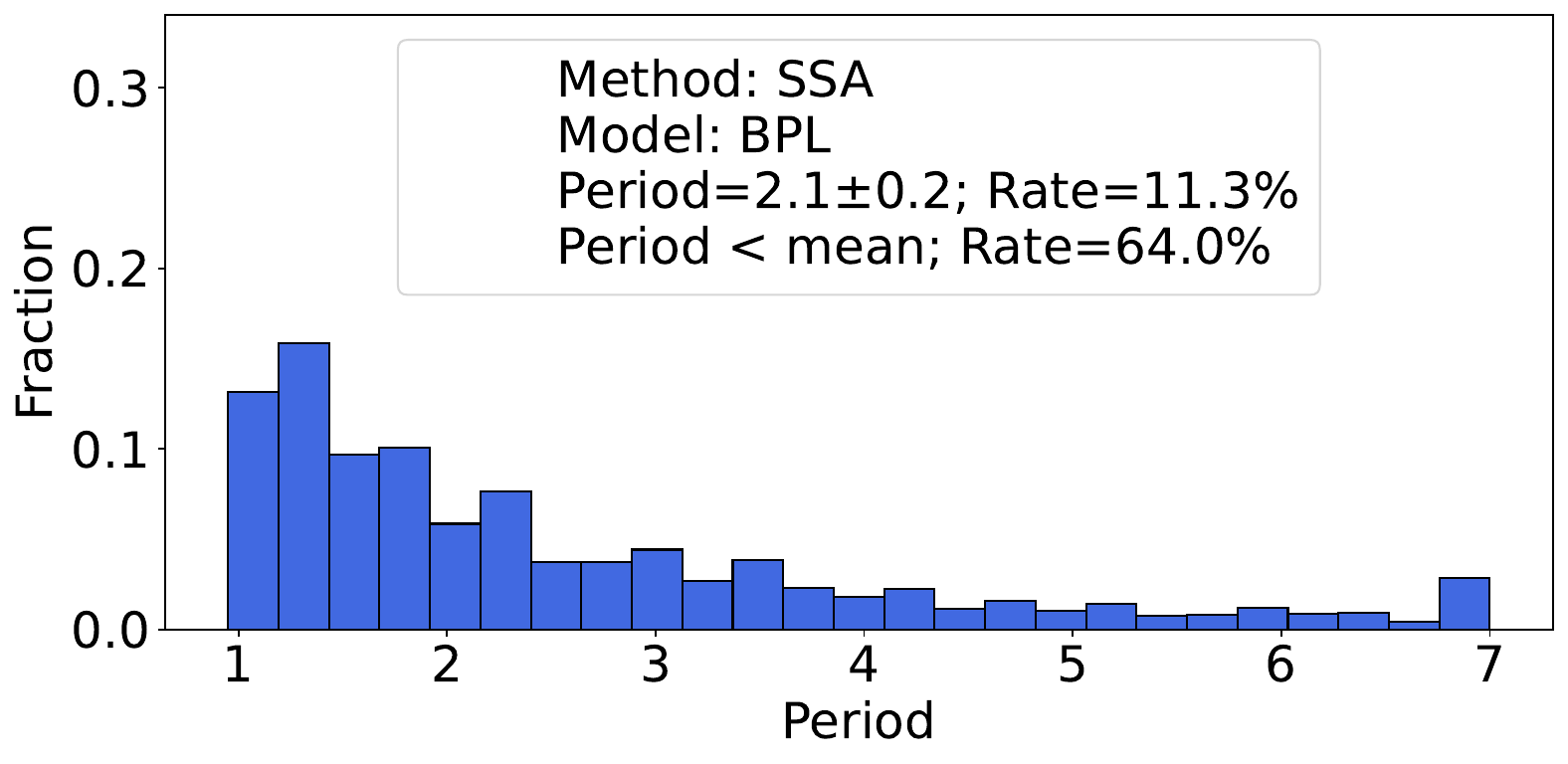}
         \includegraphics[scale=0.285]{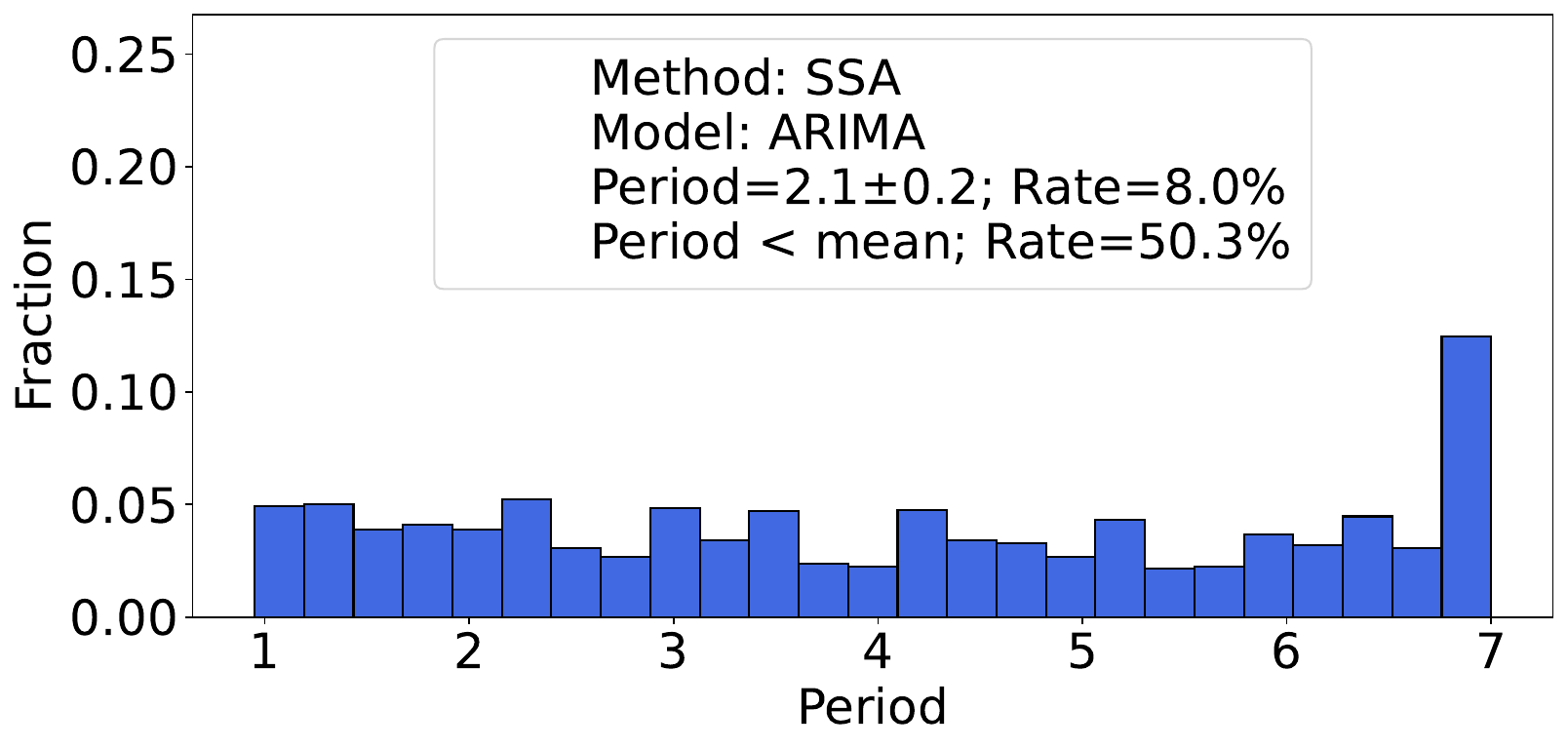}
         \includegraphics[scale=0.285]{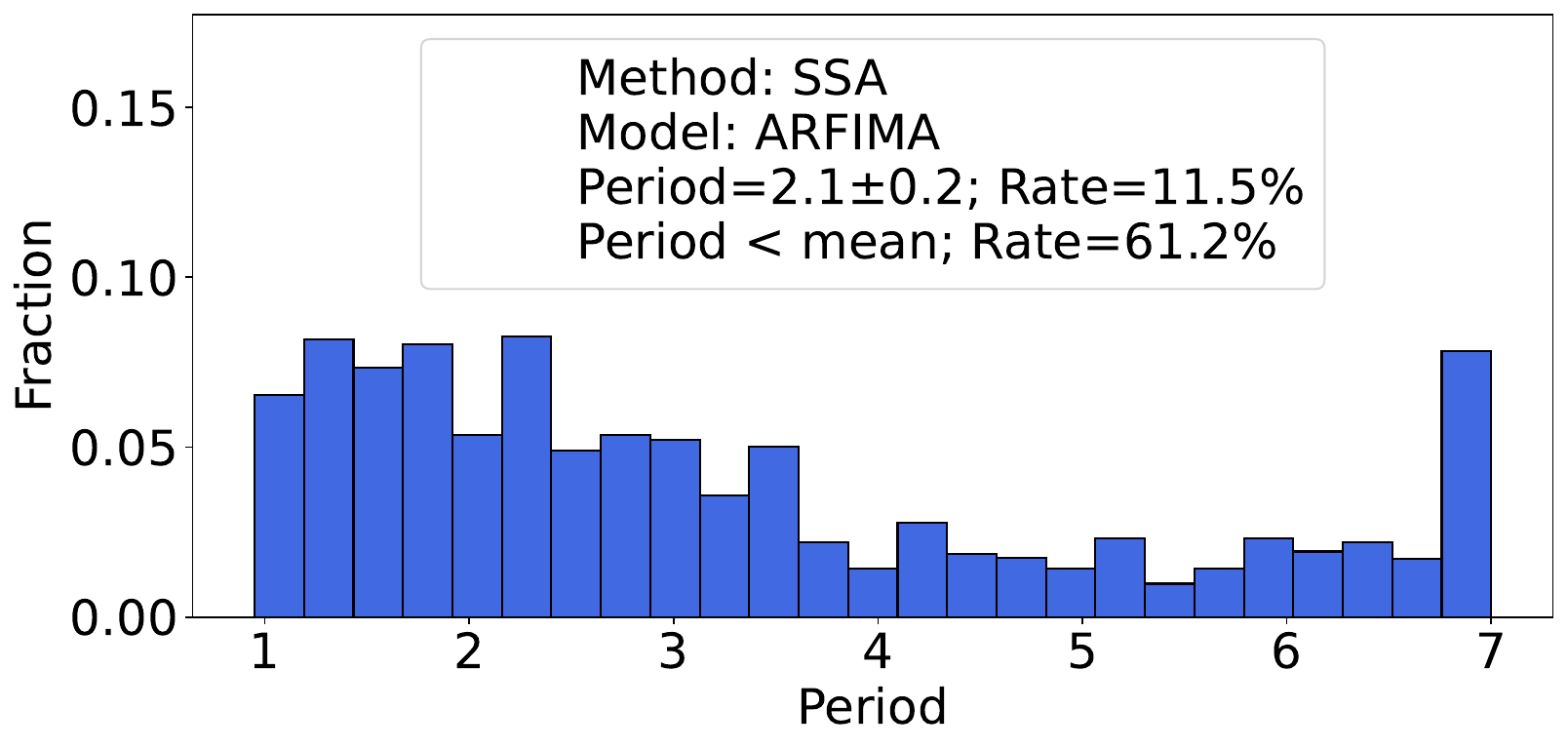}
         \caption{Distribution of periods recovered from artificial LCs that, after applying SSA+LSP, yield a maximum LSP power higher than that measured in the original LC of PG 1553+113. Artificial light curves are generated using four stochastic-noise models: a simple power law (PL), a bending power law (BPL), ARIMA, and ARFIMA. For the PL and ARIMA ensembles, the recovered periods are approximately uniformly distributed across the searched range, aside from an excess in the final period bin attributable to red-noise-dominated realizations and boundary effects at the longest periods. In contrast, the BPL and ARFIMA ensembles show a clear concentration of recovered periods at $\lesssim$ the ensemble mean period, indicating that high-power SSA+LSP peaks in these models preferentially arise at shorter characteristic periods. The period is in years. \label{fig:power_distribution_tests_ssa}}
\end{figure*}

Figures~\ref{fig:period_distribution_tests}--\ref{fig:power_distribution_tests_ssa} summarize the behavior of the artificial light curves used to estimate the test statistics for PG~1553+113. These diagnostics are included to verify whether the artificial light curves generated under each null hypothesis reproduce broad features of the period-search output, and to assess whether high-power peaks are uniformly distributed across the searched period range or preferentially concentrated at specific timescales.

\subsubsection{Dependence of the SSA reconstruction on window length}
\begin{figure*}
	\centering
         \includegraphics[scale=0.225]{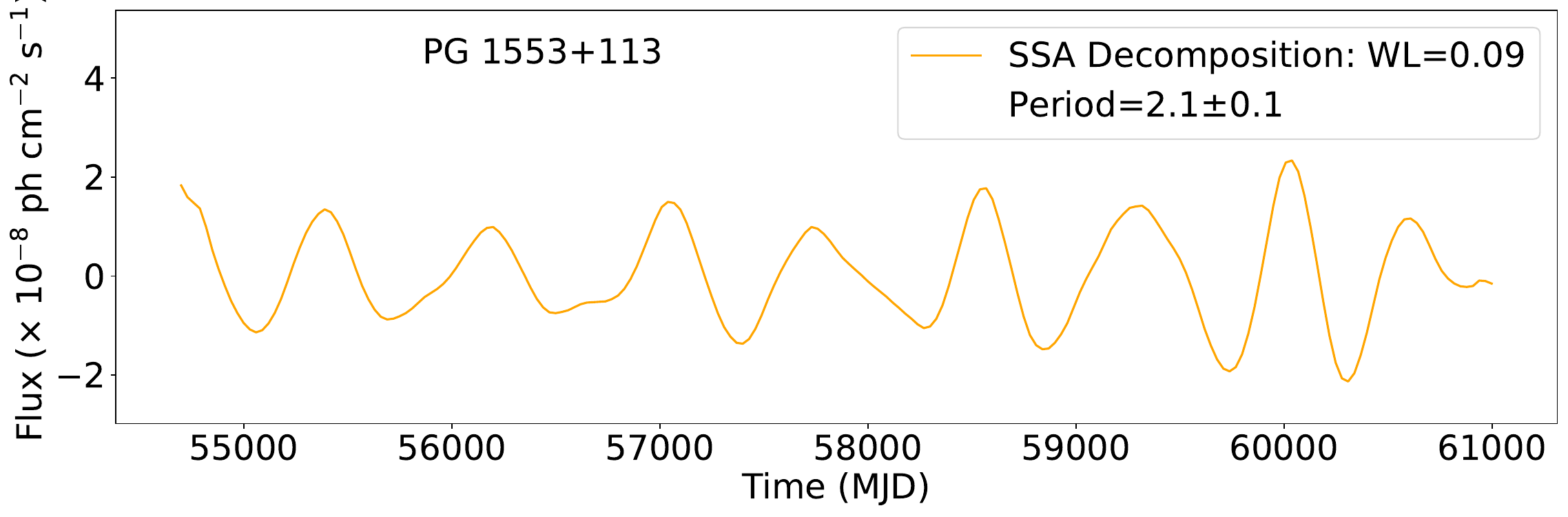}
         \includegraphics[scale=0.225]{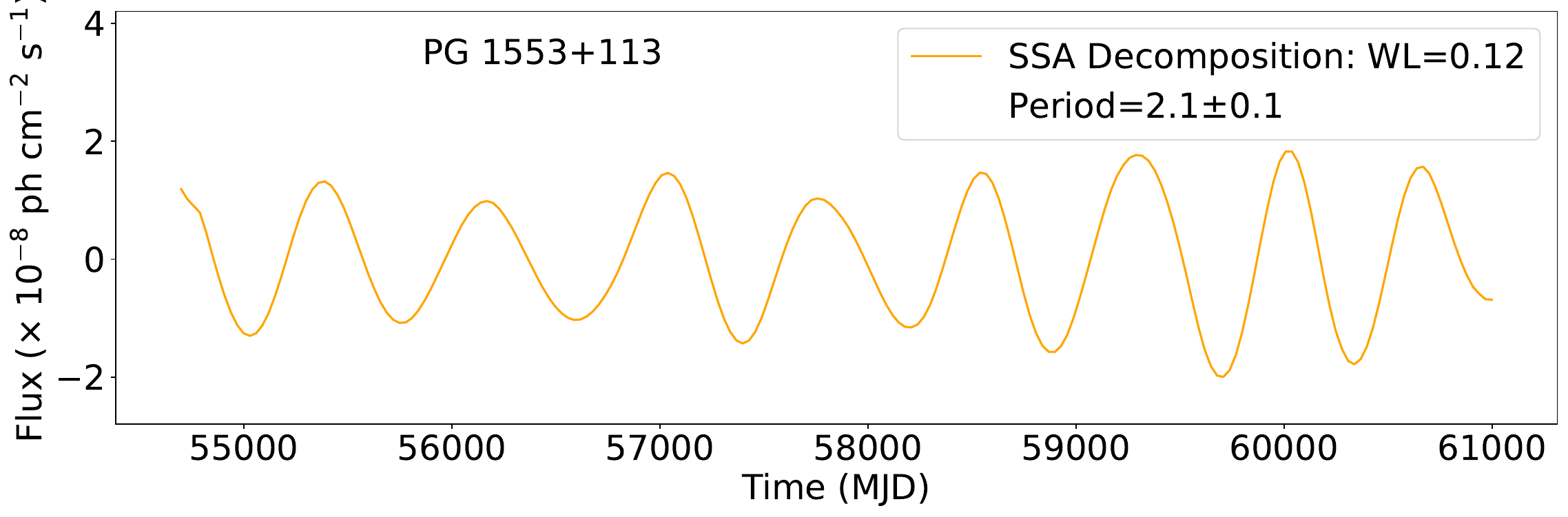}
         \includegraphics[scale=0.225]{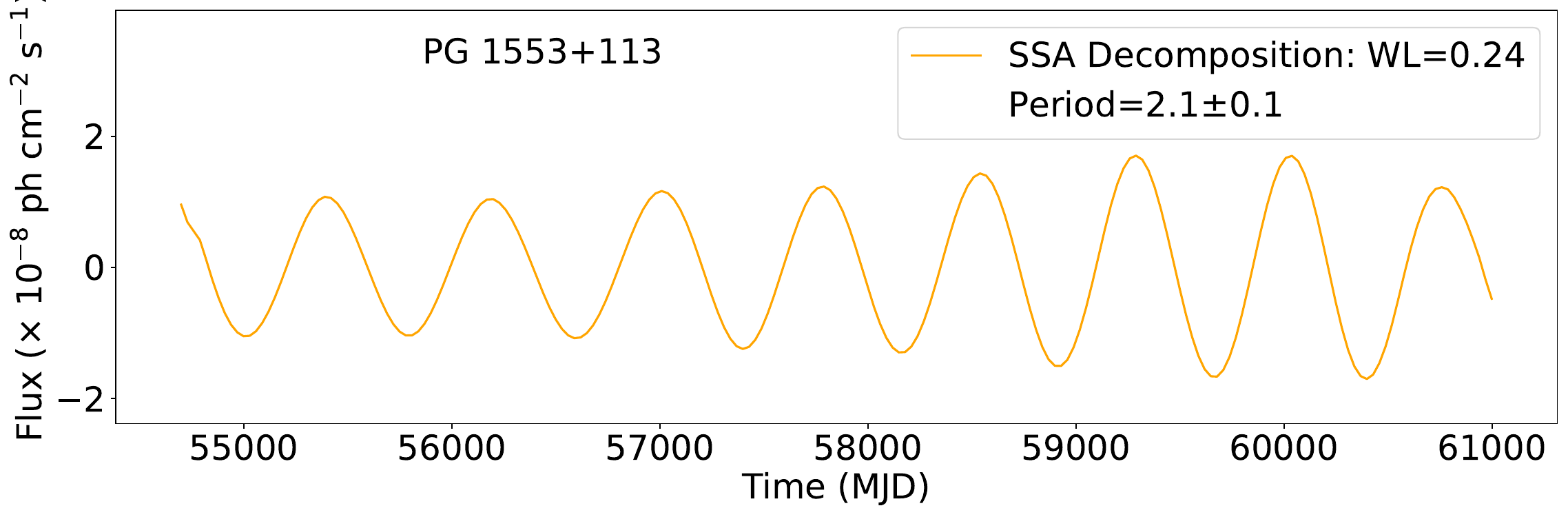}
         \includegraphics[scale=0.225]{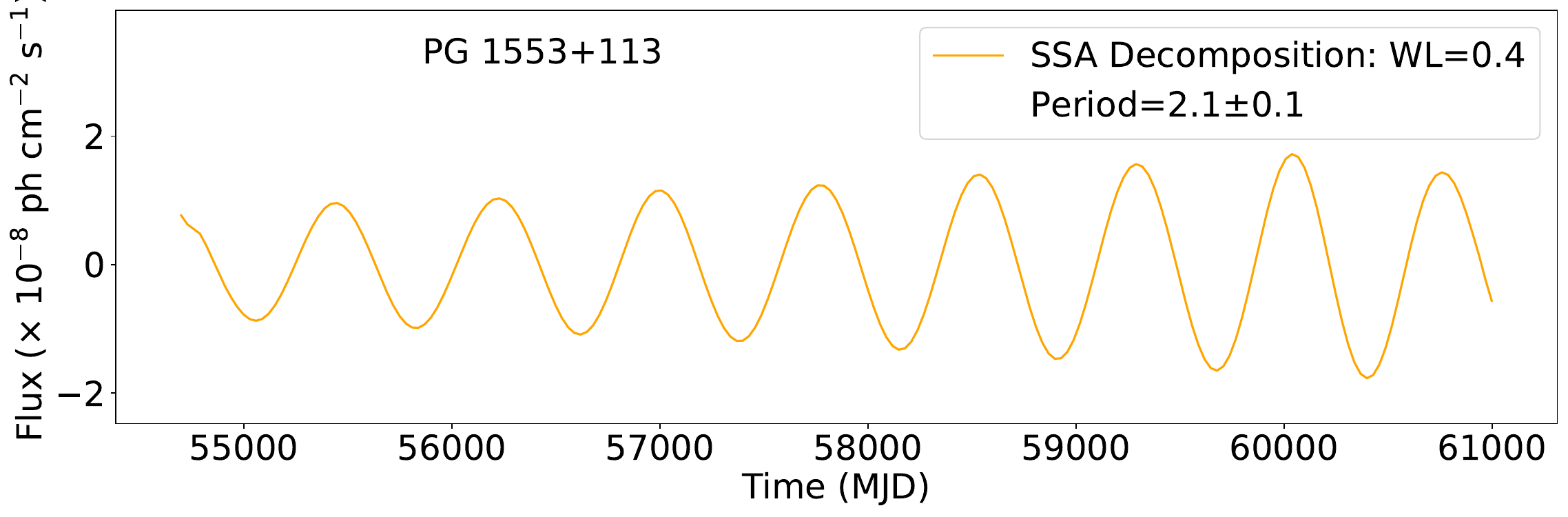}
         \includegraphics[scale=0.225]{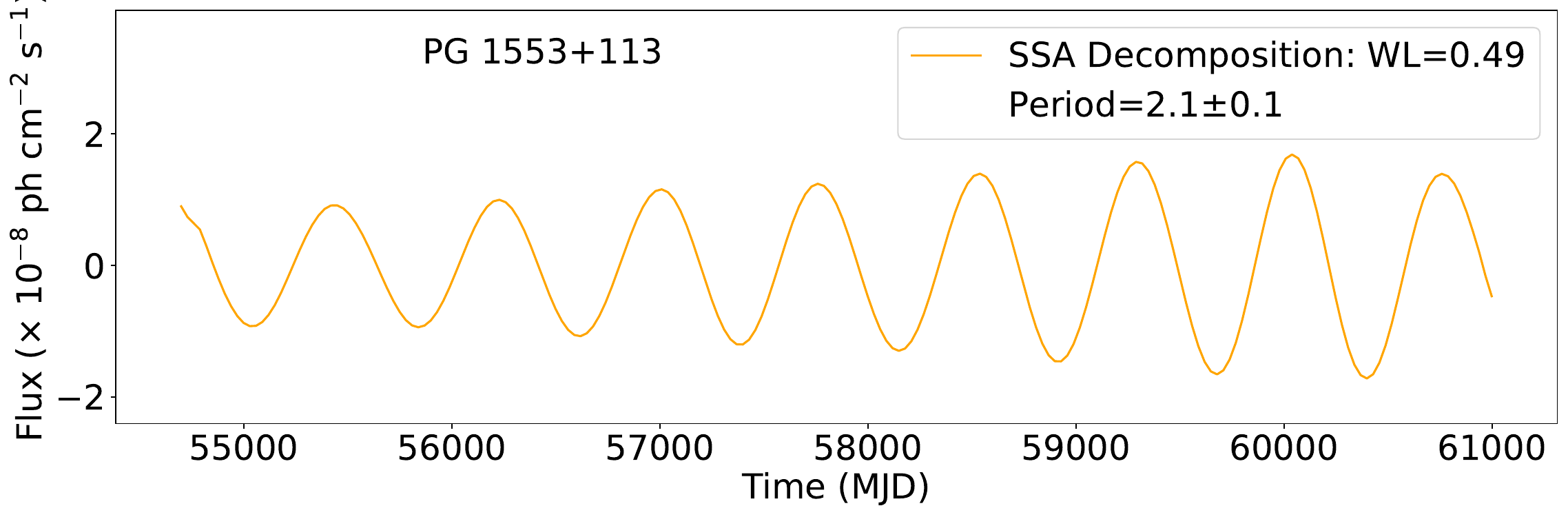}
        \caption{SSA decomposition of the oscillatory component for PG~1553+113, shown for five different window lengths ($WL$): 0.09 and 0.49, corresponding to the limits of the explored range; 0.12 and 0.24, being the optimal $WLs$; and 0.4, the value used to estimate the test statistics reported in Table~\ref{tab:periodicity_results}. The decomposition varies significantly with $WL$: smaller values enhance high-frequency features relative to the case of $WL=0.49$. \label{fig:ssa_descompositions_pg1553}}         
\end{figure*}

\begin{figure*}
	\centering
         \includegraphics[scale=0.225]{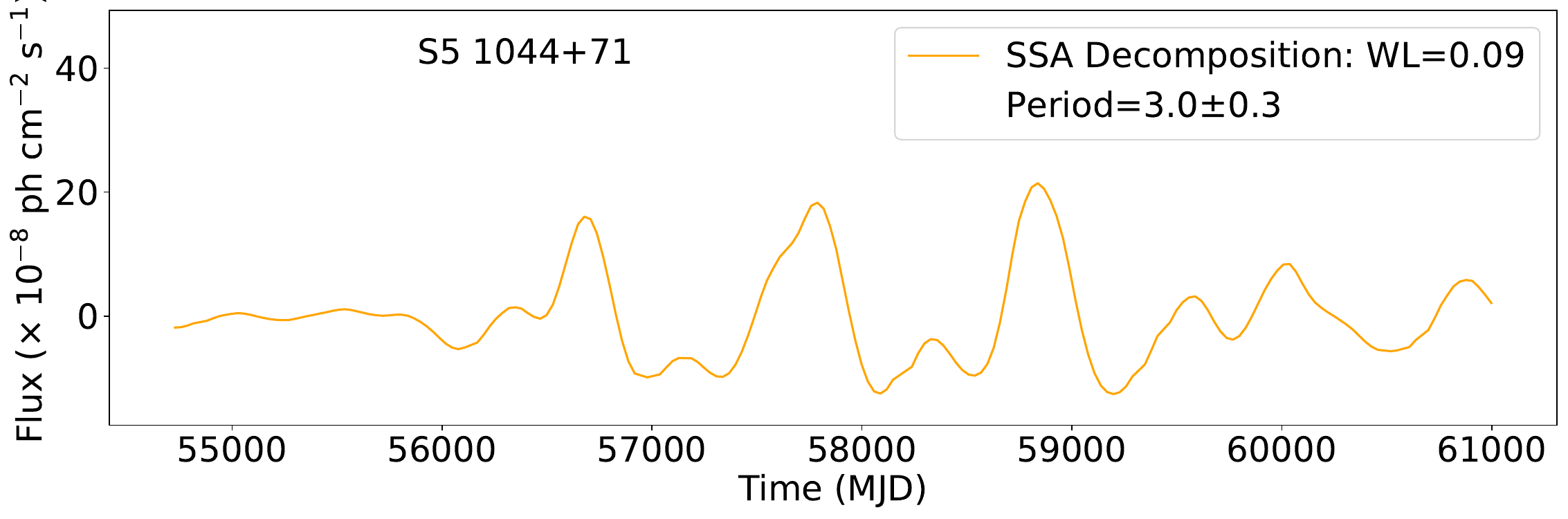}
         \includegraphics[scale=0.225]{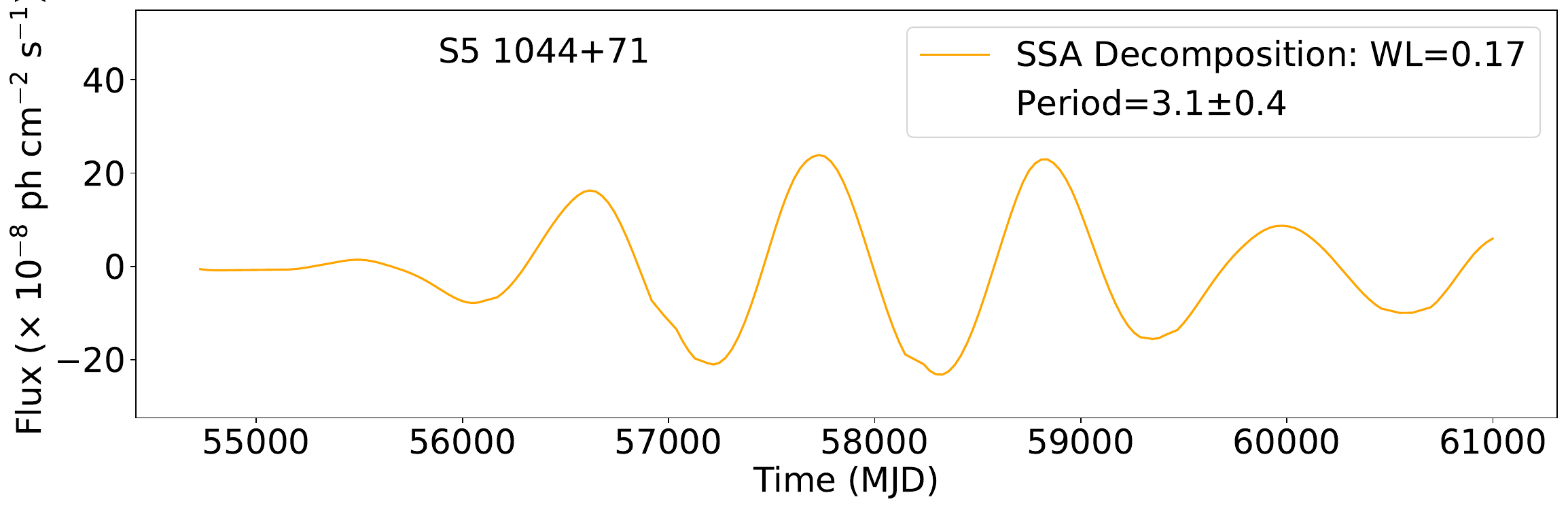}
         \includegraphics[scale=0.225]{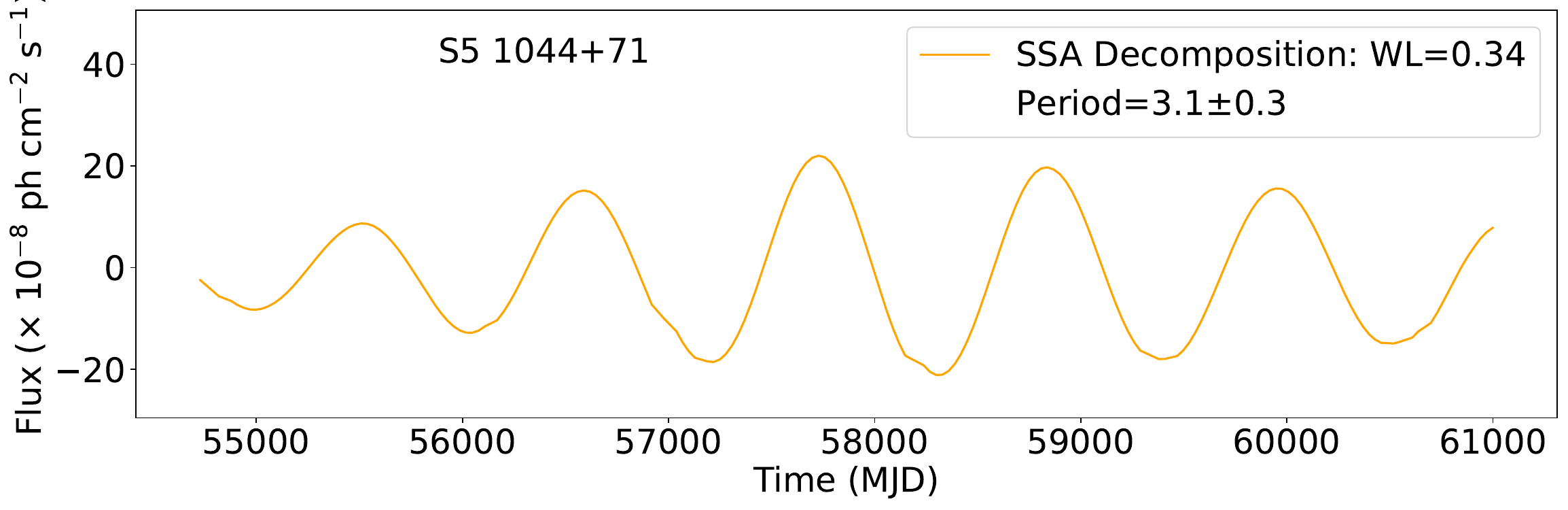}
         \includegraphics[scale=0.225]{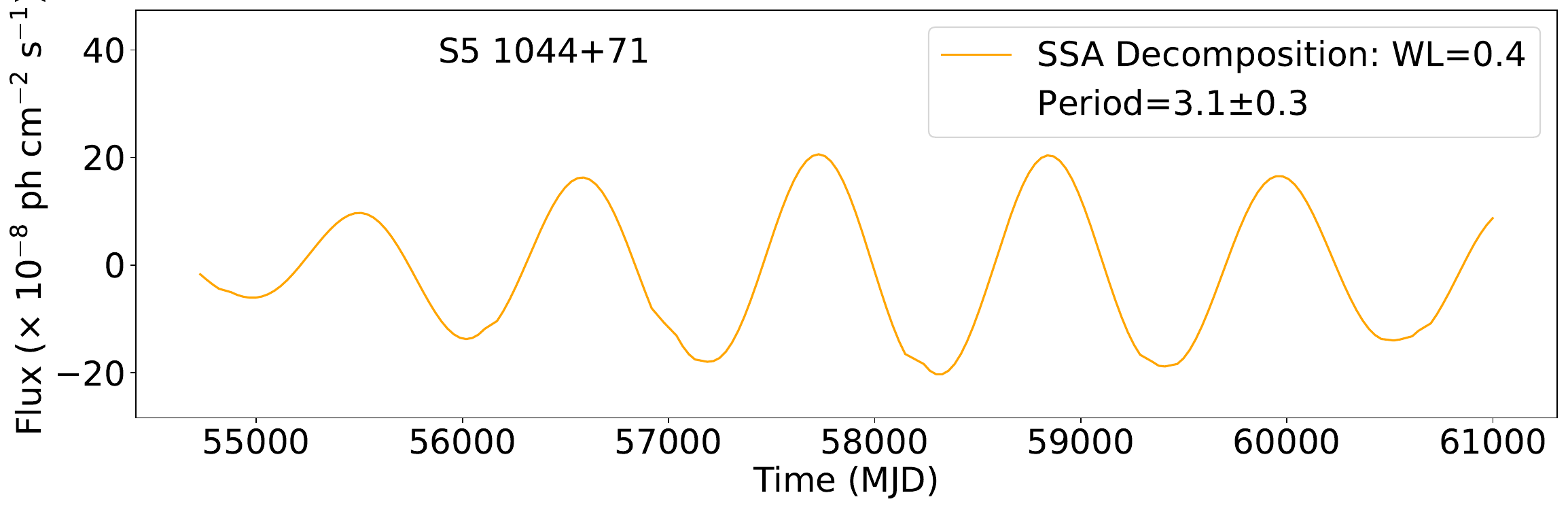}
         \includegraphics[scale=0.225]{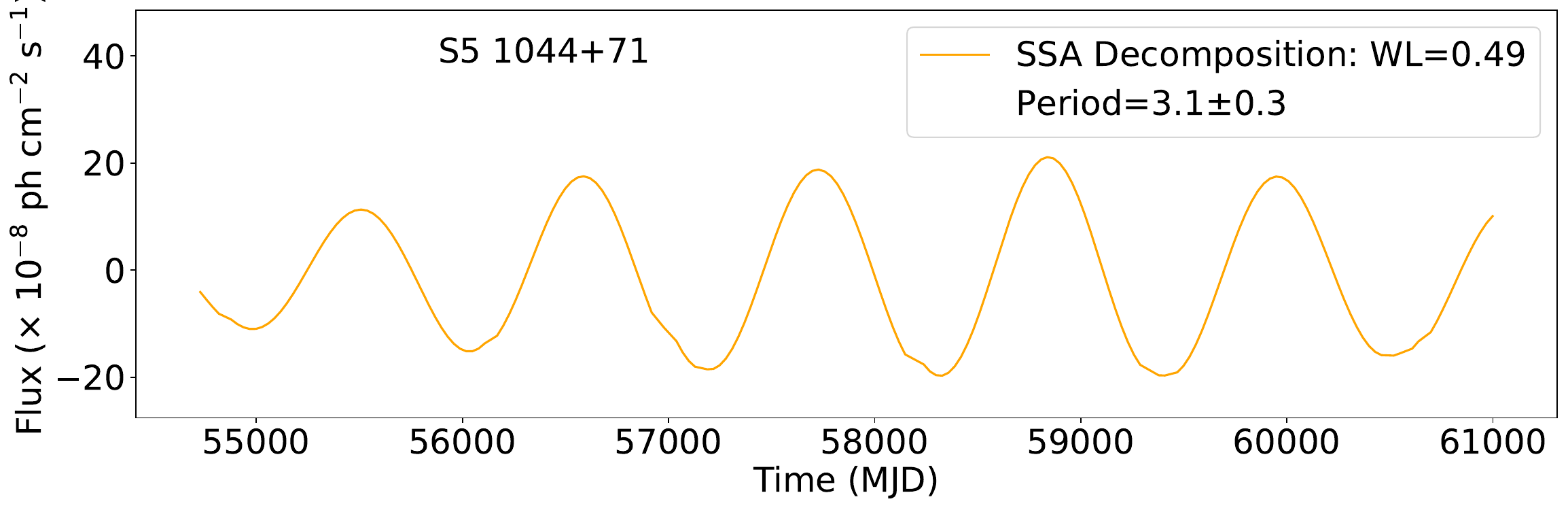}
         \caption{SSA decomposition of the oscillatory component for S5~1044+71 shown for five different window lengths ($WL$): 0.09 and 0.49, corresponding to the limits of the explored range; 0.17 and 0.34, being the optimal $WLs$; and 0.4, the value used to estimate the test statistics reported in Table~\ref{tab:periodicity_results}. \label{fig:ssa_descompositions_s51044}}
\end{figure*}

\begin{figure*}
	\centering
         \includegraphics[scale=0.225]{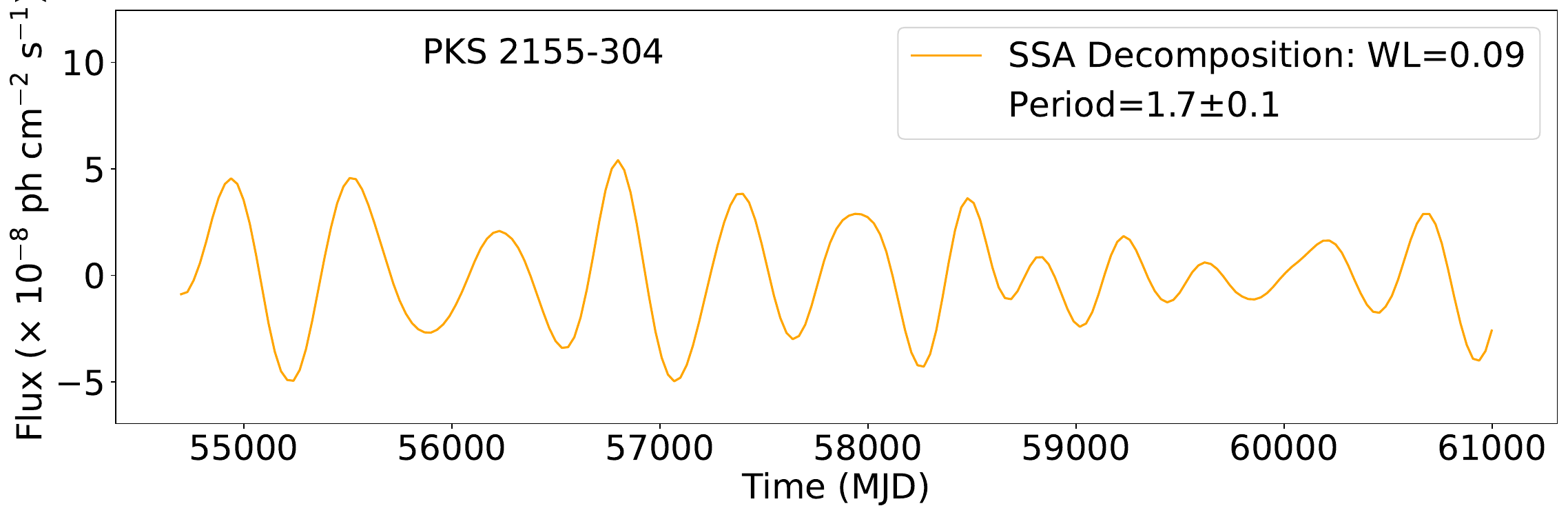}
         \includegraphics[scale=0.225]{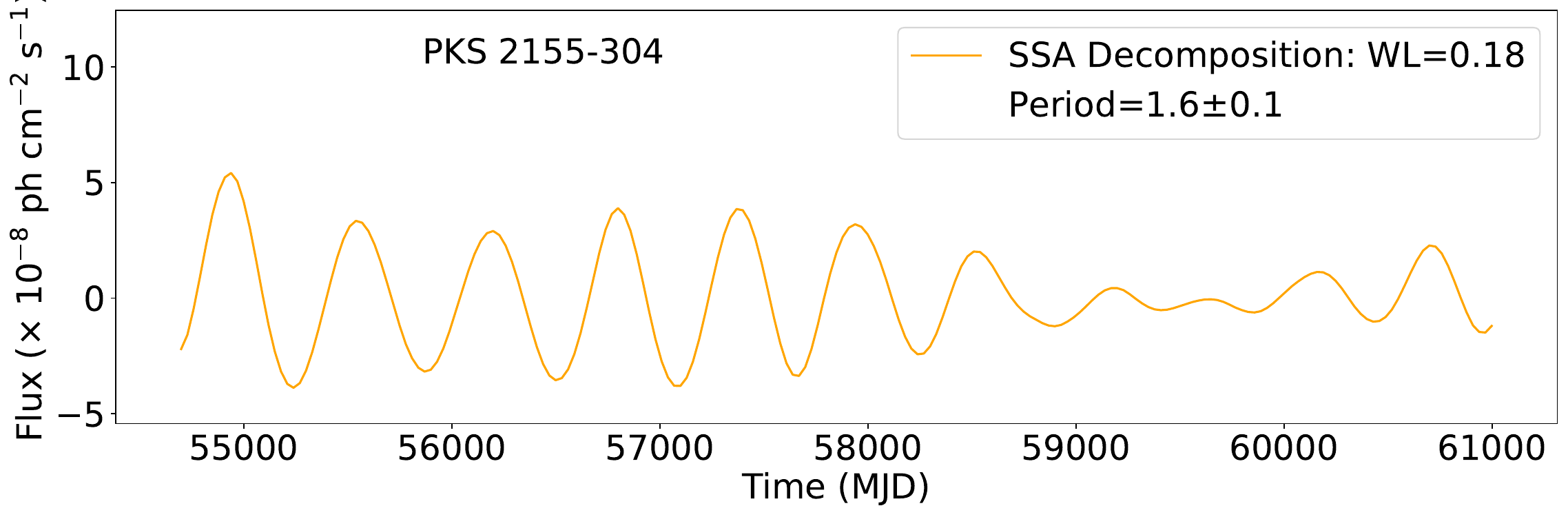}
         \includegraphics[scale=0.225]{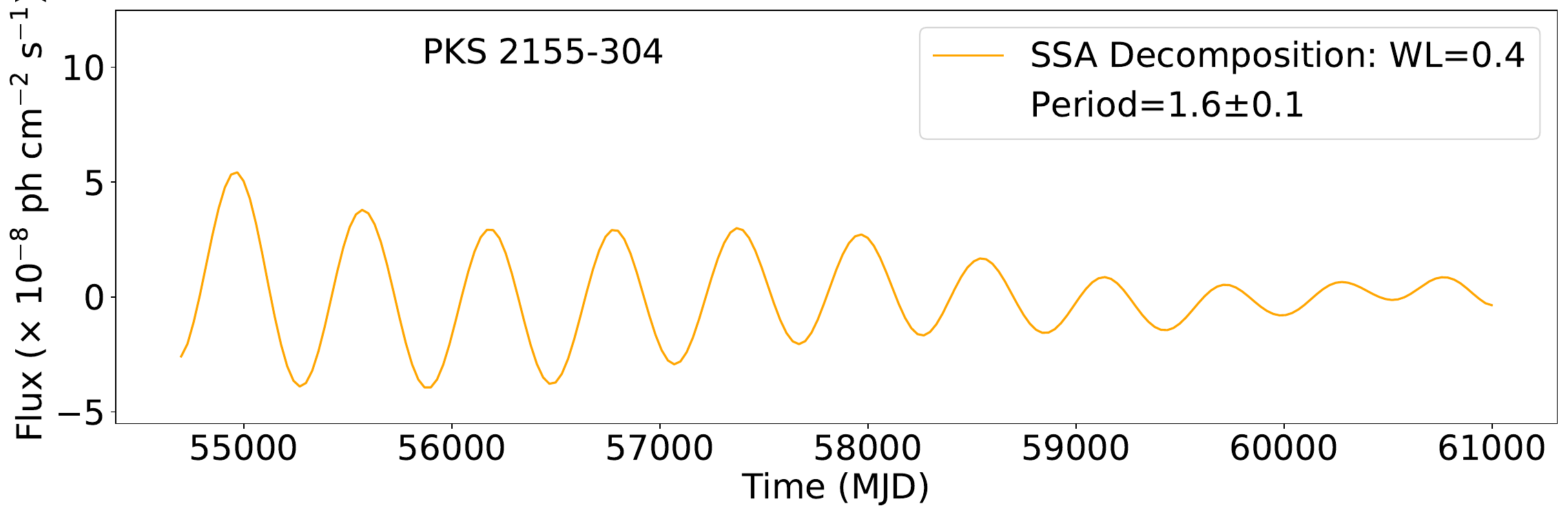}
         \includegraphics[scale=0.225]{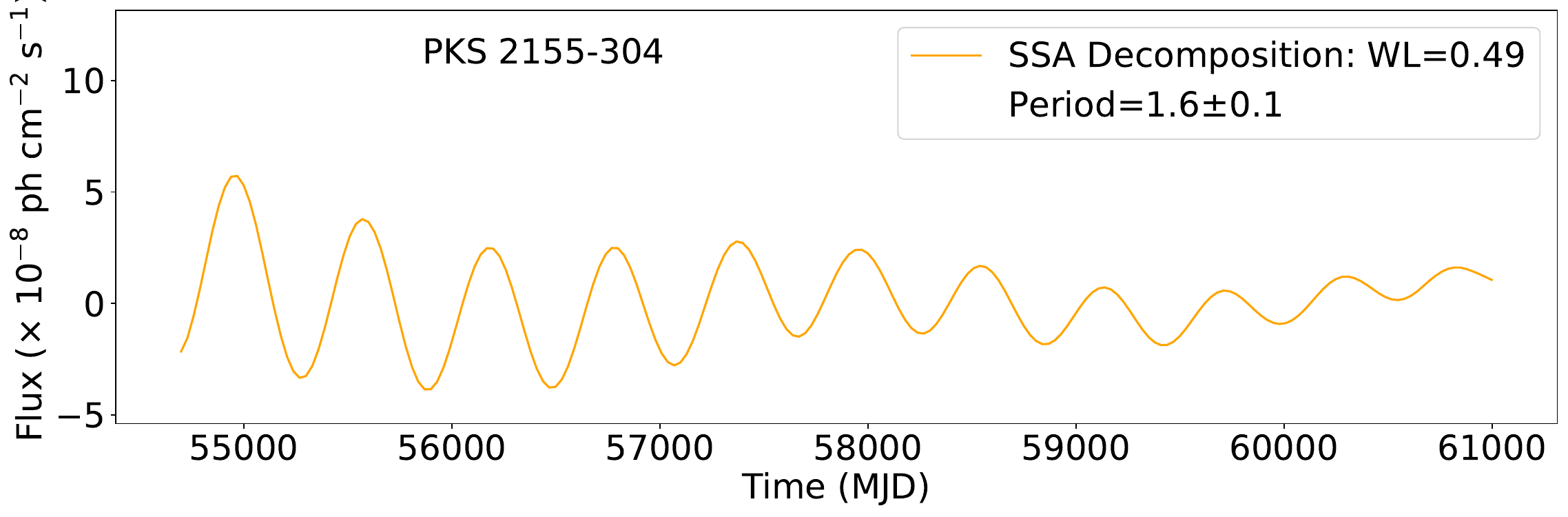}
         \caption{SSA decomposition of the oscillatory component for PKS 2155$-$304, shown for four different window lengths ($WL$): 0.49, corresponding to the limits of the explored range; 0.09 and 0.18, being the optimal $WLs$; and 0.4, the value used to estimate the test statistics reported in Table~\ref{tab:periodicity_results}. \label{fig:ssa_descompositions_pks2155}}
\end{figure*}

\begin{figure*}
	\centering
         \includegraphics[scale=0.225]{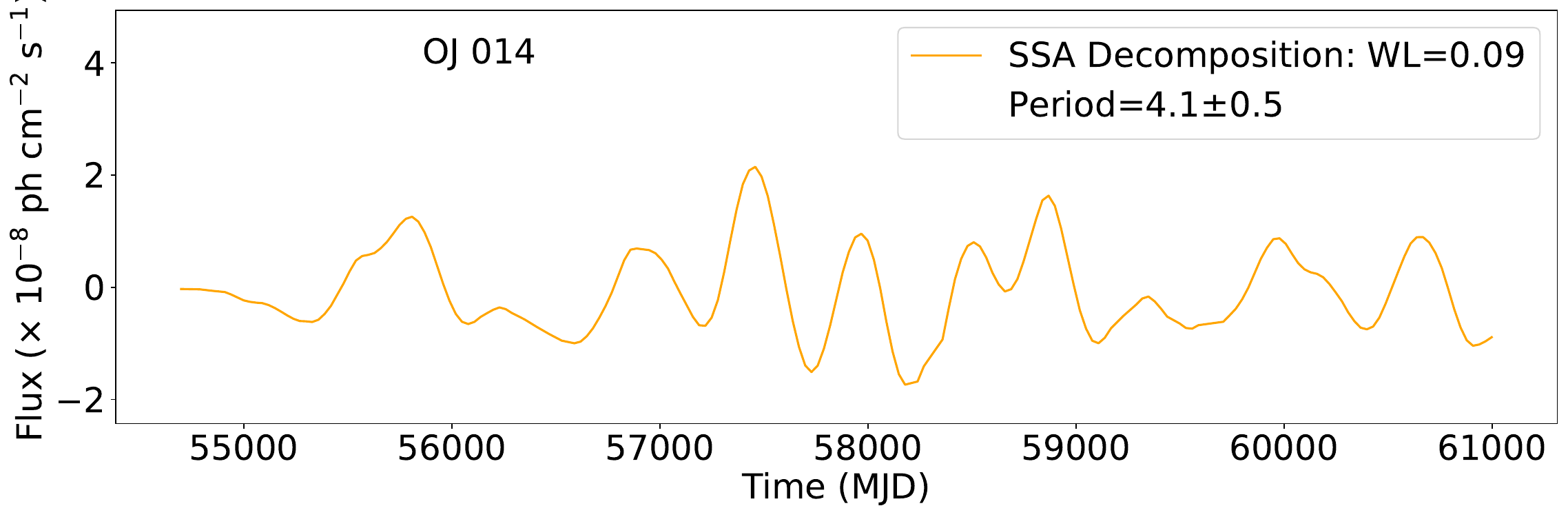}
         \includegraphics[scale=0.225]{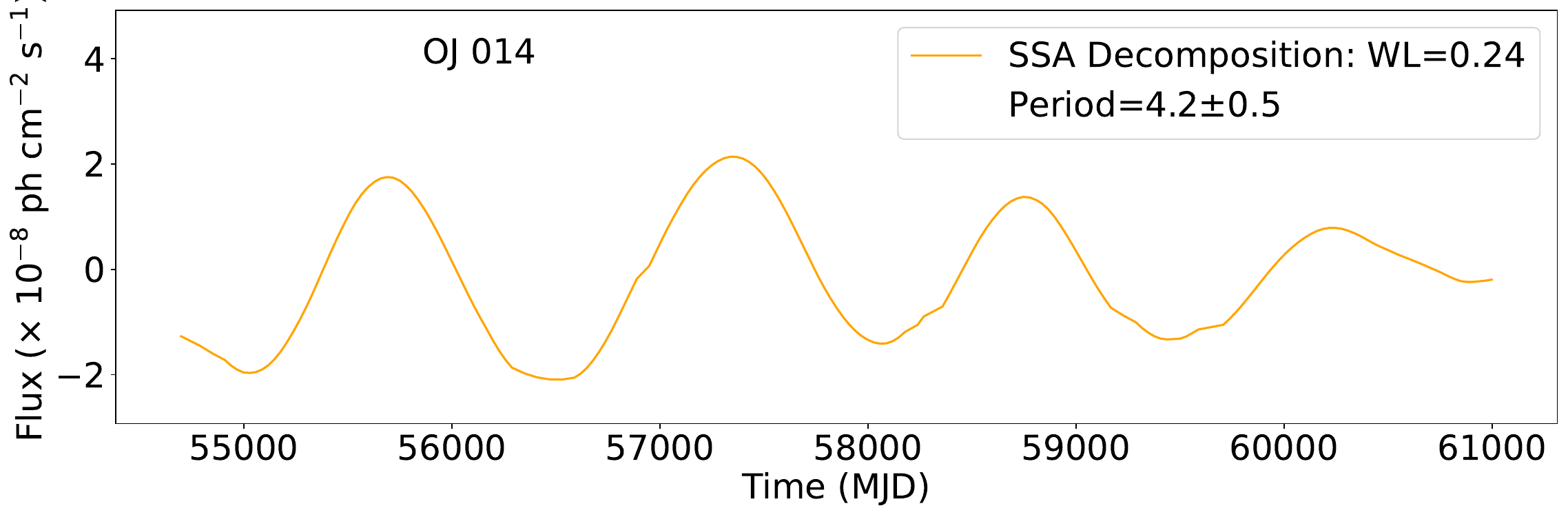}
         \includegraphics[scale=0.225]{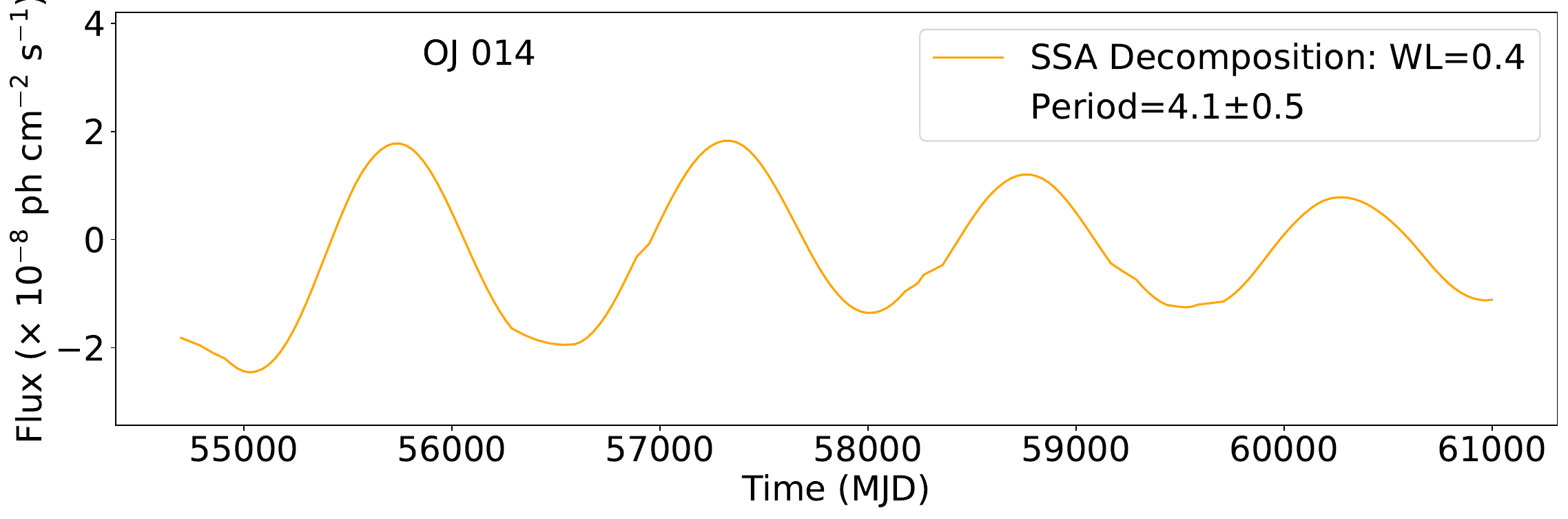}
         \includegraphics[scale=0.225]{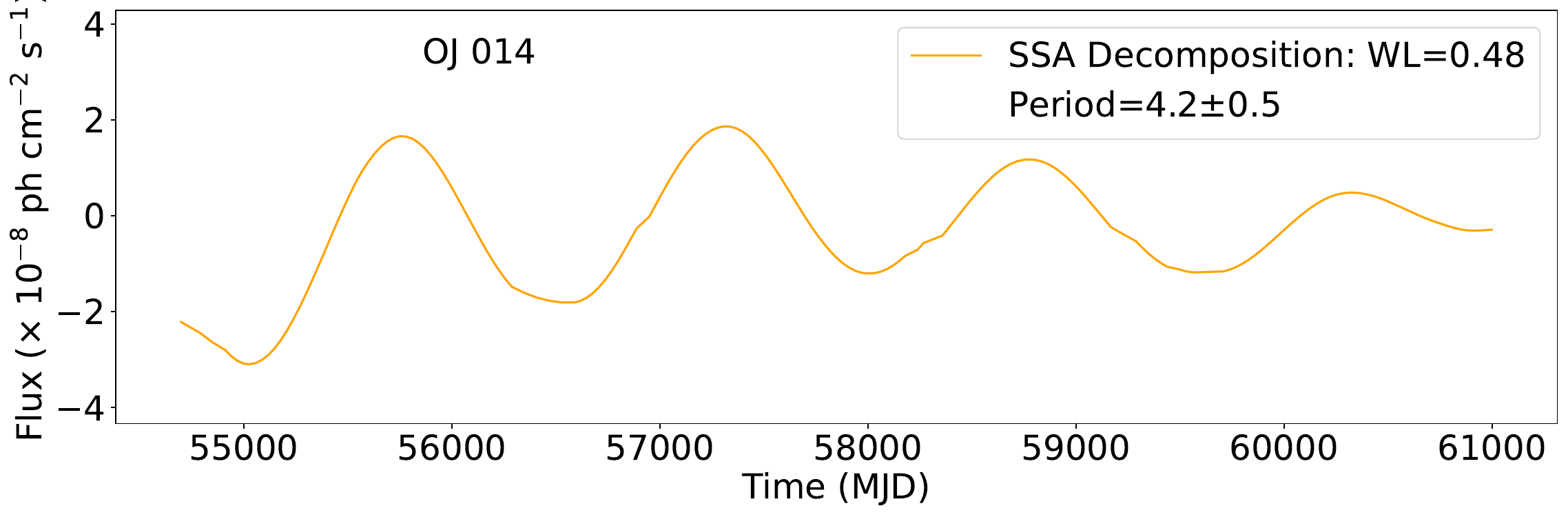}
         \caption{SSA decomposition of the oscillatory component for OJ 014, shown for four different window lengths ($WL$): 0.09, corresponding to the limit of the explored range; 0.24 and 0.48, being the optimal $WLs$; and 0.4, the value used to estimate the test statistics reported in Table~\ref{tab:periodicity_results}. \label{fig:ssa_descompositions_oj014}}
\end{figure*}

Figures~\ref{fig:ssa_descompositions_pg1553}--\ref{fig:ssa_descompositions_oj014} show examples of the SSA reconstruction of the oscillatory component for different values of the window length, $WL$. These figures are included because the choice of $WL$ affects the decomposition of the components of the signal. We therefore compare reconstructions obtained for values spanning the explored range of $WL$s, including the adopted or optimal values and the value used to estimate the test statistics. This comparison provides a visual check of the robustness of the extracted oscillatory component and helps to identify cases in which the SSA reconstruction is particularly sensitive to the adopted $WL$. The figures complement the quantitative $WL$ tests reported in the supplementary tables.

\subsubsection{Light curves of secondary candidates}
\begin{figure*}
	\centering
         \includegraphics[scale=0.2195]{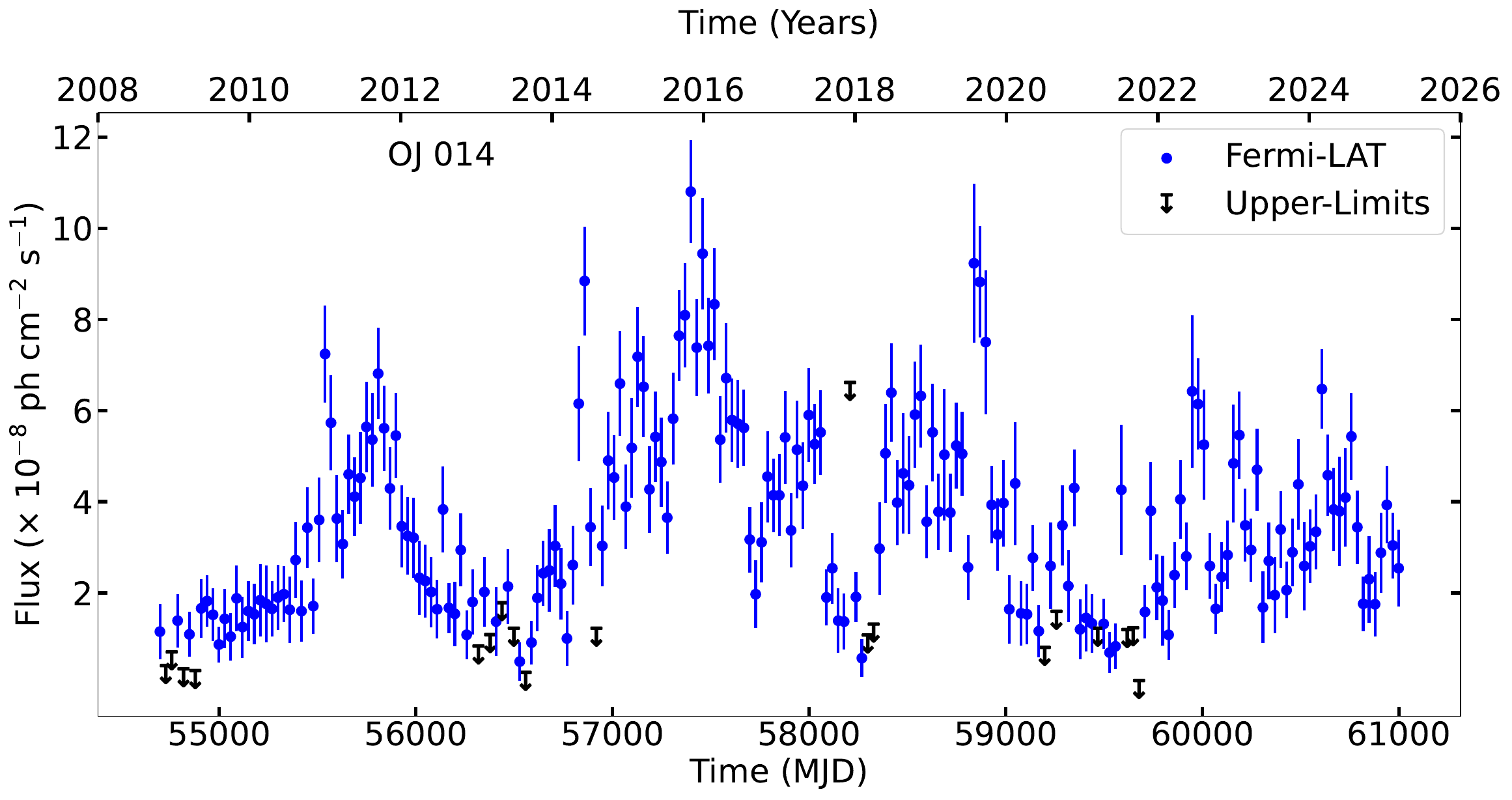}
         \includegraphics[scale=0.2195]{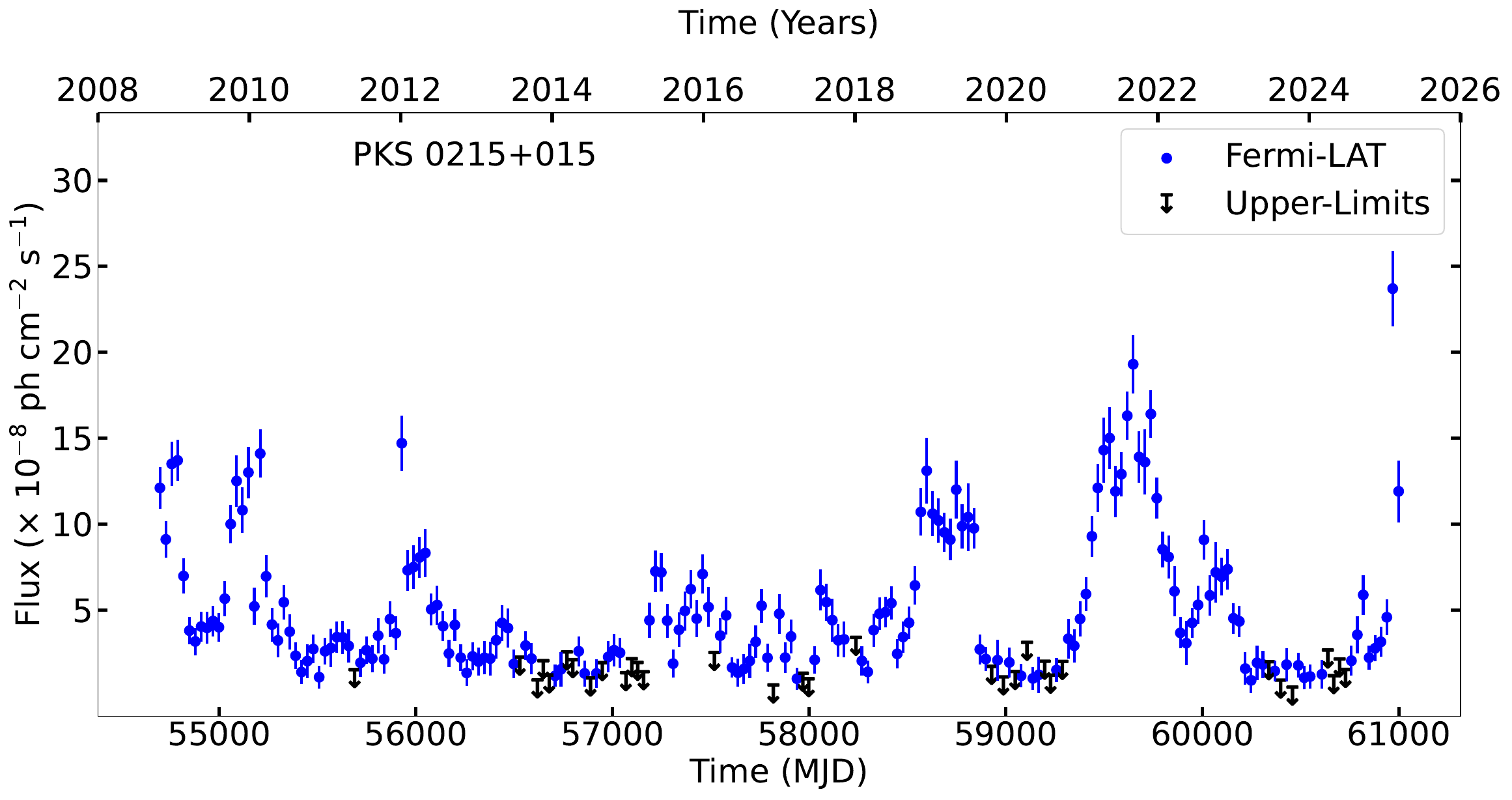}
         \includegraphics[scale=0.2195]{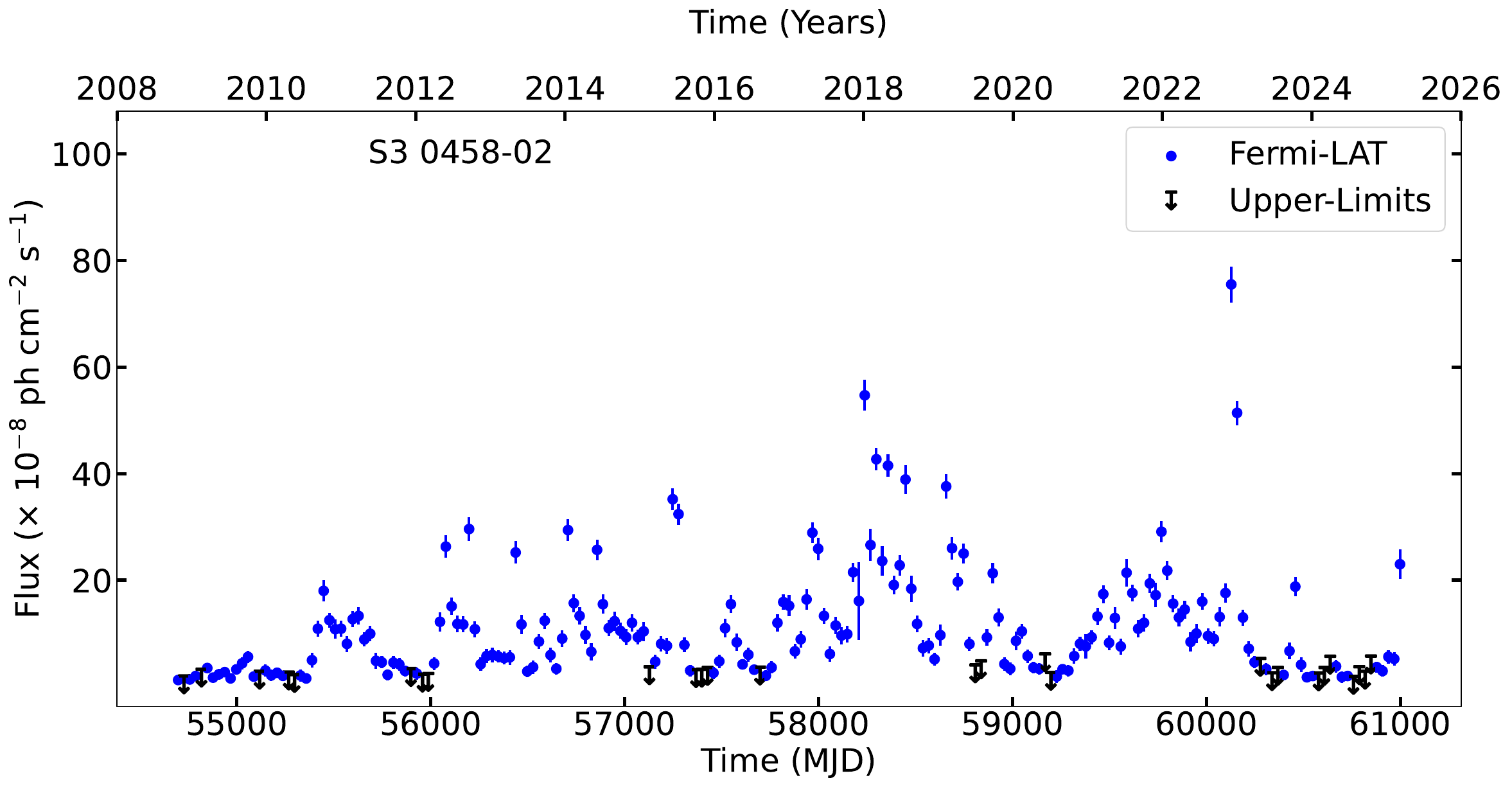}
         \includegraphics[scale=0.2195]{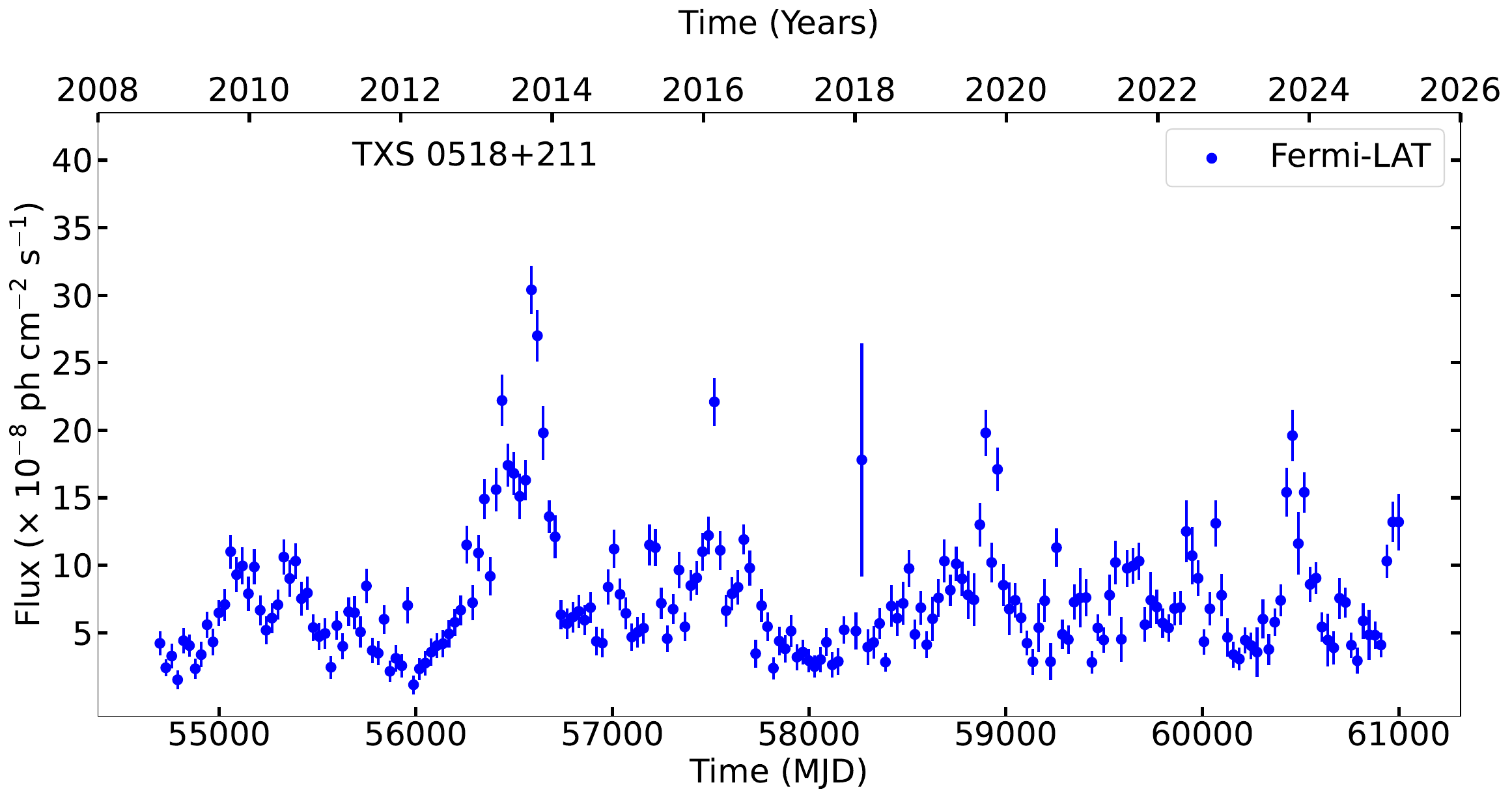}
         \includegraphics[scale=0.2195]{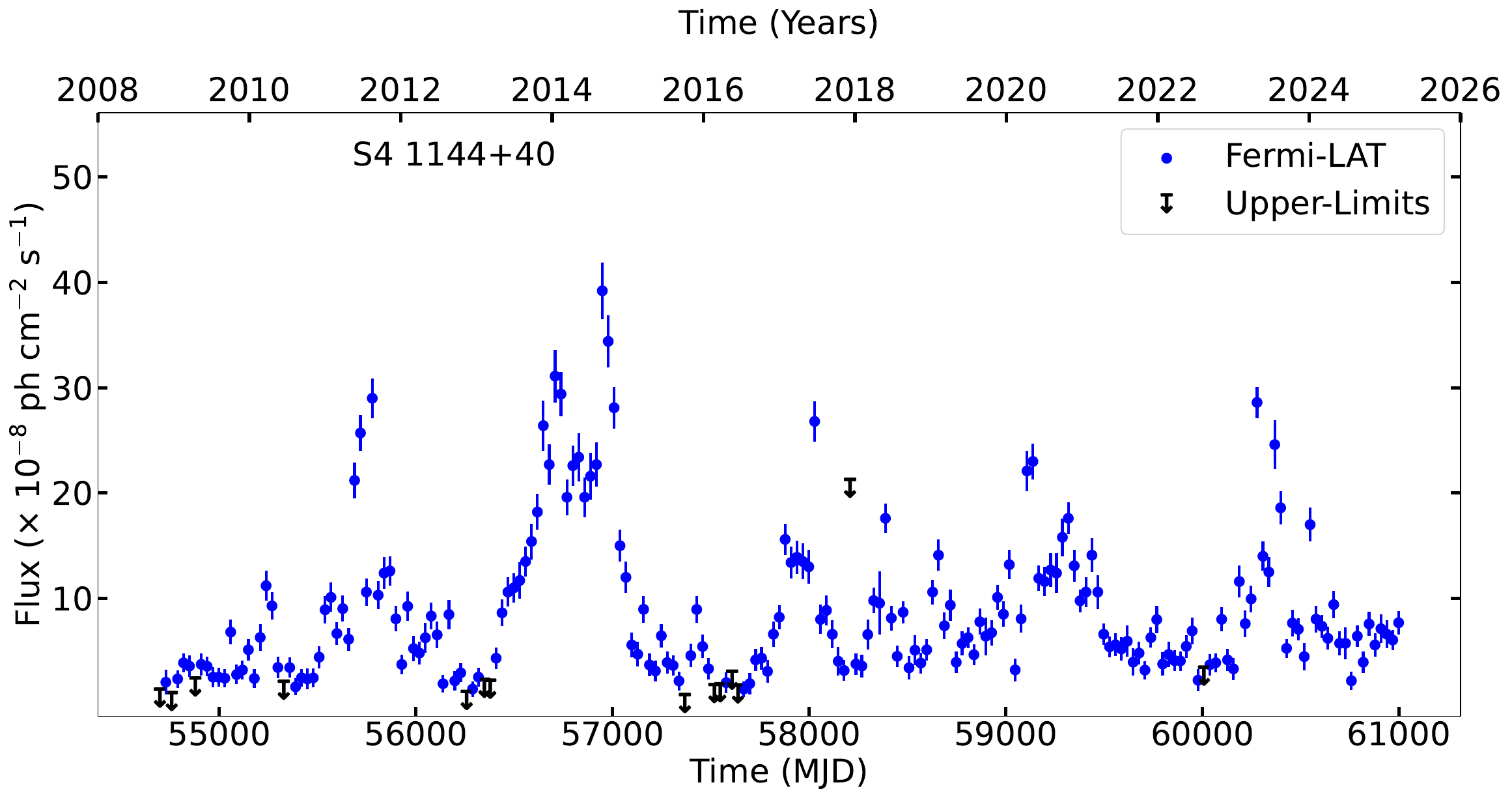}
         \includegraphics[scale=0.2195]{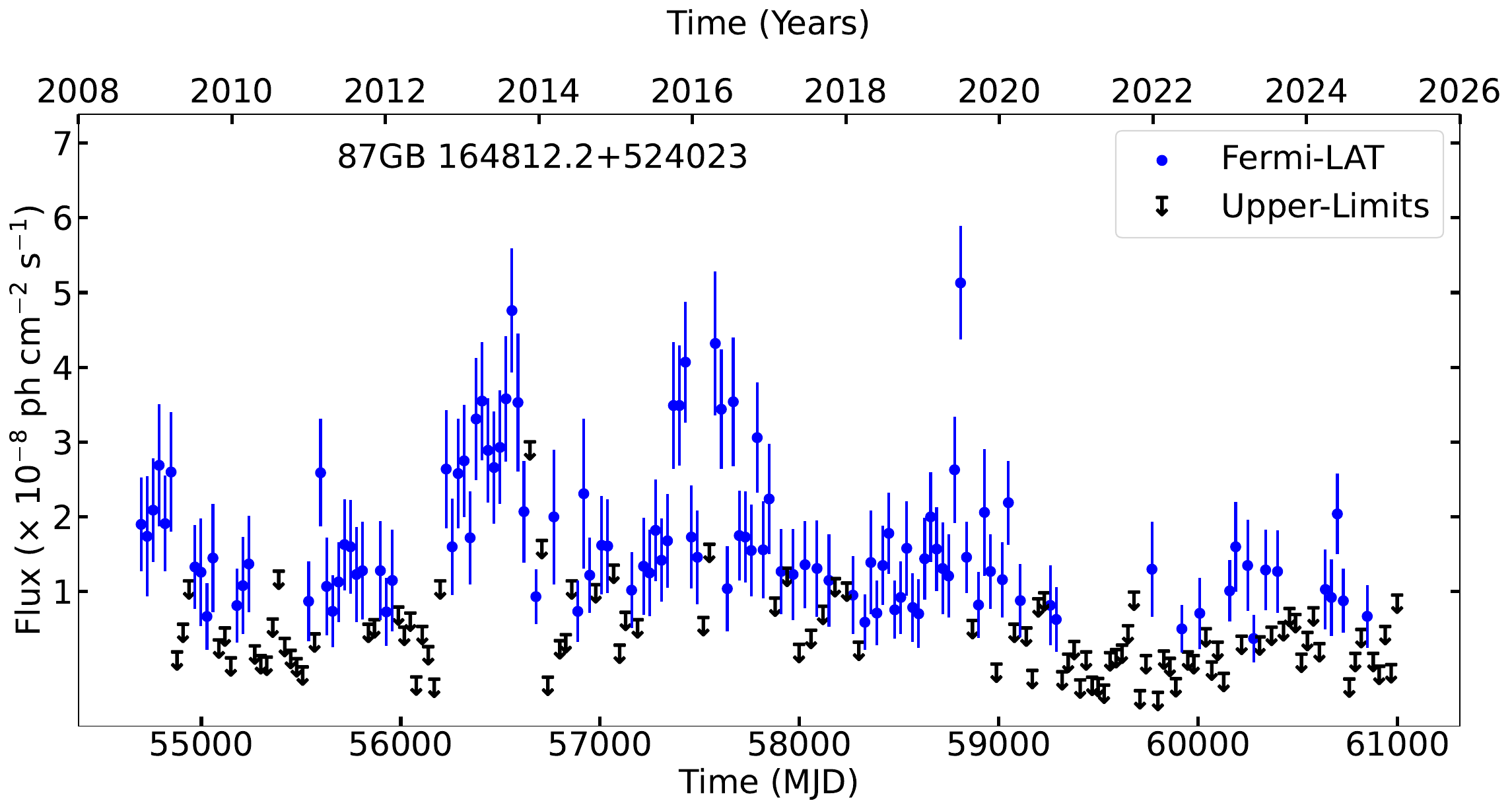}
         \caption{Light curves of the low-significance blazars ($\geq$2$\sigma$) presented in Table \ref{tab:final_candidates}. \label{fig:low_sgnificance}}
\end{figure*}

Figures~\ref{fig:low_sgnificance}  corresponds to sources classified as lower-significance candidates, for which the periodicity evidence reaches the $\geq$2$\sigma$ level but does not satisfy the threshold adopted for the most robust candidates.

\subsubsection{SSA-R forecasting examples}

\begin{figure*}
	\centering
         \includegraphics[scale=0.225]{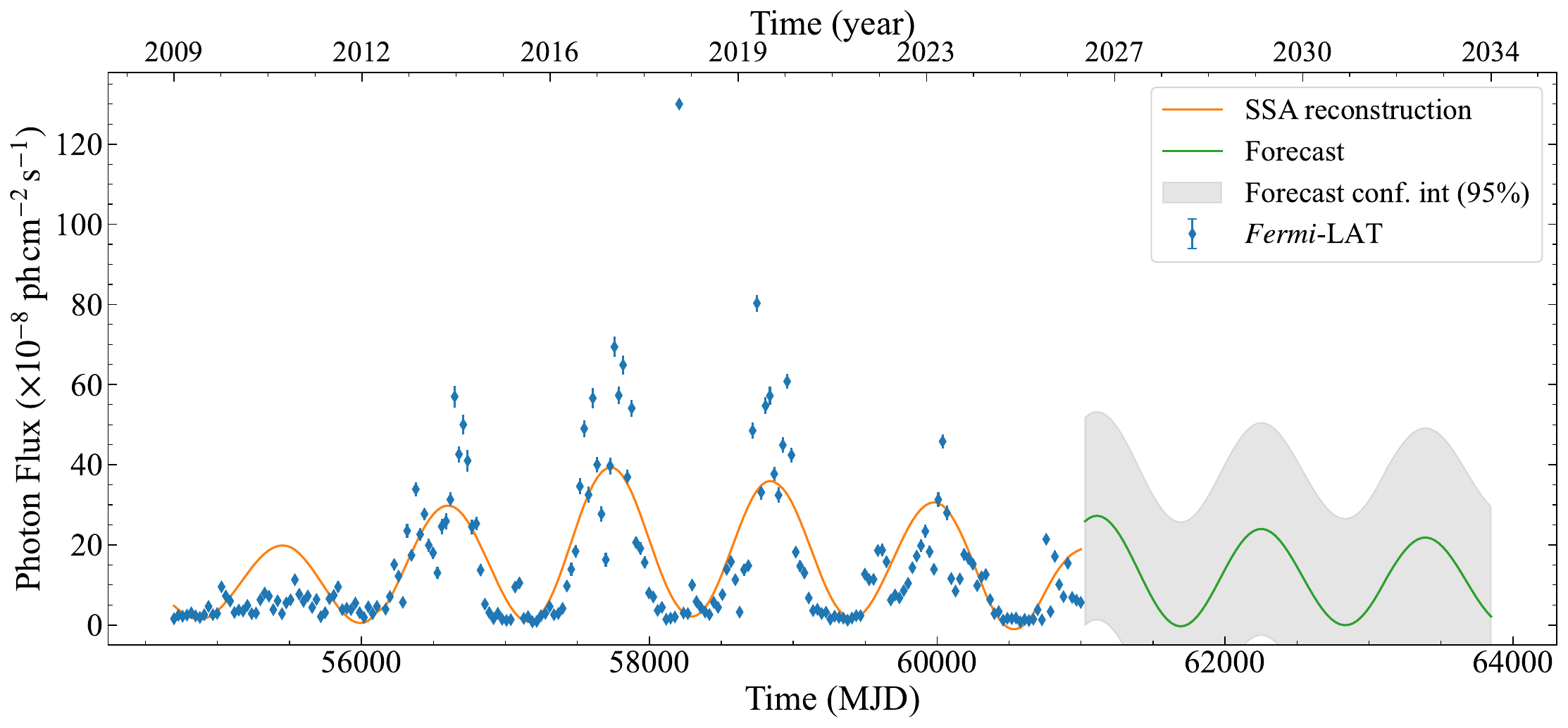}
         \includegraphics[scale=0.225]{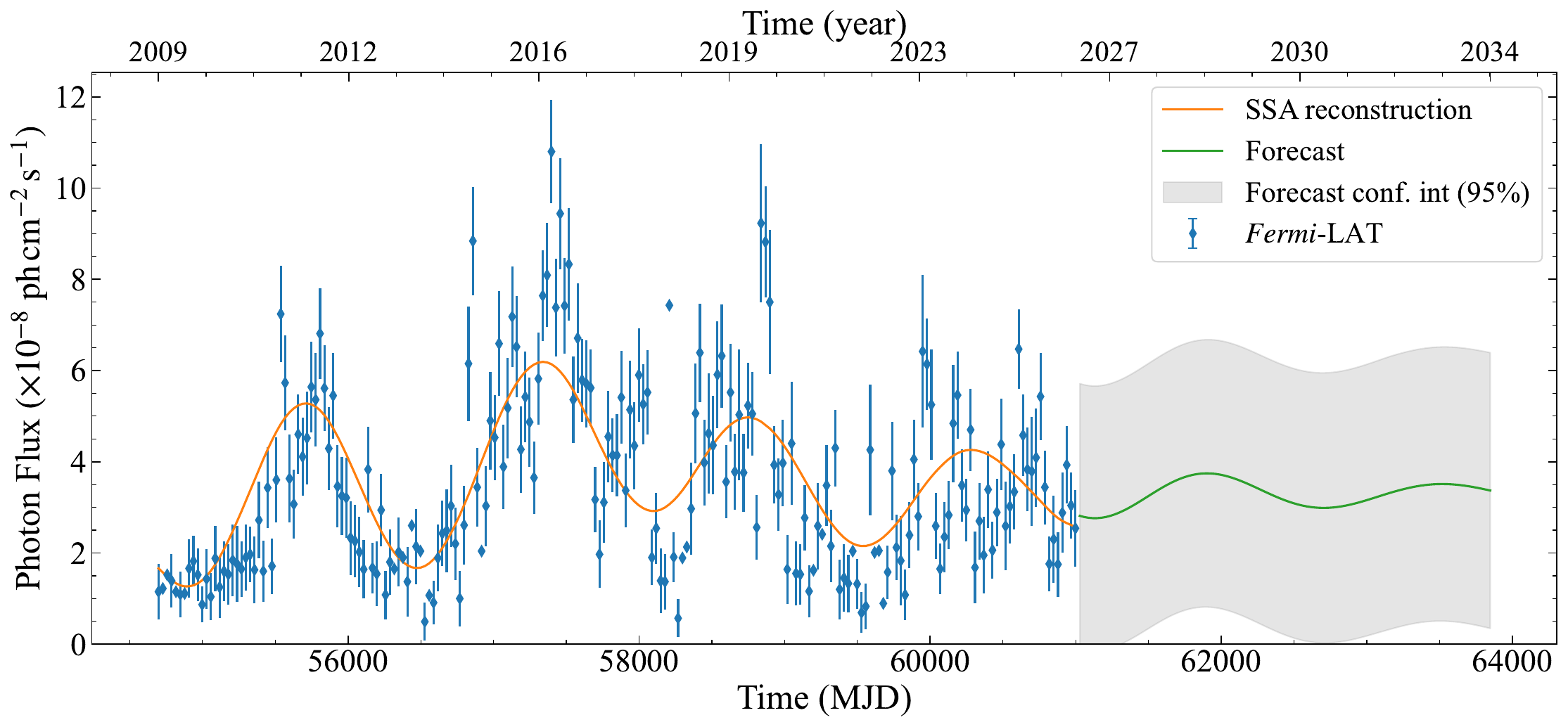}
         \includegraphics[scale=0.225]{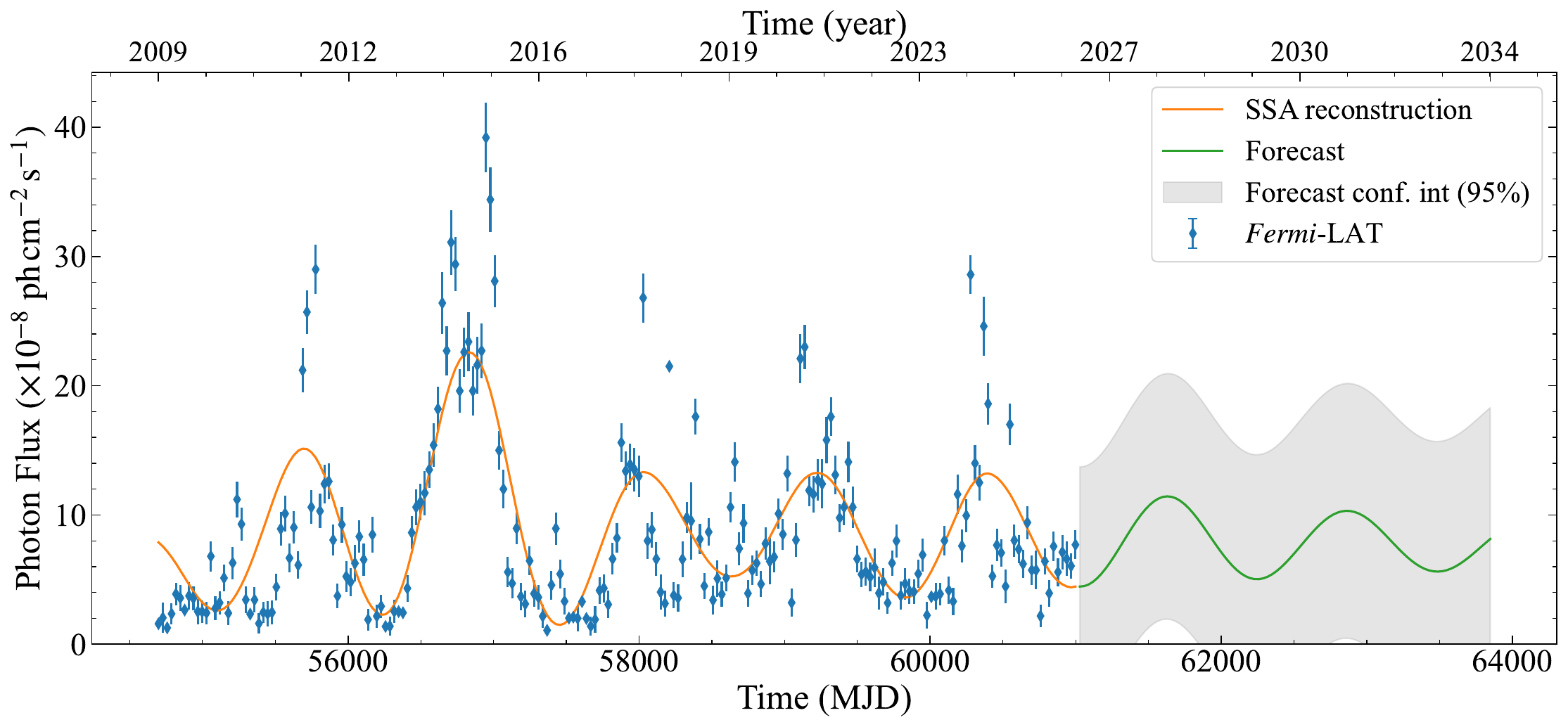}
         \includegraphics[scale=0.225]{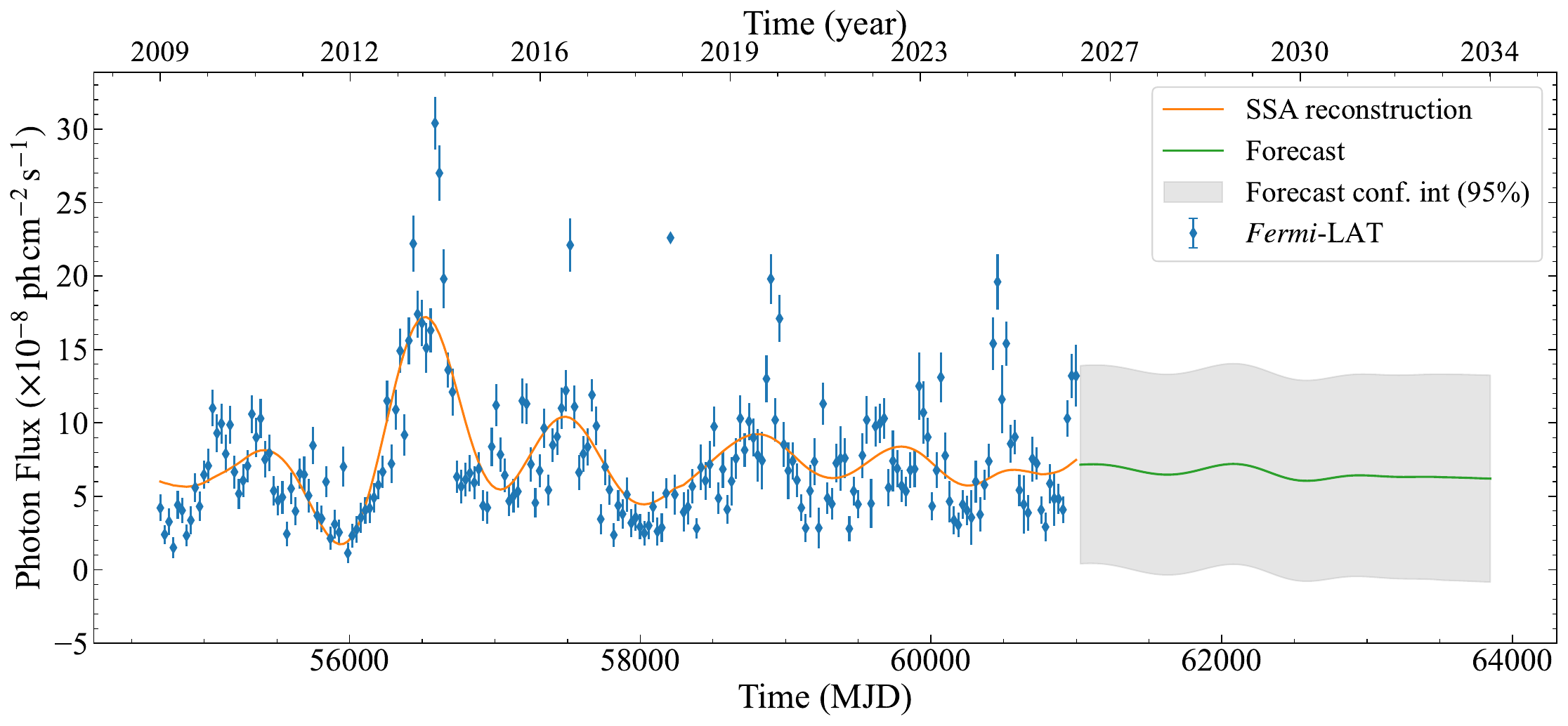}
         \caption{SSA-R forecasting results for the corresponding sources listed on Table \ref{tab:final_candidates}. The panels show, from top left to bottom right, S5 1044$+$71, OJ 014, S4 1144$+$40, and TXS 0528$+$211. In each panel, the blue points represent the {\it Fermi}-LAT light curve, the orange curve shows the SSA reconstruction of the observed behavior, and the green curve shows the forecast computed for the next 7 yr. The gray shaded region marks the corresponding 95\% forecast confidence interval. \label{fig:ssa_forecasting}}
\end{figure*}

Figure~\ref{fig:ssa_forecasting} shows examples of the SSA-R forecasting analysis for selected sources listed in Table~\ref{tab:final_candidates}. These forecasts are included to illustrate how the reconstructed oscillatory behavior evolves beyond the observed time interval and to provide a visual reference for the prediction tests discussed in the main text. 

\clearpage
\subsection{Supplementary Tables}
This section presents a set of tables that summarize some of the results obtained in this work.

\begin{table*}
\centering
\caption{Results of the correction to the test statistics using the simple power law (PL, Eq.\ref{eqn:pl_adhoc}) and the bending power law (BPL, Eq.\ref{eqn:bpl1}) variability models considering the uncertainties of each parameter for each variability model. Three different models are examined based on the fit parameters and their uncertainties: model$-$uncertainties, model, and model$+$uncertainties. The PL model is described by the
tuple (normalization [$\mathrm{rms}^2\,\mathrm{day}$], PL index, Poisson
constant [$\mathrm{rms}^2\,\mathrm{day}$]). The BPL model is described by the
tuple (normalization [$\mathrm{rms}^2\,\mathrm{day}$], bending frequency
[$\mathrm{day}^{-1}$], BPL index, Poisson constant
[$\mathrm{rms}^2\,\mathrm{day}$]). The -- denotes no values since the BPL fit did not converge. \label{tab:correction_noise_models}}
{%
\begin{tabular}{l|cccc|cccccc}
\hline
\hline
Association Name & Model (PL) & GLSP & PDM & SSA & Model (BPL) &  GLSP & PDM & SSA\\
\hline
\hline
\multirow{1}{*}{PG 1553+113} 
                & \makecell{\makecell{(3.7$\mathrm{x}10^{-2}$, 0.70, 3$\mathrm{x}10^{-2}$) \\ }\\ \makecell{(3.8$\mathrm{x}10^{-2}$, 0.72, 4$\mathrm{x}10^{-2}$) \\ }\\ \makecell{(3.9$\mathrm{x}10^{-2}$, 0.74, 5$\mathrm{x}10^{-2}$) \\ } } 
                & \makecell{3.6$\sigma$ \\ 3.7$\sigma$ \\ 3.5$\sigma$ \\ } 
                & \makecell{3.9$\sigma$ \\ 4.1$\sigma$ \\ 4.1$\sigma$ \\ } 
                & \makecell{2.0$\sigma$ \\ 2.0$\sigma$ \\ 2.1$\sigma$ \\ } 
                & \makecell{\makecell{(5.5, 2.7$\mathrm{x}10^{-3}$, 2.6, 0.7) \\ }\\ \makecell{(5.8, 2.9$\mathrm{x}10^{-3}$, 2.9, 0.8) \\ }\\ \makecell{(6.1, 3.1$\mathrm{x}10^{-3}$, 3.2, 1.0) \\ } } 
                & \makecell{4.2$\sigma$ \\ 4.1$\sigma$ \\ 4.0$\sigma$ \\ } 
                & \makecell{4.3$\sigma$ \\ 4.2$\sigma$ \\ 4.0$\sigma$ \\ } 
                & \makecell{1.9$\sigma$ \\ 1.9$\sigma$ \\ 1.8$\sigma$ \\ } 
            \\
        \hline
        \multirow{1}{*}{S5 1044+71} 
                & \makecell{\makecell{(0.88, 0.73, 0.24) \\ }\\ \makecell{(0.99, 0.75, 0.44) \\ }\\ \makecell{(1.11, 0.77, 0.64) \\ } } 
                & \makecell{4.2$\sigma$ \\ 4.0$\sigma$ \\ 4.1$\sigma$ \\ } 
                & \makecell{4.3$\sigma$ \\ 4.0$\sigma$ \\ 4.2$\sigma$ \\ } 
                & \makecell{2.2$\sigma$ \\ 2.2$\sigma$ \\ 2.3$\sigma$ \\ } 
                & \makecell{\makecell{(96.8, 8$\mathrm{x}10^{-4}$, 5.6, 16.0) \\ }\\ \makecell{(102.4, 1.3$\mathrm{x}10^{-3}$, 6.1, 16.7) \\ }\\ \makecell{(108.0, 1.8$\mathrm{x}10^{-3}$, 6.6, 17.4) \\ } } 
                & \makecell{3.7$\sigma$ \\ 3.6$\sigma$ \\ 3.6$\sigma$ \\ } 
                & \makecell{4.2$\sigma$ \\ 4.0$\sigma$ \\ 4.0$\sigma$ \\ } 
                & \makecell{2.0$\sigma$ \\ 2.0$\sigma$ \\ 2.0$\sigma$ \\ } 
            \\
        \hline
                \multirow{1}{*}{OJ 014} 
                & \makecell{\makecell{(1$\mathrm{x}10^{-3}$, 1.1, 0.81) \\ }\\ \makecell{(2$\mathrm{x}10^{-3}$, 1.2, 0.94) \\ }\\ \makecell{(3$\mathrm{x}10^{-3}$, 1.3, 1.07) \\ } } 
                & \makecell{3.0$\sigma$ \\ 2.6$\sigma$ \\ 2.0$\sigma$ \\ } 
                & \makecell{2.7$\sigma$ \\ 2.5$\sigma$ \\ 1.8$\sigma$ \\ } 
                & \makecell{2.0$\sigma$ \\ 2.2$\sigma$ \\ 2.1$\sigma$ \\ } 
                & \makecell{\makecell{(24.0, 7.6$\mathrm{x}10^{-4}$, 7.7, 2.5) \\ }\\ \makecell{(25.7, 7.8$\mathrm{x}10^{-4}$, 7.9, 2.6) \\ }\\ \makecell{(27.4, 8$\mathrm{x}10^{-4}$, 8.1, 2.7) \\ } } 
                & \makecell{2.4$\sigma$ \\ 2.5$\sigma$ \\ 2.3$\sigma$ \\ } 
                & \makecell{2.4$\sigma$ \\ 2.4$\sigma$ \\ 2.2$\sigma$ \\ } 
                & \makecell{1.7$\sigma$ \\ 1.8$\sigma$ \\ 1.6$\sigma$ \\ } 
            \\
        \hline
                \multirow{1}{*}{S4 1144+40} 
                & \makecell{\makecell{(0.10, 0.70, 0.35) \\ }\\ \makecell{(0.12, 0.75, 0.46) \\ }\\ \makecell{(0.14, 0.80, 0.57) \\ } } 
                & \makecell{3.0$\sigma$ \\ 2.8$\sigma$ \\ 2.7$\sigma$ \\ } 
                & \makecell{4.3$\sigma$ \\ 3.9$\sigma$ \\ 4.2$\sigma$ \\ } 
                & \makecell{1.8$\sigma$ \\ 1.9$\sigma$ \\ 1.9$\sigma$ \\ } 
                & \makecell{\makecell{(57.0, 5$\mathrm{x}10^{-4}$, 3.0, 2.6) \\ }\\ \makecell{(60.7, 1.3$\mathrm{x}10^{-3}$, 3.2, 2.7) \\ }\\ \makecell{(64.4, 2$\mathrm{x}10^{-3}$, 3.4, 2.8) \\ } } 
                & \makecell{1.9$\sigma$ \\ 1.8$\sigma$ \\ 1.8$\sigma$ \\ } 
                & \makecell{2.8$\sigma$ \\ 2.4$\sigma$ \\ 2.5$\sigma$ \\ } 
                & \makecell{1.7$\sigma$ \\ 1.7$\sigma$ \\ 1.6$\sigma$ \\ } 
            \\
        \hline
                \multirow{1}{*}{PKS 0215+015} 
                & \makecell{\makecell{(1.5$\mathrm{x}10^{-2}$, 0.89, 5$\mathrm{x}10^{-2}$) \\ }\\ \makecell{(1.9$\mathrm{x}10^{-2}$, 0.97, 0.14) \\ }\\ \makecell{(2.3$\mathrm{x}10^{-2}$, 1.05, 0.23) \\ } } 
                & \makecell{2.1$\sigma$ \\ 2.1$\sigma$ \\ 1.8$\sigma$ \\ } 
                & \makecell{2.2$\sigma$ \\ 2.2$\sigma$ \\ 2.0$\sigma$ \\ } 
                & \makecell{1.4$\sigma$ \\ 1.7$\sigma$ \\ 1.5$\sigma$ \\ } 
                & \makecell{\makecell{(31.2, 6$\mathrm{x}10^{-4}$, 2.8, 2.3) \\ }\\ \makecell{(33.1, 1.2$\mathrm{x}10^{-3}$, 3.0, 2.4) \\ }\\ \makecell{(35.0, 1.9$\mathrm{x}10^{-3}$, 3.2, 2.5) \\ } } 
                & \makecell{2.0$\sigma$ \\ 2.0$\sigma$ \\ 1.9$\sigma$ \\ } 
                & \makecell{2.3$\sigma$ \\ 2.0$\sigma$ \\ 1.7$\sigma$ \\ } 
                & \makecell{1.4$\sigma$ \\ 1.6$\sigma$ \\ 1.3$\sigma$ \\ } 
            \\
        \hline
                \multirow{1}{*}{S3 0458$-$02} 
                & \makecell{\makecell{(3$\mathrm{x}10^{-2}$, 0.92, 3.8) \\ }\\ \makecell{(4$\mathrm{x}10^{-2}$, 0.97, 4.5) \\ }\\ \makecell{(5$\mathrm{x}10^{-2}$, 1.02, 5.2) \\ } } 
                & \makecell{2.0$\sigma$ \\ 2.0$\sigma$ \\ 1.5$\sigma$ \\ } 
                & \makecell{2.0$\sigma$ \\ 2.0$\sigma$ \\ 1.5$\sigma$ \\ } 
                & \makecell{1.2$\sigma$ \\ 1.4$\sigma$ \\ 1.3$\sigma$ \\ }
                & \makecell{\makecell{(106.5, 3$\mathrm{x}10^{-4}$, 0.9, 1.1) \\  }\\ \makecell{(119.4, 4$\mathrm{x}10^{-4}$, 1.0, 2.2) \\ }\\ \makecell{(132.3, 5$\mathrm{x}10^{-4}$, 1.1, 3.3) \\ } }
                & \makecell{1.7$\sigma$ \\ 2.0$\sigma$ \\ 1.8$\sigma$ \\ } 
                & \makecell{1.7$\sigma$ \\ 2.1$\sigma$ \\ 1.8$\sigma$ \\ } 
                & \makecell{1.1$\sigma$ \\ 1.5$\sigma$ \\ 1.2$\sigma$ \\ } 
            \\
        \hline
        \multirow{1}{*}{TXS 0518+211} 
                & \makecell{\makecell{(1.8$\mathrm{x}10^{-2}$, 0.91, 5$\mathrm{x}10^{-2}$) \\ }\\ \makecell{(2.1$\mathrm{x}10^{-2}$, 0.96, 8$\mathrm{x}10^{-2}$) \\ }\\ \makecell{(2.4$\mathrm{x}10^{-2}$, 1.01, 0.11) \\ } } 
                & \makecell{1.9$\sigma$ \\ 1.9$\sigma$ \\ 1.7$\sigma$ \\ } 
                & \makecell{1.9$\sigma$ \\ 2.0$\sigma$ \\ 1.7$\sigma$ \\ } 
                & \makecell{1.6$\sigma$ \\ 1.8$\sigma$ \\ 1.6$\sigma$ \\ } 
                & \makecell{\makecell{(54.5, 3$\mathrm{x}10^{-4}$, 2.6, 1.6) \\  }\\ \makecell{(56.6, 1$\mathrm{x}10^{-3}$, 2.8, 1.7) \\ }\\ \makecell{(58.7, 1.7$\mathrm{x}10^{-3}$, 3.0, 1.8) \\ } } 
                & \makecell{2.0$\sigma$ \\ 1.7$\sigma$ \\ 1.4$\sigma$ \\ } 
                & \makecell{2.2$\sigma$ \\ 1.8$\sigma$ \\ 1.3$\sigma$ \\ } 
                & \makecell{1.6$\sigma$ \\ 1.7$\sigma$ \\ 1.6$\sigma$ \\ } 
            \\
        \hline
        \multirow{1}{*}{87GB 164812.2+524023} 
            & \makecell{\makecell{(7$\mathrm{x}10^{-3}$, 0.69, 0.28) \\ }\\ \makecell{(1.1$\mathrm{x}10^{-2}$, 0.75, 0.36) \\ }\\ \makecell{(1.5$\mathrm{x}10^{-2}$, 0.81, 0.44) \\ } } 
            & \makecell{1.0$\sigma$ \\ 1.3$\sigma$ \\ 0.6$\sigma$ \\ } 
            & \makecell{2.2$\sigma$ \\ 2.3$\sigma$ \\ 1.5$\sigma$ \\ } 
            & \makecell{1.8$\sigma$ \\ 2.0$\sigma$ \\ 1.6$\sigma$ \\ } 
            & \makecell{\makecell{-- \\ }\\ \makecell{-- \\ }\\ \makecell{-- \\ } } 
            & \makecell{-- \\ -- \\ -- \\ } 
            & \makecell{-- \\ -- \\ -- \\ } 
            & \makecell{-- \\ -- \\ -- \\ } 
        \\
\hline
\hline
\end{tabular}%
}
\end{table*}
 
\begin{table*}
\centering
\caption{List of sources analyzed using ARIMA and ARFIMA models. The column Period reports the most significant periodicities identified from the analysis of the residuals of the best-fit ARIMA/ARFIMA models. We also report $\Delta\mathrm{BIC}$, defined as $\Delta\mathrm{BIC} = \mathrm{BIC_{ARFIMA}}-\mathrm{BIC_{ARIMA}}$. Positive values of $\Delta\mathrm{BIC}$ indicate that the ARFIMA model provides a better description of the LC, while negative values favor the ARIMA model.\label{tab:arima_arfima}}
{%
\begin{tabular}{c|cccccccccc}
\hline
\hline
Name & ARIMA & Period & ARFIMA & Period & $\Delta$BIC \\
     &       & [year] &       & [year] &            \\
\hline
\hline
S2 0109+22 & (0, 0, 4) & 2.6 (1.8$\sigma$) & (0, 0.41, 0) & 2.9 (2.0$\sigma$) & 364.02 \\
PKS 0208$-$512 & (1, 0, 2) & 2.6 (1.5$\sigma$) & (1, 0.68, 0) & 1.7 (1.4$\sigma$) & 65.51 \\
3C 66A & (2, 0, 1) & 2.0 (1.7$\sigma$) & (0, 0.41, 0) & 2.2 (1.6$\sigma$) & 242.51 \\
PKS 0301$-$243 & (3, 0, 4) & 2.0 (1.5$\sigma$) & (1, 0.82, 0) & 2.0 (1.5$\sigma$) & 8.63 \\
PKS 0426$-$380 & (1, 0, 0) & 3.0 (2.1$\sigma$) & (1, 0.19, 0) & 1.5 (1.8$\sigma$) & 214.52 \\
PKS 0447$-$439 & (0, 0, 5) & 2.1 (1.7$\sigma$) & (0, 0.19, 4) & 2.1 (1.6$\sigma$) & 115.21 \\
PKS 0454$-$234 & (0, 0, 5) & 4.7 (1.3$\sigma$) & (1, 0.19, 0) & 2.1 (1.4$\sigma$) & 181.89 \\
TXS 0518+211 & (1, 0, 2) & 3.4 (1.9$\sigma$)  & (0, 0.41, 5) & 3.0 (1.7$\sigma$) & 304.24 \\
S5 0716+714 & (2, 0, 0) & 4.5 (1.9$\sigma$) & (3, 0.19, 1) & 1.5 (1.5$\sigma$) & 162.12 \\
S4 1250+53 & (1, 0, 3) & 1.3 (2.5$\sigma$) & (0, 0.46, 0) & 1.3 (1.6$\sigma$) & 300.65 \\
PKS 1424$-$328 & (3, 0, 1) & 3.0 (2.3$\sigma$) & (2, 0.19, 5) & 1.4 (1.4$\sigma$) & 176.14 \\
TXS 1452+516 & (1, 0, 0) & 4.8 (1.7$\sigma$)  & (0, 0.63, 0) & 1.5 (1.0$\sigma$) & 176.69 \\
PKS 1510$-$089 & (1, 0, 6) & 3.8 (2.4$\sigma$) & (2, 0.19, 5)  & 4.0 (2.4$\sigma$) & 294.99 \\
PG 1553+113 & (3, 0, 0) & 2.8 (2.9$\sigma$) & (0, 0.19, 3) & 1.1 (2.3$\sigma$) & 90.41\\
TXS 1902+556 & (1, 0, 1) & 1.9 (1.7$\sigma$) & (0, 0.59, 0) & 1.1 (1.9$\sigma$) & 251.77\\
MH 2136$-$428 & (0, 0, 7) & 2.0 (2.2$\sigma$) & -- & -- & -- \\
PKS 2155$-$304 & (1,0, 0) & 1.6 (2.9$\sigma$) & (1, 0.19, 0) & 3.9 (2.1$\sigma$) & 215.89 \\
BL Lacertae & (1, 0, 2) & 1.5 (2.7$\sigma$) & (0, 0.19, 1) & 1.6 (2.5$\sigma$) & 94.06 \\
\hline
\hline
\end{tabular}%
}
\end{table*}
\begin{table*}
\centering
\caption{List of periodicity results obtained using the Generalized Lomb–Scargle Periodogram (GLSP), the Phase Dispersion Minimization method (PDM), and the Singular Spectrum Analysis-based methodology (SSA). Periods are reported in years. The table also lists the corresponding test statistics derived from simulations of 150,000 artificial LCs, generated using three different approaches: (i) a simple power-law (PL): (ii) bending power-law power spectral density (BPL), both combined with the method of \citet{emma_lc}, as was presented in $\S$\ref{sec:methodology}; (iii) ARIMA models; and (iv) ARFIMA models for the sources analyzed in \S\ref{sec:arima_arfima} and summarized in Table \ref{tab:arima_arfima}. The order of the test statistics is \textit{(GLSP, PDM, SSA)}. The results of SSA are obtaining by using a window length $WL$ of $0.4$ ($S$\ref{sec:methods_methodology}). The symbol $\dagger$ denotes the PDM results that present the harmonic effect described in \citet{penil_2020, penil_gaps_2025}. \label{tab:periodicity_results}}
{%
\begin{tabular}{c|ccc||cccccc}
\hline
\hline
Name &  GLSP  &  PDM   &  SSA  & PL & BPL & ARIMA & ARFIMA \\
     & [year] & [year] & [year] & (S/N) & (S/N) & (S/N) & (S/N) \\
 \hline
\hline
GB6 J0043+3426 & 2.3$\pm$0.3 & 2.3$\pm$0.3 & 2.2$\pm$0.3 & (0.6$\sigma$, 1.2$\sigma$, 0.5$\sigma$) & (0.7$\sigma$, 0.9$\sigma$, 1.3$\sigma$) & -- & -- \\
S2 0109+22 & 2.7$\pm$0.2 & 1.9$\pm$0.2 & 1.7$\pm$0.2 & (0.8$\sigma$, 1.1$\sigma$, 1.1$\sigma$) & (0.8$\sigma$, 0.9$\sigma$, 0.9$\sigma$) & (1.7$\sigma$, 1.7$\sigma$, 2.3$\sigma$) & (0.8$\sigma$, 1.0$\sigma$, 1.2$\sigma$) \\
OC 457 & 2.0$\pm$0.1 & 1.9$\pm$0.2 & 1.7$\pm$0.4 & (0.7$\sigma$, 1.7$\sigma$, 0.8$\sigma$) & (0.8$\sigma$, 1.6$\sigma$, 0.9$\sigma$) & -- & -- \\
PKS 0208$-$512 & 2.0$\pm$0.3 & 3.9$\pm$0.4 & 3.3$\pm$0.3 & (0.3$\sigma$, 1.0$\sigma$, 0.9$\sigma$) & (0.4$\sigma$, 0.8$\sigma$, 1.0$\sigma$) & (0.8$\sigma$, 1.5$\sigma$, 1.3$\sigma$) & (0.4$\sigma$, 0.7$\sigma$, 0.5$\sigma$) \\
PKS 0215+015 & 3.4$\pm$0.3 & 3.3$\pm$0.3 & 3.3$\pm$0.3 & (2.1$\sigma$, 2.2$\sigma$, 1.7$\sigma$) & (2.0$\sigma$, 2.0$\sigma$, 1.6$\sigma$) & -- & -- \\
3C 66A & 2.2$\pm$0.3 & 1.9$\pm$0.2 & 2.3$\pm$0.2 & (0.5$\sigma$, 0.3$\sigma$, 1.8$\sigma$) & (0.4$\sigma$, 0.4$\sigma$, 1.8$\sigma$) & (1.4$\sigma$, 0.9$\sigma$, 2.4$\sigma$) & (0.6$\sigma$, 0.5$\sigma$, 1.8$\sigma$) \\
PKS 0250$-$225 & 1.2$\pm$0.4 & 1.2$\pm$0.3 & 1.3$\pm$0.4 & (0.4$\sigma$, 1.7$\sigma$, 1.0$\sigma$) & (0.5$\sigma$, 1.2$\sigma$, 0.7$\sigma$) & -- & -- \\
PKS 0301$-$243 & 2.2$\pm$0.3 & $\dagger$4.2$\pm$0.4 & 2.1$\pm$0.2 & (0.8$\sigma$, 0.6$\sigma$, 0.8$\sigma$) & (0.8$\sigma$, 0.6$\sigma$, 0.6$\sigma$) & (1.2$\sigma$, 1.0$\sigma$, 1.9$\sigma$) & (0.6$\sigma$, 0.5$\sigma$, 1.6$\sigma$) \\
PKS 0405$-$385 & 2.6$\pm$0.3 & 2.6$\pm$0.4 & 2.9$\pm$0.4 & (0.9$\sigma$, 0.9$\sigma$, 1.4$\sigma$) & (0.8$\sigma$, 0.7$\sigma$, 1.3$\sigma$) & -- & -- \\
PMN J0427$-$3900 & 2.2$\pm$0.2 & 1.7$\pm$0.2 & 1.0$\pm$0.4 & (0.7$\sigma$, 0.6$\sigma$, 0.8$\sigma$) & -- & -- & -- \\
PKS 0426$-$380 & 3.6$\pm$0.4 & 3.4$\pm$0.4 & 3.5$\pm$0.5 & (1.5$\sigma$, 2.0$\sigma$, 1.2$\sigma$) & (1.1$\sigma$, 1.6$\sigma$, 1.1$\sigma$) & (3.0$\sigma$, 3.1$\sigma$, 2.2$\sigma$) & (1.5$\sigma$, 2.0$\sigma$, 1.3$\sigma$) \\ 
PKS 0447$-$439 & 1.2$\pm$0.3 & 2.4$\pm$0.4 & 4.1$\pm$0.5 & (0.5$\sigma$, 1.1$\sigma$, 0.8$\sigma$) & (0.6$\sigma$, 1.2$\sigma$, 0.8$\sigma$) & (1.4$\sigma$, 1.9$\sigma$, 1.2$\sigma$) & (1.0$\sigma$, 1.6$\sigma$, 0.9$\sigma$) \\ 
PKS 0454$-$234 & 3.4$\pm$0.4 & 3.3$\pm$0,3 & 3.5$\pm$0.5 & (2.4$\sigma$, 1.9$\sigma$, 0.9$\sigma$) & (1.2$\sigma$, 1.4$\sigma$, 1.1$\sigma$) & (3.2$\sigma$, 2.6$\sigma$, 1.6$\sigma$) & (2.7$\sigma$, 2.0$\sigma$, 1.2$\sigma$) \\
S3 0458$-$02 & 4.0$\pm$0.4 & 3.9$\pm$0.4 & 4.2$\pm$0.2 & (2.0$\sigma$, 2.0$\sigma$, 1.4$\sigma$) & (2.0$\sigma$, 2.1$\sigma$, 1.5$\sigma$) & -- & -- \\        
TXS 0518+211 & 3.1$\pm$0.3 & 3.2$\pm$0.3 & 3.1$\pm$0.3 & (1.9$\sigma$, 2.0$\sigma$, 1.8$\sigma$) & (1.7$\sigma$, 1.8$\sigma$, 1.7$\sigma$) & (2.8$\sigma$, 2.5$\sigma$, 2.4$\sigma$) & (2.0$\sigma$, 2.2$\sigma$, 1.8$\sigma$) \\
PKS 0524$-$485 & 1.9$\pm$0.2 & $\dagger$3.9$\pm$0.4 & 2.1$\pm$0.2 & (0.6$\sigma$, 2.7$\sigma$, 1.0$\sigma$) & (0.7$\sigma$, 2.0$\sigma$, 0.8$\sigma$) & -- & -- \\
PMN J0533$-$5549 & 2.1$\pm$0.2 & 2.1$\pm$0.3 & 4.0$\pm$0.4 & (0.4$\sigma$, 0.9$\sigma$, 0.3$\sigma$) & (0.6$\sigma$, 0.5$\sigma$, 0.4$\sigma$) & -- & --  \\
B2 0716+33 & 2.1$\pm$0.3 & 2.2$\pm$0.3 & 2.1$\pm$0.3 & (0.4$\sigma$, 0.7$\sigma$, 0.4$\sigma$) & (0.5$\sigma$, 1.0$\sigma$, 0.5$\sigma$) & -- & -- \\
S5 0716+714 & 2.5$\pm$0.1 & 2.5$\pm$0.2 & 2.7$\pm$0.3 & (0.6$\sigma$, 0.9$\sigma$, 1.2$\sigma$) & (0.7$\sigma$, 0.8$\sigma$, 1.1$\sigma$) & (1.7$\sigma$, 1.8$\sigma$, 2.1$\sigma$) & (1.6$\sigma$, 1.6$\sigma$, 2.0$\sigma$) \\
OJ 014 & 4.2$\pm$0.4 & 4.2$\pm$0.4 & 4.2$\pm$0.4 & (2.6$\sigma$, 2.5$\sigma$, 2.2$\sigma$) & (2.5$\sigma$, 2.4$\sigma$, 1.8$\sigma$) & -- & -- \\
PMN J0948+0022$\dagger$ & 1.2$\pm$0.5 & 2.4$\pm$0.5 & 1.4$\pm$0.2 & (0.5$\sigma$, 0.6$\sigma$, 1.0$\sigma$) & (0.6$\sigma$, 0.5$\sigma$, 1.1$\sigma$) & -- & --\\
S4 1030+41 & 2.2$\pm$0.3 & 2.4$\pm$0.3 & 2.4$\pm$0.3 & (1.7$\sigma$, 1.9$\sigma$, 1.5$\sigma$) & (1.9$\sigma$, 2.0$\sigma$, 1.6$\sigma$) & -- & -- \\
S4 1030+61 & 4.1$\pm$0.4 & 4.3$\pm$0.5 & 3.8$\pm$0.4 & (2.1$\sigma$, 1.8$\sigma$, 1.4$\sigma$) & (1.5$\sigma$, 1.0$\sigma$, 1.3$\sigma$) & -- & -- \\
S5 1039+81 & 3.4$\pm$0.3 & 2.0$\pm$0.4 & 1.2$\pm$0.5 & (0.4$\sigma$, 1.1$\sigma$, 0.5$\sigma$) & (0.6$\sigma$, 1.2$\sigma$, 0.6$\sigma$) & -- & -- \\
S5 1044+71 & 3.1$\pm$0.3 & 3.2$\pm$0.3 & 3.1$\pm$0.3 & (4.0$\sigma$, 4.0$\sigma$, 2.2$\sigma$) & (3.6$\sigma$, 4.0$\sigma$, 2.0$\sigma$) & -- & -- \\
S4 1144+40 & 3.3$\pm$0.3 & 3.2$\pm$0.3 & 3.2$\pm$0.3 & (2.8$\sigma$, 3.9$\sigma$, 1.9$\sigma$) & (1.8$\sigma$, 2.4$\sigma$, 1.7$\sigma$) & -- & -- \\
4C +04.42 & 2.6$\pm$0.5 & 2.5$\pm$0.6 & 2.1$\pm$0.2 & (1.2$\sigma$, 1.6$\sigma$, 0.6$\sigma$) & (0.7$\sigma$, 0.8$\sigma$, 1.3$\sigma$) & -- & -- \\
S5 1221+80 & 2.7$\pm$0.3 & 2.6$\pm$0.4 & 2.6$\pm$0.5 & (1.2$\sigma$, 1.8$\sigma$, 0.8$\sigma$) & (1.2$\sigma$, 1.6$\sigma$, 0.7$\sigma$) & -- & -- \\
S4 1250+53 & 2.2$\pm$0.1 & 2.2$\pm$0.2 & 2.3$\pm$0.3 & (1.5$\sigma$, 0.9$\sigma$, 1.6$\sigma$) & (1.4$\sigma$, 1.1$\sigma$, 1.3$\sigma$) & (2.0$\sigma$, 1.6$\sigma$, 2.7$\sigma$) & (1.3$\sigma$, 1.1$\sigma$, 1.9$\sigma$)\\ 
OP 313 & 1.4$\pm$0.3 & 5.4$\pm$0.6 & 1.3$\pm$0.3 & (0.3$\sigma$, 0.9$\sigma$, 0.4$\sigma$) & (0.4$\sigma$, 0.8$\sigma$, 0.3$\sigma$) & -- & -- \\
PKS B1310$-$041 & 2.3$\pm$0.1 & 4.5$\pm$0.4 & 4.6$\pm$0.6 & (1.4$\sigma$, 3.0$\sigma$, 1.3$\sigma$) & (1.0$\sigma$, 3.0$\sigma$, 1.4$\sigma$) & -- & -- \\
TXS 1318+225 & 5.1$\pm$0.3 & 5.0$\pm$0.4 & 4.9$\pm$0.3 & (1.5$\sigma$, 2.5$\sigma$, 1.4$\sigma$) & (1.1$\sigma$, 2.6$\sigma$, 1.2$\sigma$) & -- & -- \\
PKS 1424$-$328 & 1.5$\pm$0.1 & 1.5$\pm$0.1 & 1.5$\pm$0.1 & (1.7$\sigma$, 2.0$\sigma$, 1.2$\sigma$) & (1.2$\sigma$, 1.1$\sigma$, 1.5$\sigma$) & (2.0$\sigma$, 1.7$\sigma$, 1.9$\sigma$) & (1.5$\sigma$, 1.4$\sigma$, 1.6$\sigma$)\\
TXS 1452+516 & 2.2$\pm$0.5 & 2.1$\pm$0.4 & 5.0$\pm$0.4 & (0.6$\sigma$, 0.5$\sigma$, 0.4$\sigma$) & (0.4$\sigma$, 0.3$\sigma$, 0.2$\sigma$) & (0.9$\sigma$, 1.0$\sigma$, 0.7$\sigma$) & (0.3$\sigma$, 0.5$\sigma$, 0.2$\sigma$)\\
PKS 1510$-$089 & 1.2$\pm$0.2 & 3.8$\pm$0.4 & 3.4$\pm$0.3 & (0.4$\sigma$, 1.0$\sigma$, 1.4$\sigma$) & (0.3$\sigma$, 0.8$\sigma$, 0.3$\sigma$) & (0.5$\sigma$, 1.4$\sigma$, 1.0$\sigma$) & (0.4$\sigma$, 1.3$\sigma$, 0.9$\sigma$)\\
B2 1520+31 & 1.2$\pm$0.1 & 3.7$\pm$0.4 & 1.8$\pm$0.1 & (1.0$\sigma$, 1.0$\sigma$, 1.5$\sigma$) & (0.3$\sigma$, 1.6$\sigma$, 1.0$\sigma$) & -- & -- \\
TXS 1530$-$131 & 1.4$\pm$0.1 & 2.8$\pm$0.3 & 2.5$\pm$0.3 & (0.3$\sigma$, 3.2$\sigma$, 0.6$\sigma$) & (0.3$\sigma$, 2.9$\sigma$, 0.5$\sigma$) & -- & -- \\ 
PG 1553+113 & 2.1$\pm$0.1 & 2.1$\pm$0.1 & 2.1$\pm$0.1 & (3.7$\sigma$, 4.1$\sigma$, 2.0$\sigma$) & (4.1$\sigma$, 4.2$\sigma$, 1.9$\sigma$) & (3.4$\sigma$, 3.3$\sigma$, 3.1$\sigma$) & (4.0$\sigma$, 4.0$\sigma$, 2.7$\sigma$)\\
87GB 164812.2+524023 & 3.2$\pm$0.4 & 3.1$\pm$0.3 & 3.3$\pm$0.6 & (1.3$\sigma$, 2.3$\sigma$, 2.0$\sigma$) & -- & -- & -- \\
4C +48.41 & 5.4$\pm$0.4 & 4.9$\pm$0.4 & 5.1$\pm$0.7 & (0.4$\sigma$, 1.7$\sigma$, 1.1$\sigma$) & (0.5$\sigma$, 2.0$\sigma$, 1.0$\sigma$) & -- & -- \\
PKS 1716$-$771 & 2.2$\pm$0.2 & 2.5$\pm$0.3 & 2.2$\pm$0.2 & (1.0$\sigma$, 2.0$\sigma$, 1.5$\sigma$) & (0.8$\sigma$, 1.8$\sigma$, 1.3$\sigma$) & -- & -- \\
TXS 1902+556 & 3.2$\pm$0.3 & 3.2$\pm$0.3 & 3.1$\pm$0.3 & (1.6$\sigma$, 1.7$\sigma$, 1.9$\sigma$) & (1.2$\sigma$, 1.2$\sigma$, 1.1$\sigma$) & -- & -- \\
PKS 1903$-$80 & 3.5$\pm$0.3 & 4.2$\pm$0.4 & 4.0$\pm$0.6 & (1.2$\sigma$, 1.5$\sigma$, 1.1$\sigma$) & (0.3$\sigma$, 0.8$\sigma$, 0.9$\sigma$) & (1.4$\sigma$, 1.6$\sigma$, 2.0$\sigma$) & (0.5$\sigma$, 1.0$\sigma$, 1.1$\sigma$)\\
7C 2010+4619 & 2.6$\pm$0.2 & 2.7$\pm$0.3 & 3.7$\pm$0.7 & (0.7$\sigma$, 1.1$\sigma$, 0.5$\sigma$) & -- & -- & -- \\
MH 2136$-$428 & 3.5$\pm$0.2 & 3.6$\pm$0.3 & 1.9$\pm$0.1 & (0.6$\sigma$, 1.2$\sigma$, 1,1$\sigma$) & (0.6$\sigma$, 0.9$\sigma$, 0.6$\sigma$) & (1.6$\sigma$, 1.9$\sigma$, 1.7$\sigma$) & -- \\
PKS 2155$-$304 & 1.7$\pm$0.1 & 1.7$\pm$0.1 & 1.6$\pm$0.1 & (1.7$\sigma$, 1.7$\sigma$, 1.1$\sigma$) & (1.6$\sigma$, 1.8$\sigma$, 1.0$\sigma$) & (2.0$\sigma$, 1.8$\sigma$, 2.1$\sigma$) & (1.5$\sigma$, 1.6$\sigma$, 1.6$\sigma$)\\
PKS 2155$-$83 & 4.7$\pm$0.5 & 2.7$\pm$0.3 & 4.5$\pm$0.4 & (1.0$\sigma$, 0.5$\sigma$, 1.6$\sigma$) & (0.8$\sigma$, 0.5$\sigma$, 1.4$\sigma$) & -- & -- \\	
BL Lacertae & 5.7$\pm$0.3 & 5.6$\pm$0.5 & 5.4$\pm$0.4 & (0.8$\sigma$, 0.8$\sigma$, 0.8$\sigma$) & (0.9$\sigma$, 0.8$\sigma$, 0.7$\sigma$) & (2.4$\sigma$, 2.0$\sigma$, 1.6$\sigma$) & (2.5$\sigma$, 2.2$\sigma$, 0.9$\sigma$)\\ 
B2 2234+28A & 1.4$\pm$0.3 & 2.9$\pm$0.4 & 2.2$\pm$0.1 & (0.3$\sigma$, 1.0$\sigma$, 0.6$\sigma$) & (0.4$\sigma$, 0.5$\sigma$, 0.7$\sigma$) & -- & --\\   
\hline
\hline
\end{tabular}%
}
\end{table*}
\begin{table*}
\centering
\caption{Test statistics of applying the approach SSA+LSP, considering different window length (WL) for the SSA oscillatory decomposition, and for different models for the noise: simple power-law (PL), and bending power-law (BPL). The result shows the impact of the WL selected for inferring the test statistics associated to the period of S5 1044+71. The $WL$ in bold is the ${\rm WL}_{\rm opt}$, which is the optimal $WL$ according to the properties of the LC and the period of the signal.\label{tab:ssa_wl_s51044}}
{%
\begin{tabular}{c|cccccccccc}
\hline
\hline
WL & PL & BPL  \\
   & [S/N] & [S/N] \\ 
\hline
\hline
0.09 & 2.2$\sigma$ & 1.9$\sigma$  \\
0.11 & 2.6$\sigma$ & 2.2$\sigma$  \\
0.13 & 2.7$\sigma$ & 2.3$\sigma$  \\
0.15 & 2.5$\sigma$ & 2.3$\sigma$  \\
\textbf{0.17} & 2.3$\sigma$ & 2.1$\sigma$  \\
0.21 & 2.1$\sigma$ & 1.8$\sigma$  \\
0.24 & 1.9$\sigma$ & 1.7$\sigma$  \\
0.27 & 2.0$\sigma$ & 1.8$\sigma$  \\
0.30 & 2.3$\sigma$ & 2.0$\sigma$  \\
\textbf{0.34} & 2.4$\sigma$ & 2.2$\sigma$  \\
0.36 & 2.3$\sigma$ & 2.1$\sigma$  \\
0.39 & 2.2$\sigma$ & 2.0$\sigma$  \\
0.40 & 2.2$\sigma$ & 2.0$\sigma$  \\
0.43 & 2.1$\sigma$ & 1.9$\sigma$  \\
0.46 & 2.3$\sigma$ & 2.1$\sigma$  \\
0.49 & 2.5$\sigma$ & 2.3$\sigma$  \\
\hline
\hline
\end{tabular}%
}
\end{table*}
\begin{table*}
\centering
\caption{Test statistics of applying the approach SSA+LSP, considering different window length (WL) for the SSA oscillatory decomposition, and for different models for the noise: simple power-law (PL), bending power-law (BPL), ARIMA, and ARFIMA. The result shows the impact of the WL selected for inferring the test statistics associated to the period of PKS 2155$-$304. The $WL$ in bold is the ${\rm WL}_{\rm opt}$, which is the optimal $WL$ according to the properties of the LC and the period of the signal. \label{tab:ssa_pks2155}}
{%
\begin{tabular}{c|cccccccccc}
\hline
\hline
WL & PL & BPL & ARIMA & ARFIMA  \\
   & [S/N] & [S/N] & [S/N] & [S/N] \\ 
\hline
\hline
\textbf{0.09} & 1.3$\sigma$ & 1.2$\sigma$ & 2.2$\sigma$ & 1.7$\sigma$  \\
0.11 & 1.2$\sigma$ & 1.1$\sigma$ & 2.2$\sigma$ & 1.7$\sigma$ \\
0.13 & 1.1$\sigma$ & 1.0$\sigma$ & 2.2$\sigma$ & 1.6$\sigma$ \\
0.15 & 1.1$\sigma$ & 1.0$\sigma$ & 2.2$\sigma$ & 1.6$\sigma$ \\
\textbf{0.18} & 0.9$\sigma$ & 0.8$\sigma$ & 2.0$\sigma$ & 1.5$\sigma$ \\
0.21 & 0.8$\sigma$ & 0.7$\sigma$ & 1.9$\sigma$ & 1.4$\sigma$ \\
0.24 & 0.9$\sigma$ & 0.8$\sigma$ & 2.0$\sigma$ & 1.5$\sigma$ \\
0.27 & 0.8$\sigma$ & 0.7$\sigma$ & 1.9$\sigma$ & 1.4$\sigma$ \\
0.30 & 0.6$\sigma$ & 0.5$\sigma$ & 1.6$\sigma$ & 1.2$\sigma$ \\
0.33 & 0.9$\sigma$ & 0.7$\sigma$ & 2.0$\sigma$ & 1.4$\sigma$ \\
0.36 & 1.0$\sigma$ & 0.9$\sigma$ & 2.1$\sigma$ & 1.6$\sigma$ \\
0.39 & 1.1$\sigma$ & 0.9$\sigma$ & 2.1$\sigma$ & 1.6$\sigma$ \\
0.40 & 1.1$\sigma$ & 1.0$\sigma$ & 2.1$\sigma$ & 1.6$\sigma$ \\
0.43 & 1.1$\sigma$ & 0.9$\sigma$ & 2.2$\sigma$ & 1.7$\sigma$ \\
0.46 & 1.0$\sigma$ & 0.9$\sigma$ & 2.1$\sigma$ & 1.6$\sigma$ \\
0.49 & 1.0$\sigma$ & 0.9$\sigma$ & 2.1$\sigma$ & 1.6$\sigma$ \\
\hline
\hline
\end{tabular}%
}
\end{table*}
\begin{table*}
\centering
\caption{Test statistics of applying the approach SSA+LSP, considering different window length (WL) for the SSA oscillatory decomposition, and for different models for the noise: simple power-law (PL), and bending power-law (BPL). The result shows the impact of the WL selected for inferring the test statistics associated to the period of OJ 014. The $WL$ in bold is the ${\rm WL}_{\rm opt}$, which is the optimal $WL$ according to the properties of the LC and the period of the signal. \label{tab:ssa_wl_oj014}}
{%
\begin{tabular}{c|cccccccccc}
\hline
\hline
WL & PL & BPL  \\
   & [S/N] & [S/N] \\ 
\hline
\hline
0.09 & 0.4$\sigma$ & 0.8$\sigma$  \\
0.11 & 1.5$\sigma$ & 1.5$\sigma$  \\
0.13 & 2.0$\sigma$ & 1.7$\sigma$  \\
0.15 & 2.2$\sigma$ & 1.8$\sigma$  \\
0.18 & 2.2$\sigma$ & 1.8$\sigma$  \\
0.21 & 2.3$\sigma$ & 1.8$\sigma$  \\
\textbf{0.24} & 2.3$\sigma$ & 1.8$\sigma$  \\
0.27 & 2.2$\sigma$ & 1.7$\sigma$  \\
0.30 & 2.0$\sigma$ & 1.6$\sigma$  \\
0.33 & 2.2$\sigma$ & 1.7$\sigma$  \\
0.36 & 2.3$\sigma$ & 1.8$\sigma$  \\
0.39 & 2.3$\sigma$ & 1.8$\sigma$  \\
0.40 & 2.2$\sigma$ & 1.7$\sigma$  \\
0.43 & 1.9$\sigma$ & 1.4$\sigma$  \\
0.46 & 1.7$\sigma$ & 1.2$\sigma$  \\
\textbf{0.48} & 1.6$\sigma$ & 1.2$\sigma$  \\
\hline
\hline
\end{tabular}%
}
\end{table*}
\begin{table*}
\centering
\caption{Evaluation of previously published prediction. The column “Previous Predictions” lists the number of peaks forecast in \citet{alba_ssa,penil_2025_4fgl} that were expected to occur within the time span of the LC analyzed here. “Evaluation” indicates whether those forecasts are supported by the newly available data, denoted by \redcheck or they do not (\redxmark).\label{tab:predictions}}
{%
\begin{tabular}{c|cccccccccc}
\hline
\hline
Name & Previous Predictions & Evaluation \\
                 &             &            \\ 
\hline
\hline
GB6 J0043+3426 & (2021-04, 2023-07, 2025-09) & (\redxmark, \redxmark, \redxmark)  \\
S2 0109+22 & (2022-08, 2025-03) & (\redxmark, \redcheck) \\
3C 66A & (2022-01, 2024-05) & (\redxmark, \redxmark)  \\
PKS 0405$-$385 & (2022-03, 2025-04) & (\redxmark, \redxmark) \\
PKS 0447$-$439 & (2022-09, 2024-08) &(\redxmark, \redcheck) \\ 
PKS 0454$-$234 & (2024-01) & (\redcheck) \\
PKS 0524$-$485 & (2022-10, 2024-11) &(\redcheck, \redxmark) \\
S5 0716+714 & (2023-09, 2025-09) & (\redxmark, \redcheck) \\
OJ 014 & (2024-08) & (\redcheck) \\
S4 1030+41 & (2021-01, 2023-04, 2025-07) &(\redxmark, \redxmark, \redxmark) \\
S4 1030+61 & (2025-04) & (\redxmark) \\
S5 1039+81 & (2025-07) & (\redxmark) \\
S5 1044+71 & (2023-01) & (\redcheck) \\
S4 1144+40 & (2021-05, 2024-10) & (\redcheck, \redcheck) \\
4C +04.42 & (2023-02, 2025-04) & (\redcheck, \redxmark) \\
S5 1221+80 & (2023-05, 2025-12) &(\redcheck, \redxmark) \\
S4 1250+53 & (2021-04, 2023-07, 2025-11) & (\redxmark, \redxmark, \redxmark) \\ 
OP 313 & (2025-09) &(\redcheck) \\
TXS 1452+516 & (2022-10, 2025-03) & (\redxmark, \redxmark) \\
PKS 1510$-$089 & (2021-12, 2023-07, 2025-05) & (\redxmark, \redxmark, \redxmark) \\
TXS 1530$-$131 & (2021-03, 2022-08-29, 2024-01, 2025-06) & (\redxmark, \redxmark, \redxmark, \redxmark) \\ 
PG 1553+113 & (2021-03, 2023-05, 2025-06) & (\redcheck, \redcheck, \redcheck) \\
87GB 164812.2+524023 & (2025-08) & (\redxmark) \\
PKS 1716$-$771 & (2023-02, 2025-09) &(\redxmark, \redxmark)  \\
PKS 1903$-$80 & (2022-08, 2025-09) & (\redxmark, \redxmark) \\
MH 2136$-$428 & (2021-08, 2023-07, 2025-06) & (\redxmark, \redcheck, \redxmark)  \\
PKS 2155$-$304 & (2022-08, 2024-05) & (\redxmark, \redcheck) \\
PKS 2155$-$83 & (2024-05) & (\redxmark) \\	
BL Lacertae & (2022-07) & (\redxmark) \\ 
B2 2234+28A & (2021-02, 2024-03) & (\redcheck, \redcheck) \\
\hline
\hline
\end{tabular}%
}
\end{table*}

% Don't change these lines
\bsp	% typesetting comment
\label{lastpage}
\end{document}